\documentclass[article,11pt]{amsart}
\usepackage{xr}
\usepackage{amsaddr}
\usepackage{amssymb, amsthm, amsfonts, amsgen, stmaryrd}
\usepackage{slashed}
\usepackage[english]{babel}
\usepackage{fullpage}
\usepackage[centertags]{amsmath}
\usepackage{indentfirst}
\usepackage{fancyhdr}
\usepackage[dvips]{graphicx}
\usepackage{psfrag}
\usepackage{enumerate}
\numberwithin{equation}{section}
\usepackage[latin1]{inputenc}

\usepackage{hyperref}
\hypersetup{
pdftitle={},%
pdfauthor={},%
pdfsubject={},%
pdfkeywords={},%
colorlinks=true,%
linkcolor=blue,%
citecolor=red,%
linktocpage=true,%
hyperfootnotes=true,%
pageanchor=true
}
\usepackage{color}

\newcommand{\blu}[1]{ \textcolor{blue}{{#1}}}

\title{Symmetries and reduction \\ Part II -- Lagrangian and Hamilton-Jacobi picture} 
\date{11 august 2021}
\author{Giuseppe Marmo} 
\address{Dipartimento di Fisica ``E. Pancini'', Universit\`a di Napoli Federico II; \\
INFN - Sezione di Napoli \\ Via Cintia - 80126 Napoli, Italy}
\email{marmo@na.infn.it}
\author{Luca Schiavone}
 \address{Dipartimento di Matematica ed Applicazioni ``R. Caccioppoli'', Universit\`a di Napoli Federico II \\ 
Via Cintia - 80126 Napoli, Italy; \\ Departamento de Matem\'aticas, Univ. Carlos III de Madrid \\ Av.da de la Universidad 30 -- 28911 Legan\'es, Madrid, Spain}
\email{luca.schiavone@unina.it}
\author{Alessandro Zampini}
 \address{Dipartimento di Matematica ed Applicazioni ``R. Caccioppoli'', Universit\`a di Napoli Federico II; \\ 
 INFN - Sezione di Napoli \\ Via Cintia - 80126 Napoli, Italy } 
 \email{alessandro.zampini@unina.it}

\newtheorem{theo}{Theorem}[section]
\newtheorem{lemm}[theo]{Lemma}
\newtheorem{prop}[theo]{Proposition}

\newtheorem{rema}[theo]{Remark}
\newtheorem{example}{Example}[section]

\newcommand{\nn}{\nonumber}

\newcommand{\dd}{{\rm d}}

\newcommand{\cL}{\mathcal{L}}

\newcommand{\tL}{\theta_{\mathcal L}}
\newcommand{\oL}{\omega_{\mathcal L}}
\newcommand{\EL}{E_{\cL}}
\newcommand{\cH}{\mathcal H}

\newcommand{\figureheight}{8cm}
\newcommand{\putfig}[2]{\begin{figure}[htp]
        \special{isoscale c:/itex/texfig/#1.wmf, \the\hsize \figureheight}
        \vspace{\figureheight}
        \caption{#2}\label{fig:#1}
        \end{figure}}
\newcommand{\pictureheight}{4cm}
\newcommand{\putpicture}[2]{\begin{figure}[htp]
        \special{isoscale c:/itex/texfig/#1.wmf, \the\hsize \pictureheight}
        \vspace{\pictureheight}
        \caption{#2}\label{fig:#1}
        \end{figure}}

\newcommand{\beqa}{\begin{eqnarray}}
\newcommand{\eeqa}{\end{eqnarray}}
\newcommand{\beq}{\begin{equation}}
\newcommand{\eeq}{\end{equation}}

\newcommand{\del}{\partial}

\newcommand{\R}{{\mathbb{R}}}

\newcommand{\C}{\mathbb{C}}

\newcommand{\D}{\mathcal{D}}

\newcommand{\A}{\mathcal{A}}
\newcommand{\E}{\mathcal{E}}

\newcommand{\mD}{\mathfrak D}
\newcommand{\FL}{\Phi_{\cL}}
\newcommand{\cV}{\mathcal V}
\newcommand{\gh}{\mathrm g_h}

\begin{document}

\thispagestyle{empty}

\begin{abstract}
Following the analysis we have presented  in a previous paper (that we refer to as [I]), we describe a Noether theorem related to symmetries, with the associated reduction procedures, for classical dynamics within the Lagrangian and the Hamilton-Jacobi formalism.  
\end{abstract}


\maketitle
\tableofcontents

\section{Introduction}
\label{2-intro}
In a previous paper (i.e. \cite{Uno}, hereafter referred to as [I]) we have dealt with the notion of symmetry and reduction for classical dynamical systems described, as an appropriate  limit of quantum dynamical systems,  within the Poisson and the Hamiltonian formulation. In this paper we aim to deal  with the Lagrangian formulation for the dynamics of a classical (point, i.e. test particles, we do not intend to analyse field theories) system, as well as with the Hamilton-Jacobi (HJ) formalism for a symplectic dynamics on a cotangent bundle manifold.

We begin by noticing  that the Lagrangian formalism is not equivalent to the Hamiltonian (or the Newtonian) one. In order to illustrate this point, we recall that, if a point particle system evolves with the time $t$ on the position space $\R^N$ (whose points are labelled by a  coordinate system given by $\{x^a\}_{a=1,\dots,N}$) the Euler-Lagrange equations of the motions are 
\begin{align}
\frac{\dd x^a}{\dd t}&=\,v^a \label{30.2} \\ \frac{\dd}{\dd t}\left(\frac{\del\cL}{\del v^a}\right)&=\,\frac{\del \cL}{\del x^a},
\label{ll1}
\end{align}
where $\{v^a\}_{a=1,\dots,N}$ denote  the velocity coordinates and $\cL=\cL(x,v,t)$ provides a Lagrangian function for the system. Such equations give the stationary points (for a suitable class of variations) of the action functional 
\beq
\label{ll2}
S\,=\,\int_{t_0}^{t_1}\dd t\,\cL.
\eeq
While both in the Poisson and Hamiltonian formalism the dynamical datum is given by a vector field on a smooth manifold, i.e. an \emph{explicit} system of (in general) first order ordinary differential equations, the Euler-Lagrange  equations of the motions are \emph{implicit} and of \emph{second order}. The second order condition is clearly given by \eqref{30.2}.  If we indeed expand \eqref{ll1}, we get the implicit equation
\beq
\label{30.1}
 H_{ks}a^k\,=\,\left(\frac{\del \cL}{\del x^s}\,-\,\frac{\del^2\cL}{\del v^s\del x^k}\,v^k\right)
\eeq
  for the acceleration $a^j=\dd v^j/\dd t$, with $$H_{ks}=\frac{\del^2\cL}{\del v^k\del v^s}.$$  When the Lagrangian is regular, i.e.  when the matrix $H_{ks}$ is invertible, accelerations can be solved for, and we recover the Newton equations of the motion. When the matrix $H_{ks}$ is not invertible, contracting the relations in   \eqref{30.1} with null eigenvectors of $H_{ks}$  provides local  relations  not involving accelerations.  These relations provide a set of constraints on the allowed Cauchy data. 

The second order character of the equations has a geometrical description  on a carrier space which is the second order jet bundle over a configuration space $Q$.   When $\cL$ is regular, the \eqref{ll1} can be written in terms of a unique vector field which turns to have a symplectic description on $TQ$ via  a symplectic 2-form suitably defined by $\cL$. When  the Lagrangian function does not satisfy such a condition, the problem of formulating the equations in terms of  vector fields can be studied\footnote{This was elaborated by P.A.M. Dirac and P.G. Bergman when studying the quantization of the electromagnetic and the gravitational field. Nowadays this approach is more generally adopted to analyse quantization of gauge theories.}, provided the Lagrangian satisfies a weak regularity condition,  by further  developing the presymplectic formalism  introduced in [I].

That the Euler-Lagrange equations are implicit reflects also in another crucial difference with respect to the Poisson and to the Hamiltonian equations. In the latter case, 
when the vector field is complete, the evolution is described by a flow, i.e. a one parameter group of transformations\footnote{This reflects in  the problem of considering self-adjoint extensions of symmetric operators  in quantum mechanics, where the evolution is always described in terms of a one-parameter group of unitary transformations.}.
Even when the Euler-Lagrange equations of the motions are integrable (i.e. solutions exist), it is in general not possible to claim that the evolution is described in terms of a one-parameter group of transformations.

For these reasons, we begin our analysis by describing a geometric setting for implicit first and second order ordinary differential equations, with a suitable definition of symmetries and constants of the motion. 
We shall not review a complete theory, but focus on implicit equations of Lagrangian type. 
After an intrinsic characterization of a tangent bundle manifold in terms of a partial linear structure and of a soldering tensor, we analyse first the 
Lagrangian formalism in the regular case, with a 
 Noether theorem for Newtonian and Newtonoid symmetries. Specific examples of reduction driven by such symmetries show the analogies and the differences with respect to the reduction within the symplectic scheme as described in [I]. We then move
to  the equations associated to a singular Lagrangian only when they can written in terms of a presymplectic structure. When describing such formalism, we shall compare it with the Dirac-Bergmann theory of constrained systems, analysing the notion of gauge symmetry and the associated reduction. 

Coherently with the approach taken in this paper, we start the last section by describing how the Hamilton-Jacobi equation on a cotangent bundle manifold $T^*Q$ comes as a suitable semi-classical approximation of the quantum mechanical Schr\"odinger equation on the configuration manifold $Q$, and how the relations between differential operators on $T^*Q$ and symbols allow to cast in a geometric setting the connections between a class of non linear p.d.e.'s and a suitable class of Hamiltonian vector fields. Analysing the properties of Lagrangian submanifolds of a cotangent bundle we describe a Noether theorem for the Hamilton-Jacobi 
and for the generalised Hamilton-Jacobi problem, and then show how the usual method of solution by separation of variables can be read as an example of symplectic reduction.


\section{Implicit equations of Lagrangian type}
\label{sec:boh}
Following our introduction, in order to analyse symmetries and conservation laws for dynamics described by the Euler-Lagrange equations with singular Lagrangian we describe a geometric setting for implicit ordinary differential equations, which allows to define a notion of symmetry and constants of the motions which reduce to the standard one when such implicit equations can be formulated as explicit. Our exposition closely follows \cite{bg2, bgm97, bgm99,  mmt92, mmt95, mmt97, me-tu78}.

\subsection{A geometric setting for first order implicit differential equations}
\label{subsec:imp}
When $M$ is a smooth $N$-dimensional manifold,  
its tangent bundle $\pi:TM\to M$ is the vector bundle whose fiber at each point $m\in M$ can be suitably identified with the $N$-dimensional real linear space $T_mM$ of vectors which are tangent to curves through $m$. If $\{x^a\}_{a=1,\dots,N}$ is a local chart for $U\subset M$, then $\{x^a,v^a\}_{a=1,\dots,N}$ is a local chart for the fiber product $U\times \R^N$ which locally trivializes $TM$.

We have repeatedly considered that a vector field $\Gamma\in\mathfrak X(M)$ on a $n$-dimensional manifold $M$ defines a first order o.d.e., which we locally write  as 
\beq
\label{fodee}
\dot{x}^a\,=\,\Gamma^a(x),
\eeq
if the local expression for the vector field is $\Gamma\,=\,\Gamma^a\del_a$. A vector field is indeed a section of the tangent bundle manifold on $M$, i.e. $\Gamma\,:\,M\,\to\,TM$, whose graph is the submanifold in $TM$  which can be represented as the zero level set $\mathfrak Z$ given by 
\beq
\psi^a\,=\,v^a\,-\,\Gamma^a(x)\,=\,0
\label{zerols}
\eeq
with $\psi^a\in\mathcal F(TQ)$. 
It is natural then to define a first order (ordinary) differential equation (i.e. a f.o.d.e.) on a differentiable manifold $M$ as a subset $\mathfrak Z$ in $TM$, usually assumed to be 
a submanifold. If there exists a vector field $\Gamma$ on $M$ such that  submanifold $\mathfrak Z$ is the graph ${\rm Im}(\Gamma)$, then the f.o.d.e. is called \emph{explicit}. If there is no vector field on $M$ whose graph coincides with $\mathfrak Z$, then the f.o.d.e. is called \emph{implicit}. 

Such a definition is \emph{local}. The condition that $\mathfrak Z$ is a global manifold embedded in $TM$ can be relaxed, and corresponds to define $\mathfrak Z\subset TM$ as the zero level set given by 
$$
\psi^a=0
$$
with $\{\psi^a\}_{a=1,\dots,k}\in\mathcal F(TM)$ such that the rank of the k-form $\dd\psi^1\wedge\ldots\wedge\dd\psi^k$ is not necessarily invariant, i.e. it may depend on the point 	in $M$.
\begin{example}
\label{ex-25.2}
Consider for example  $M=\R^2$, with $TM=\R^4$ on which a global coordinate system is $\{x^a,v^a\}_{a=1,2}$. The condition $$(v^1)^2+(v^2)^2-f(x^1,x^2)=0$$ defines, if $\dd f\neq 0$,  a submanifold $\mathfrak Z\hookrightarrow TM$. Since $\mathfrak Z$ is 3-dimensional, it can not represent the graph of any vector field on $\R^2$, so it represents an implicit first order differential equation. If we write $v^1=\pm\sqrt{\alpha}$ and $v^2\,=\,\pm\sqrt{f(x)-\alpha}$ we see that $\mathfrak Z$ can be locally represented as a family of vector fields depending on a function $\alpha$ fulfilling  specific conditions. 
\end{example}

A function $$\gamma\,:\,I\subseteq\,\R\,\to\,M$$ is called a \emph{solution}, or an \emph{integral curve} for the given f.o.d.e. if $T\gamma\,\in\,\mathfrak Z$ for any $t\in \,I$. This condition has a local version. The curve $\gamma\,:\,I\subseteq\,\R\,\to\,M$ is a solution if 
\beq
\label{25.3}
(x=\gamma(t), v=\dot{\gamma}(t))
\eeq
is an element in $\mathfrak Z$ for any $t\in I$. 
A f.o.d.e. $\mathfrak Z\subset TM$ is called \emph{integrable} if, for any $z\in\mathfrak Z$, there exists a solution $\gamma\,:\,I\,\to\,M$ such that $T\gamma(t)=z$ for a given $t\in I$.

\begin{example}
\label{ex-25.2}Consider the manifold  $M=\R^2$  so that $TM=\R^4$ as above. It is immediate to see that the f.o.d.e.
$$
\mathfrak Z\,=\,\{(x^1,x^2=0,v^1=0,v^2=\alpha)\}.
$$
 is not integrable for $\alpha\neq0$, while the equation defined  on $M={\rm S}^1$, with $TM={\rm S^1}\times\R$, as the image $\mathfrak Z$ of the map $\epsilon\,:\,\R\,\to\,T{\rm S}^1$ given by 
 $$
\alpha\,\mapsto\,(\cos\alpha, \sin\alpha,\alpha)
$$ 
is integrable. Solutions are given by motion in a circle ${\rm S}^1$ with constant acceleration. For any $\alpha\in\rm S^1$ there is an infinity of solutions $\gamma$ such that $\gamma(0)=\alpha$, since $\mathfrak Z$ is an elix, i.e. only locally the the image of a vector field\footnote{For a more complete study of the integrability of first order differential equations, see \cite{mmt95}}. 
\end{example}
\subsubsection{\blu{Symmetries and constants of the motion}}
\label{subsec:sy}
A differentiable function $f\in\mathcal F(M)$ is called a \emph{constant of the motion} for the integrable f.o.d.e. $\mathfrak Z\subset TM$ if the composition $f\circ\gamma$ is a constant function for any solution $\gamma$ of the equation $\mathfrak Z$. This is easily seen equivalent to the condition 
\beq
\label{condsym}
(\dd f)(z)_{\mid z\in\mathfrak Z}\,=\,0
\eeq
since $\dd f$, as a section $M\,\to\,T^*M$, is a fiberwise linear function.  A diffeomorphism $\varphi:M\to M$ is called a \emph{symmetry} for the integrable differential equation $\mathfrak Z$ if 
\beq
\label{condsym1}
T\varphi(\mathfrak Z)=\mathfrak Z. 
\eeq
This condition has a local version. When one has   $\varphi\,:\,x\,\mapsto\,x'=\varphi(x)$ written as $x'^a=\varphi^a(x)$, then $T\varphi\,:\,TM\,\to\,TM$ is the diffeomorphism (called the \emph{tangent lift} of the diffeomorphism $\varphi$) which can be written as    
\beq
\label{25.4}
T\varphi\,:\,(x^a, v^a)\,\mapsto\,(x'^a=\varphi^a(x), v'^a=v^b\frac{\del \varphi^a}{\del x^b}).
\eeq
These  conditions  encompass the conditions \eqref{I-defc00}[I] and \eqref{I-defsymm}[I] which are valid for explicit first order differential equations. 

\begin{rema}
\label{rag}
Although the notions of constant of the motions and of symmetry have been introduced for integrable f.o.d.e., the conditions \eqref{condsym} e \eqref{condsym1} 
do not depend on the integrability of $\mathfrak Z$. One can immediately see that 
$$
f\,:\,(x^1,x^2)\,\mapsto\,\,\tilde f(x^2)
$$
satisfies the condition \eqref{condsym} for any real valued $\tilde f$ with respect to the f.o.d.e. $\mathfrak Z$ introduced in the previous example \ref{ex-25.2}.
\end{rema}

A vector field $X\in\mathfrak X(M)$ is called an \emph{infinitesimal symmetry} for the f.o.d.e. $\mathfrak Z$ if it generates a (local) diffeomorphism which is a symmetry for $\mathfrak Z$. If $\varphi_s:M\to M$ is the one parameter group of local diffeomorhisms generated by the vector field $X$ on $M$ with $X=X^a\del_{x^a}$,  then the infinitesimal generator of the one parameter group of diffeomorphisms $T\varphi_s:TM\to TM$ is the vector field $X^{(N)}$ on $TM$ given by 
\beq
\label{25.5}
X^{(N)}\,=\,X^a\frac{\del}{\del x^a}\,+\,v^b\frac{\del X^a}{\del x^b}\frac{\del}{\del v^a}
\eeq
(where we have identified $X^a$ and $\del_{x^a}X^b$ with $\pi^*(X^a)$ and $\pi^*(\del_{x^a}X^b)$ as functions on $TM$, with respect to the tangent bundle projection), which we call the \emph{tangent lift} of the vector field $X$ on $M$. 
One can prove that, given an integrable  first order differential equation $\mathfrak Z\subset TM$,  the vector field $X$ on $M$ is an infinitesimal symmetry for $\mathfrak Z$ if and only if its tangent lift $X^{(N)}$ on $TM$   is tangent to $\mathfrak Z$, i.e. if and only if $X^{(N)}\in\mathfrak X(\mathfrak Z)$. If the submanifold defining the first order differential equation $\mathfrak Z\subset TM$ is given as the zero level set of a set of functions, i.e. 
$$
\mathfrak Z\,=\,\{z\,\in\,TM\,:\,\psi^a(z)\,=\,0\}
$$
with $\psi^a\in\mathcal F(TM)$, then $X$  turns to be an infinitesimal symmetry for $\mathfrak Z$ if and only if 
\beq
\label{new01}
L_{X^{(N)}}\psi^a\,=\,A^{a}_b\psi^b
\eeq
with $A^a_b\in\mathcal F(TM)$. If 
$$
\{f\,:\,f_{\mid \mathfrak Z}=0\}\,=\,\mathcal F_{\mathfrak Z}\subset\mathcal F(TM)
$$
 denotes the ideal of functions on $TM$ which vanish on $\mathfrak Z$, then $X$ is proven to be an infinitesimal symmetry for $\mathfrak Z$ if and only if 
\beq
\label{new02}
L_{X^{(N)}}f\,\in\,\mathcal F_{\mathfrak Z}
\eeq
for any $f\in\mathcal F_{\mathfrak Z}$, i.e. the tangent lift $X^{(N)}$ is a derivation for the ideal $\mathcal F_{\mathfrak Z}$. 

When the first order differential equation $\mathfrak Z\subset TM$ is explicit, i.e. $\mathfrak Z={\rm Im}(\Gamma)$ with $\Gamma\in\mathfrak X(M)$, then the above definitions \eqref{new01}-\eqref{new02} are proven equivalent to the condition $[X,\Gamma]=0$ (see \eqref{I-defsymmi}[I]), while the relation \eqref{condsym} can be written as
\beq
\label{imp01}
(\dd_Nf)_{\mid\mathfrak Z}=0,
\eeq
in terms of the derivation operator $\dd_N:\Lambda^k(M)\to\Lambda^{k}(TM)$  defined in analogy to  a Lie derivative by the sum 
\beq
\label{imp03}
\dd_N\,=\,i_N\dd\,+\,\dd i_N,
\eeq
where each term is given upon composing the exterior differential  $\dd\,:\,\Lambda^{k}(M)\to\Lambda^{k+1}(M)$ with  the degree (-1) derivation $i_N\,:\,\Lambda^k(M)\to\Lambda^{k-1}(TM)$, whose action is defined by the requirement that it annihilates elements $\Lambda^0(M)=\mathcal F(M)$, it reads 
\beq
\label{imp02}
i_N\,:\,\alpha=\alpha_j\dd x^j\quad\mapsto\quad v^j\alpha_j 
\eeq
on 1-forms and is extended to higher order form via the graded Leibniz rule\footnote{Both the operators $\dd_N$ and $i_N$ owe their suffix $_N$ to the fact that they are natural within the setting of Newtonian transformations, as will be showed in section \ref{ss:T}.}. It is immediate to recover that, if $f\in\mathcal F(M)$, then it is\footnote{We have again identified $f$ with $\pi^*f$ as functions on $TM$.}
$$
\dd_Nf\,=\,v^a\frac{\del f}{\del x^a}
$$
 The tangent lift \eqref{25.5} of the vector field $X=X^a\del_{x^a}$ on $M$ can then be written as
\beq
\label{eq26.s}
\mathfrak X(TM)\,\ni\,X^{(N)}\,=\,X^a\del_{x^a}\,+\,(\dd_N X^a)\del_{v^a},
\eeq
with
\begin{align}
&i_{X^{(N)}}\dd_N=\dd_Ni_{X}, \nn \\
&L_{X^{(N)}}\dd_N\,=\,\dd_N\,L_{X},  \nn \\
&i_{X^{(N)}}i_N\,=\,i_Ni_X
\label{imp06}
\end{align}
from $\Lambda(M)$ to $\Lambda(TM)$. Between these sets one can also prove that, for any diffeomorphism $\phi\,:\,M\,\to\,M$, it is 
$$
\dd_N\phi^*\,=\,(T\phi)^* \dd_N
$$ 
 in terms of the tangent maps. 

\begin{example}
\label{s3i}
Consider on $T\R^2$ the implicit differential equation $\mathfrak Z$  defined by the condition
\beq
\label{3p1}
x^2+y^2+v_x^2+v_y^2=1,
\eeq
which represents a codimension one  sphere. We begin by noticing that, since $\mathfrak Z$ is odd dimensional, it does not come as the graph of a vector field on $\R^2$ and that  the submanifold $\mathfrak Z$ is projected under the action of the tangent bundle projection $\pi\,:\,T\R^2\,\to\,\R^2$ onto a radius 1 disk in $\R^2$. 

The analysis on the solutions of this equation starts by noticing that the relations
\begin{align*}
&v_x\,=\,\pm\sqrt{1-x^2-y^2}\cos \,a, \\ & v_y\,=\,\pm\sqrt{1-x^2-y^2}\sin\,a
\end{align*}
for any arbitrary function  $a=a(x,y)$ give a family of vector fields whose graphs are contained in $\mathfrak Z$. This shows that the implicit equation $\mathfrak Z$ is integrable at any point of the disk.

In particular, the only solutions on the boundary of the disk are given   by the  curves $$\gamma\,=\,(x(s)=x_0, y(s)=y_0),$$ provided $x_0^2+y_0^2=1$.  Solutions to this equation can be cast in two different classes, namely those passing (respectively not passing) through the origin$(x=0,y=0)$. 

Among those passing through the origin there are 
$$
x(t)=\frac{\alpha}{\sqrt{2(1+\alpha^2)}}\sin t, \qquad y(t)=\frac{\pm1}{\sqrt{2(1+\alpha^2)}}\sin t
$$
for any constant  $\alpha\in\R$. A more general analysis on the  motions not crossing the origin is performed in radial coordinates $(r,\varphi)$, along which the submanifold  \eqref{3p1} is written as
\beq
\label{3p2}
r^2+r^2v_\varphi^2+v_r^2=1.
\eeq
This shows immediately that the vector field $X_\varphi\,=\,\del/\del\varphi$ is an infinitesimal symmetry for the equations, since for its Newtonian lift one has $X^{(N)}=\del/\del\varphi$ and $$L_{X^{(N)}}(r^2+r^2v_\varphi^2+v_r^2-1)=0.$$
For a generic motion, one has $v_\varphi=v_\varphi(r, v_r)$: this shows that they are
  spirals from $r=0$ to $r=1$, with 
\begin{align*}
&v_r\,=\,\pm\sqrt{1-r^2}\sin b, \\ &rv_\varphi\,=\,\pm\sqrt{1-r^2}\cos b
\end{align*}
for any arbitrary function $b=b(r,\varphi)$. If we select $b=\pi/2$ we get the radial motions with $v_\varphi=0$. If we select $b=0$ we have the circular motions, those with $v_r=0$, resulting in the submanifold
$$
\rho^2v_\varphi^2=1-\rho^2
$$
for fixed $r=\rho<1$. The integral curves can be written as $\varphi(t)=\varphi(0)\pm t\sqrt{-1+1/\rho^2}$.

We end our analysis of this example by noticing that searching a constant of the motion amounts to determine a function $f=f(x,y)$ on the disk solving the condition \eqref{imp01}, i.e.
$$
(v_x\frac{\del f}{\del x}+v_y\frac{\del f}{\del y})_{\mid \mathfrak Z}=0.
$$
This is equivalent to solve the partial differential equation
$$
\frac{\del f}{\del x}=kv_y, \qquad \frac{\del f}{\del y}=-kv_x
$$
for any $k=k(x,y)$, that can be written as
$$
x^2+y^2+(\frac{1}{k}\frac{\del f}{\del x})^2+(\frac{1}{k}\frac{\del f}{\del y})^2=1.
$$ 
This equation can be analysed as a Hamilton-Jacobi equation\footnote{We shall describe some aspects of the Hamilton-Jacobi theory in the following sections.} for $f$.

\end{example}
\begin{example}
\label{s3L}
A refinement of the previous example is given by the implicit differential equation $\mathfrak Z$ described as the zero level set
\begin{align}
&\psi_1=x^2+y^2+v_x^2+v_y^2-1=0,  \nn \\
&\psi_2=xv_y-yv_x-L=0
\label{3p3}
\end{align}
or equivalently as 
\begin{align}
&x^2+y^2+v_x^2+v_y^2-1=0,  \nn \\
&(x^2+y^2)(v_x^2+v_y^2)-(xv_x+yv_y)^2-L^2=0. 
\label{3p4}
\end{align}
The set $\mathfrak Z$ turns to be a 2-dimensional submanifold embedded in $T\R^2$ as the intersection of a sphere with a hyperboloid only if $1>4L^2$. 
For $L\neq0$ the set $\mathfrak Z$ is given, in radial coordinates, by
\begin{align}
&r^2v_\varphi=L,  \nn \\
&r^2+v_r^2+\frac{L^2}{r^2}=1; 
\label{3p5}
\end{align}
and corresponds to the graph of the vector fields 
$$
\Gamma\,=\,\frac{L}{r^2}\frac{\del}{\del\varphi}\pm(\sqrt{1-r^2-\frac{L^2}{r^2}})\frac{\del}{\del r}
$$
for 
$$
\frac{1}{2}-\sqrt{\frac{1}{4}-L^2}\,<\,r^2\,<\, \frac{1}{2}+\sqrt{\frac{1}{4}-L^2}. 
$$
\end{example} 

\subsection{A geometric setting for second order implicit differential equations}
\label{ss:2}
Given the smooth $N$-dimensional  manifold $M$, one can define the natural submersion $\tau\,:\,M'\simeq M\times \R\,\to\,\R$, so to have  that the triple $(M', \pi, \R)$ is a fiber bundle with basis $\R$ and typical fiber $M$. If  $\Gamma_t(\tau)$ denotes  the set of local sections $t\,\to\,(t, \gamma(t))$ with $\gamma\,:\,\R\to M$ (i.e. curves on $M$),  the quotient $\Gamma_t(\tau)/\sim$,  where   the equivalence relation is given by 
\beq
\label{25.6}
\gamma\sim\gamma'\quad\Leftrightarrow\quad T\gamma_{\mid t}=T\gamma'_{\mid t},
\eeq
 is proven to be a  smooth manifold. This manifold\footnote{We refer to \cite{mks_natural, krupkova, saunders} for a more complete description of this subject.}  is  called the \emph{first jet manifold} over the fiber bundle $\tau$ and is denoted by $J^1\tau$, with elements $j^1_t\gamma$ in the equivalence class $[\gamma]$ defined by the curve $\gamma$.   Moreover, it is easy to see that one has the equivalence
\beq
\label{25.8}
J^1\tau\,\simeq\,\R\times TM
\eeq
and that a local coordinate system on $J^1\tau$ is given by $(t, x^a, v^a)_{a=1,\dots,N}$. If $\gamma$ gives a curve on $M$ and then a section of the bundle $\tau$, its first order prolongation is locally written as 
\beq
\label{25.7}
j^1_t\gamma=(t, x^a=\gamma(t), v^a=\dot\gamma^a(t)).
\eeq
The equivalence \eqref{25.8}, together with the relations \eqref{25.3}, show that a first order o.d.e. (indeed non autonomous, i.e. depending on $t$) can  be defined as a subset (usually assumed to be a submanifold) $\mathfrak Z\subset J^1\tau$. The analogy of the relations \eqref{25.3} and \eqref{25.7} shows that a solution for $\mathfrak Z$ is a curve $\gamma$ whose (so called) first order prolongation $j^1_t\gamma$ to the first jet bundle is in $\mathfrak Z$. Notice that, if $\mathfrak Z$ is defined as the zero level set $\psi^a=0$ with $\psi^a\in\mathcal F(J^1\tau)$, then the equation is autonomous if $\psi^a$ does not depend on the base coordinate $t$. 

The \emph{second jet manifold} is defined in analogy. Within $\Gamma_t(\tau)$ one defines the equivalence relation 
\beq
\label{27.1}
\gamma\sim\gamma'\quad\Leftrightarrow\quad \dot\gamma=\dot\gamma', \quad \ddot\gamma=\ddot\gamma'
\eeq
in terms of the second order derivative with respect to the base coordinate in a local (adapted, as it is usually said) chart. The quotient $J^2\tau=\Gamma_t\tau/\sim$ is a smooth manifold, with a local coordinate system given by  $$(t,x^j, v^j, a^j)_{j=1,\dots,N}.$$   If $\gamma$ gives a curve on $M$ and then a section of the bundle $\tau$, its second order prolongation is locally written as 
\beq
\label{27.3}
j^2_t\gamma=(t, x^j=\gamma^j(t), v^j=\dot\gamma^j(t), a^j=\ddot\gamma^j(t)).
\eeq
Along the same lines reading \eqref{25.8} one can prove that
\beq
\label{27.2}
J^2\tau\,\simeq\,\R\times T^2M,
\eeq
where $T^2M$ is the so called second order tangent bundle to $M$. 

Such manifold can be introduced as follows. If $\pi\,:\,TM\,\to\,M$ denotes the tangent bundle to $M$, and $\pi_\mathfrak T\,:\,T(TM)\,\to\,TM$ denotes the tangent bundle to $TM$, then $T^2M$ is the subset of elements in $\xi\in T(TM)$ such that $$\pi_\mathfrak T(\xi)=T\pi(\xi).$$ If we consider a local coordinate chart on $T(TM)$ such that we can collectively write $\xi=(x,v,u,a)$,  it is 
\begin{align*}
&\pi_\mathfrak T\,:\,(x,v,u,a)\,\mapsto\,(x,v), \\ &T\pi\,:\,(x,v,u,a)\,\mapsto\,(x,u).
\end{align*}
 The elements in $\xi\in T^2M\subset T(TM)$ can be locally written as $\xi=(x,v,u=v,a)$.  One proves that $T^2M$ is a smooth manifold, actually the total space of a  subbundle in $\pi_\mathfrak T:T^2M\to TM$ and in $T\pi: T^2M\to TM$.  
 
It is then natural to say that a second order (ordinary) differential equation (i.e. s.o.d.e.) is a subset (usually assumed to be a submanifold) $\mathfrak Z$ in $J^2\tau$ (autonomous if independent on $t$).  Such equations describe the motion of point particle systems within the Newtonian formalism to mechanics. A solution for it is a section described by a curve $\gamma(t)$ on $M$ such that its second order prolongation \eqref{27.3} is in $\mathfrak Z$. A diffeomorphism $\varphi\,:\,M\,\to\,M$ whose action we write as $$x'^j=\varphi^j(x)$$ with respect to a coordinate chart provides a symmetry for $\mathfrak Z$ if its jet prolongation  maps the solution $j^2_t\gamma$ into a solution $j^2_t\gamma'$ with 
$$
j^2_t\gamma'=(t, x'^j=\gamma'(t)=\varphi^j\mid_{\gamma(t)}, v'^j=\dot\gamma^s\frac{\del \varphi'^j}{\del x^s}{}{\mid_{\gamma(t)}}, a'^j=\ddot\gamma^s\frac{\del \varphi'^j}{\del x^s}{}{\mid_{\gamma(t)}}+ \dot\gamma^s\dot\gamma^k\frac{\del^2 \varphi'^j}{\del x^s\del x^k}{\mid_{\gamma(t)}}),
$$
Second order (ordinary) autonomous differential equations  can then be considered as a specific class of first order (ordinary) autonomous differential equations on $TM$, namely those that can be written as the zero level set in $T(TM)$ given by 
\begin{align}
&u^j=v^j,  \nn \\
\label{27.4} &\psi^a=0
\end{align}
where $j=1,\dots,N$ and ${\psi^a}_{a=1,\dots, k}$ denotes a set of functions on $T(TM)$. When the rank of the form $\dd\psi^1\wedge\dots\dd\psi^k$ is constant on $T(TM)$, then the zero level set $\mathfrak Z$ is a submanifold. When the second order tangent coordinates $a^j$ can be solved for in \eqref{27.4}, then the set $\mathfrak Z$ is the graph of a section of $T^2M$ which we write as $(x^j, v^j)\,\mapsto\,(x^j, v^j, v^j, a^j(x,v))$ (the equation is supposed $t$-independent), equivalently the graph of the vector field $D$ on $TM$ which we locally write as 
\beq
\label{27.5}
D\,=\,v^j\frac{\del}{\del x^j}\,+\,a^j\frac{\del}{\del v^j}.
\eeq
Vector fields on $TM$ which are sections of $T^2M$ are called \emph{second order} vector fields.  Second order vector fields can be described also by a deeper analysis of the geometric structures intrinsically defining a tangent bundle manifold. The following section is devoted to this topic.

\subsection{The geometry of the tangent bundle}\label{ss:T}
Let us consider again the tangent bundle  $\pi:TM\to M$ with $M$ a smooth $N$-dimensional manifold.  Since each fiber $T_mM$ is a vector space, it is clear that on $TM$ it is possible to define the tensors 
\begin{align*}
&\Delta_M=v^a\frac{\del}{\del v^a}\,\in\,\mathfrak X(TM), \\ &S=\dd x^a\otimes\frac{\del}{\del v^a}\,\in\,T^1_1(TM).
\end{align*}
The vector field $\Delta_M$ gives a partial linear structure on $TM$: it is the infinitesimal generators of the fiberwise defined dilation $\phi_s:(x,v)\mapsto(x,e^sv)$,  its kernel gives the algebra $\mathcal F(M)$ of smooth functions on the basis $M$ of the tangent bundle $TM$; the tensor $S$ is usually referred to as the vertical endomorphism of $TM$ or the soldering tensor on $TM$. It is immediate to prove that the following identities hold:
\begin{enumerate}[(i)]
\item $\Delta_M\in{\rm Ker}\,S$;
\item ${\rm Ker}\,S\,=\,{\rm Im}\,S$;
\item $L_{\Delta_M}S=-S$;
\item $N_S=0$
\end{enumerate}
where the Nijenhuis tensor associated to $S$ is given by (see the Schouten-Nijenhuis bracket introduced in section \ref{I-sec:poisson}[I])
$$
N_S(A,B)=S\{S([A,B])-[S(A),B]-[A,S(B)]\}+[S(A),S(B)],
$$
for any pair $A,B\in\mathfrak X(TM)$. It is possible to prove \cite{TQteam} that if on a smooth manifold $\mathcal M$ one has a complete vector field $\Delta$ and a $(1,1)$-tensor $S$ which satisfy the relations (i)-(iv) above\footnote{With $\Delta$ replacing $\Delta_M$.}, then $\mathcal M$ can be given the structure of a tangent bundle on a manifold $M$ which is the quotient of $\mathcal M$ upon  identifying the fibers which happen to be vector spaces. 

A vector field $X\in\mathfrak{X}(TM)$ is called \emph{vertical} if $S(X)=0$. The local coordinate expression for a vertical vector field is $X\,=\,X^i\frac{\del}{\del v^i}$ with $X^i\in\mathcal F(TM)$.  It is immediate to see that the set $\mathfrak X^V(TM)$ of vertical vector fields is a Lie subalgebra in $\mathfrak X(TM)$.  Given a vector field $X\in\mathfrak X(M)$ with coordinate expression  $X\,=\,X^a\frac{\del}{\del x^a}$ (for $X^a\in\mathcal F(M)$), its \emph{vertical lift} is defined to be the vertical vector field given by $$X^{(V)}=(\pi^*X^a)\frac{\del}{\del v^a}.$$
A vector field $D\in\mathfrak X(TM)$ is called a \emph{second order} vector field if 
\beq
\label{seorco}
S(D)=\Delta_M.
\eeq
 The local coordinate expression for a second order vector field is 
\beq
\label{equ24-1}
D\,=\,v^j\frac{\del}{\del x^j}\,+\,a^j(x,v)\frac{\del}{\del v^j}
\eeq
so that the corresponding  system of first order differential equations  are written as
\begin{align}
&\dot{x}^j=v^j, \nn \\
&\dot{v}^j=a^j(x,v),
\label{eq24}
\end{align}
i.e. the Newton equations of the motions for a point particle whose configuration space is $M$, with acceleration $a^j(x,v)$. Second order vector fields are also called Newtonian vector fields.  The condition \eqref{seorco} is easily seen to be equivalent to the definition of second order vector field on $M$ given in section \ref{ss:2} in terms of section of the second order tangent bundle $T^2M$, as the comparison of the coordinate expression \eqref{27.5} with \eqref{equ24-1} shows.

It is then natural to define a diffeomorphism $$\phi:TM\to TM$$ as \emph{Newtonian} if it maps second order vector fields into second order vector fields. Upon directly using the properties of the tangent map $$T\phi:T(TM)\to T(TM)$$ it is possible to see that a diffeomorphism $\phi$ on $TM$ is Newtonian if and only if there exists a diffeomorphism $\varphi:M\to M$ such that the action of $\phi$ is given  by  (see \eqref{25.4})
\beq
\label{eq25}
\phi\,:\,(x^a,v^a)\,\mapsto\,(\varphi^a(x), v^b\frac{\del \varphi^a}{\del x^b}).
\eeq
This coordinate expression shows that a diffeomorphism $\phi$  on $TM$ is Newtonian if and only if it is given by the \emph{tangent lift} of a diffeomorphism $\varphi:M\to M$, that is $T\varphi=\phi$ as we defined in \eqref{25.4}. Notice that, given this characterization, a Newtonian diffeomorphism on $TM$ is usually referred to as a \emph{point transformation}.

This definition  has an infinitesimal counterpart. 
Denote by $\varphi_s:M\to M$ the one parameter group of (local) diffeomorphisms generated by the vector field $X=X^{a}\del/\del x^a\,\in\,\mathfrak X(M)$. We have already written in \eqref{25.5} that the infinitesimal generator $X^{(N)}$ of the  one parameter group of Newtonian diffeomorphisms $\phi_s=T\varphi_s:TM\to TM$ reads
\beq
\label{eq26}
X^{(N)}\,=\,(\pi^*X^a)\frac{\del}{\del x^a}\,+\,(L_D(\pi^*X^a))\frac{\del}{\del v^a}
\eeq
for any second order vector field $D$.  From \eqref{imp03} it is also immediate to see that the tangent lift in \eqref{eq26} can be written as 
\beq
\label{8giu-1}
X^{(N)}\,=\,(\pi^*X^a)\frac{\del}{\del x^a}\,+\,(i_NX^a)\frac{\del}{\del v^a}.
\eeq

It is indeed possible to characterize the set of Newtonian vector fields on $TM$ more intrinsically, following two different paths. 
A vector field $X\in\mathfrak X(TM)$ is Newtonian, i.e. it generates a one parameter group of (local) Newtonian diffeomorphism,  if and only if 
\beq
\label{eq26bis}
S([X,D])=0\quad\forall\,D\in\mathfrak X(TM)\,:\,S(D)=\Delta_M,
\eeq
 that is \emph{for any} second order vector field $D$ on $TM$. This turns out to be the infinitesimal characterization of a vector field that generates a fiber preserving  diffeomorphism which maps \emph{any} second order vector field into a second order vector field. 
For any $A,B\in\mathfrak X(M)$ it is possible to prove that the following relations, i.e.
\begin{align}
&[A^{(V)}, B^{(V)}]=0, \nn \\
 &[A^{(N)}, B^{(N)}]=[A,B]^{(N)}, \nn \\
 &[A^{(V)}, B^{(N)}]=[A,B]^{(V)}, \nn \\
 &L_{A^{(V)}}S=0, \nn \\
&L_{A^{(N)}}S=0,
\label{eq27}
\end{align}
hold as well as 
\beq
\label{2-uno}
 [A^{(N)},D]\,\in\,\mathfrak X^V(TM)
 \eeq
for any second order vector field $D$. The first and the third relations out of \eqref{eq27}  show that vertical lifts of vector fields on $M$ close an infinite dimensional Abelian Lie algebra, whose normaliser contains Newtonian lifts of vector fields in $M$. Vertical and Newtonian lifts  define derivations which are enough to separate one-forms on $TM$, i.e. they locally generate the tangent bundle on $TM$. 

Equivalently, one can prove that a vector field $X\in\mathfrak X(TM)$ is Newtonian if and only if it generates transformations on $TQ$ which leave invariant both the tensors $S$ and $\Delta_M$, (i.e. the tensors intrinsically characterising  the tangent bundle structure)  that is  $X\in\mathfrak X(TM)$ is Newtonian if and only if $L_XS=0$ and $L_X\Delta_M=0$.

\subsection{Euler-Lagrange equations}\label{ss:L}
We introduce now the Euler-Lagrange equations for a dynamical system on a configuration space. Following the historical tradition for analytical mechanics textbooks, we denote by $Q$ such configuration space, a smooth $N$-dimensional manifold with a local coordinate chart given by $\{q^a\}_{a=1,\dots,N}$. 

For any smooth element $\cL\in\mathcal F(TQ)$ we define the Cartan (or Lagrangian) 1-form $$\tL=S(\dd\cL)$$ and the (Lagrangian) 2-form $$\oL=-\dd\tL.$$ Clearly one has that $\oL$ is closed: the Lagrangian $\cL$ is called  \emph{regular} 
if $\oL$ is a symplectic structure on $TQ$, i.e. if $\oL$ is non degenerate. In local coordinates $(q^a,v^a)$ we have 
\begin{align}
&\tL\,=\,\frac{\del\cL}{\del v^a}\,\dd q^a, \nn \\
&\oL\,=\,\frac{\del^2\cL}{\del v^a\del v^b}\,\dd q^a\wedge\dd v^b\,+\,\frac{1}{2}\left(\frac{\del^2\cL}{\del v^a\del q^b}-\frac{\del^2\cL}{\del q^a\del v^b}\right)\dd q^a\wedge\dd q^b,
\label{eq28}
\end{align}
which means that $\cL$ is regular if and only if the Hessian matrix 
\beq
\label{H0}
H_{ab}=\frac{\del^2\cL}{\del v^a\del v^b}
\eeq
 is of maximum rank. The mapping $\cL\mapsto\oL$ is not injective: upon using the properties of the soldering tensor $S$ it is possible to prove that a Lagrangian $\cL$ is mapped into an identically vanishing 2-form $\oL$ if and only if (for any second order vector field\footnote{Any 1-form $\alpha$ on $Q$ defines a fiberwise linear function $i_N\alpha$ on $TQ$, recalling \eqref{imp02}.} 
 $D$)  
 \beq
 \label{dtop}
 \cL=\pi^*f\,+\,i_D(\pi^*\alpha)
 \eeq with $f\in\mathcal F(Q)$ and $\alpha\in\Lambda^1(Q)$ such that $\dd\alpha=0$. Such expression can also be written as 
 $$\cL=\pi^*f\,+\,i_N\alpha.$$
 A Lagrangian $\cL=\pi^*f$ for $f\in\mathcal F(Q)$ is usually said to be a \emph{pure potential}, a Lagrangian $\cL=i_N\alpha$ is usually called a \emph{pure gauge}.

We say that a second order dynamics $D$ on $TQ$ has a Lagrangian formulation  if there exists a Lagrangian function $\cL\in\mathcal F(TQ)$ such that\footnote{It is indeed easy  to see that $\dd q^a\{\frac{\dd}{\dd t}\left(\frac{\del \cL}{\del v^a}\right)-\frac{\del\cL}{\del q^a}\}\,=\,L_D\tL-\dd\cL$.} 
\beq
\label{eq29}
L_D\tL-\dd\cL\,=0.
\eeq
In such a case, we say $\cL$ is an admissible Lagrangian for the second order dynamics $D$ on $TQ$.
Moreover, from the identity 
\beq
\label{eq30}
L_D\tL-\dd\cL\,=\,-i_D\oL+\dd\EL
\eeq
with the energy function being defined by 
\beq
\EL\,=\,i_{\Delta_Q}\dd\cL-\cL\,=\,i_D\tL-\cL
\label{eq31}
\eeq
for any second order vector field $D$, we see that the Euler-Lagrange equations \eqref{ll1}, whose integral curves give  the stationary points of the action functional \eqref{ll2} for a suitable class of variations (see \cite{saunders}), can be written as 
\beq
\label{eq32}
i_D\oL\,=\,\dd\EL
\eeq
or, in local coordinates $(q,v,a)$ on $T^2Q$, as 
\begin{align}
\label{ll3}
 H_{ks}a^k\,=\,\left(\frac{\del \cL}{\del q^s}\,-\,\frac{\del^2\cL}{\del v^s\del q^k}\,v^k\right).
\end{align}
This is in general a system of second order implicit ordinary differential equations  (notice that $D$ is given as the solution of a relation analogue to   \eqref{I-ps1}[I]), namely a submanifold (if the rank of $H_{ks}$ is constant) $\mathfrak Z_{\cL}$ in $T^2M$.  If $\cL$ is regular, then the accelerations can be solved for: the Euler-Lagrange equations are explicit (see \eqref{eq24}) and can be written in terms of a second order vector field, since for any any $D$ fulfilling the relations  \eqref{eq31}-\eqref{eq32} it is $$S(D)=\Delta_Q.$$  The Lagrangian function defines its own symplectic structure $\oL$ on $TQ$,  the vector field $D$ is therefore Hamiltonian with Hamiltonian function $\EL$ and symplectic structure $\oL$.  

We find it interesting to remark that a  regular Lagrangian function determines both a symplectic structure on $TQ$ (i.e. a \emph{kinematical} datum) and a Hamiltonian function generating the time evolution (i.e. a \emph{dynamical} datum).

We shall start our analysis by assuming the Lagrangian to be regular and describe the case of a singular Lagrangian in the following sections. 
The general problem of determining an \emph{admissible} Lagrangian function $\cL$ for a given second order field $D$ is usually referred to as the inverse problem in the calculus of variations \cite{inv}. 

Without dwelling here into such a theme, we limit ourselves to notice  that if $\cL$ is an admissible Lagrangian for $D$, then from the analysis described above the function $\cL'=\cL+i_N\alpha$ with $\alpha\in\Lambda^1(Q)$ and $\dd\alpha=0$ is also admissible for $D$. Moreover, apart from this gauge ambiguities, there exist second order dynamics with admissible Lagrangians resulting in different Lagrangian symplectic forms and different energy functions. 
This comes by noticing that the $\mathcal F(TQ)$-linear map
$\cL\,\mapsto\,\EL$
defined by \eqref{eq31} is not injective, its kernel being given by Lagrangians $\cL=i_N\beta$ with $\beta\in\Lambda^1(Q)$ (i.e. not necessarily closed). There exist then different (regular) Lagrangians reading the same symplectic structure $\oL$ but different Energy functions $\EL$. 
This aspect is indeed relevant in the present paper since, as we already described in [I] within the Poisson and the Hamiltonian formalism, alternative Lagrangian descriptions for a given dynamics allow for the construction of different sets of constants of the motions\footnote{and, in a more general setting, to different quantized versions of the same classical dynamics.}. Analogously, it is immediate to see that regular Lagrangians differing by a (pure) potential term provide the same symplectic structure but different Hamiltonians.

\section{Symmetries  for a regular Lagrangian dynamics}
\label{ss:lag}
In order to consider the relations between infinitesimal symmetries and constants of the motion within the Lagrangian formalism for regular Lagrangians, we start by noticing that the Cartan 1-form $\theta_{\cL}$ allows to define a map 
$$
\tilde{\delta}\,:\,X\,\in\,\mathfrak X(TQ)\,\mapsto\,i_X\theta_{\cL}\,\in\,\mathcal F(TQ).
$$
 Such a map can not  be inverted: it is  for instance  $i_{X}\theta_{\cL}=0$ for any  vertical vector field $X$. Using the Cartan identity and the equations of motions \eqref{eq29} we write, for any $X\in\mathfrak X(TQ)$, 
\begin{align}
& L_D(i_X\tL)-i_XL_D\tL\,=\,i_{[D,X]}\tL \nn \\ & L_D(i_X\tL)-L_X\cL\,=\,i_{[D,X]}\tL.
\label{eq34}
\end{align}
This shows that, if $X$ is an infinitesimal symmetry for $D$, i.e. $[D,X]=0$, then the function $\varphi=i_X\tL-u$ is invariant along the dynamics if and only if there exists a function $u\in\mathcal F(TQ)$ such that  $L_X\cL=L_Du$. 

Another form of a correspondence between infinitesimal symmetries and constants of the motion comes by noticing that  the relations  \eqref{eq29} and \eqref{eq32}  show that $\oL$ provides an invariant 2-form under the (regular) Lagrangian dynamics $D$. Via such symplectic structure on $TQ$, a suitable analogue to the maps $\tau$ and its inverse (see \eqref{I-eq5}-[I] and \eqref{I-eq33}-[I]) is defined.  We have $$L_Di_X\oL\,=\,i_{[D,X]}\oL.$$ The 1-form $i_X\oL$ is exact (thus reading a constant of the motion) if and only if $i_X\oL=\dd f$, i.e. if and only if $X$ is $\oL$-Hamiltonian on $TQ$. It is clear that what we have described is a Hamiltonian form (see the proposition \ref{I-noesymp2} in section \ref{I-sec:hamilton}-[I])  of the Noether correspondence on $TQ$. 

\subsection{Newtonian symmetries}
\label{subsubL2}
We have not considered so far that  $D$ is a second order field on a  tangent bundle manifold $TQ$.  We aim to  inspect the possibility of having a Noether correspondence which more naturally fits into such a setting,  therefore consider a Newtonian vector field $X^{(N)}$, and assume it is $\oL$-Hamiltonian, with 
$$i_{X^{(N)}}\oL=\dd \varphi$$ for a suitable $\varphi\in\mathcal F(TQ)$. From the second line out of  the identities
\begin{align}
&L_{X^{(N)}}\tL=\theta_{L_{X^{(N)}}\cL}, \nn \\
&L_{X^{(N)}}\oL=\omega_{L_{X^{(N)}}\cL},
\label{eq35}
\end{align}
we have  that $\omega_{L_{X^{(N)}}\cL}=0$, and this means  that there exists a function $\alpha\in\mathcal F(Q)$ and a closed 1-form $\beta\in\Lambda^1(Q)$ such that  $$L_{X^{(N)}}\cL=\pi^*\alpha+i_D(\pi^*\beta).$$ For the difference $u\,=\,i_{X^{(N)}}\tL-\varphi$, we have 
\begin{align}
\dd u&=\,\dd(i_{X^{(N)}}\tL)-\dd\varphi \nn \\
&=\,L_{X^{(N)}}\tL\,-\,i_{X^{(N)}}\dd\tL\,-\,i_{X^{(N)}}\oL\nn \\
&=\,L_{X^{(N)}}\tL\nn \\ &=\,\theta_{L_{X^{(N)}}\cL},
\label{eq35p1}
\end{align}
where the last line comes from the first relation out of \eqref{eq35}, while a direct computation reads $\theta_{L_{X^{(N)}}\cL}=\pi^*\beta$. This means that  $L_Vu=0$ for any vertical element $V\in\mathfrak X(TQ)$ and $\pi^*\beta=\dd u$, thus giving  (compare it with \eqref{dtop})
$$L_{X^{(N)}}\cL=\pi^*\alpha+L_Du.$$  

\noindent We have then  proven that if the correspondence $\tilde{\tau}:\mathfrak X(TQ)\to\Lambda^1(TQ)$ (see \eqref{I-intro01}[I]) maps a Newtonian vector field $X^{(N)}$ into the exact 1-form $i_{X^{(N)}}\oL=\dd\varphi$, then the function $$u=i_{X^{(N)}}\tL-\varphi$$ satisfies the identities \begin{align}&L_Vu=0, \nn \\ &L_V(L_{X^{(N)}}\cL-L_Du)=0 \label{Lag01} \end{align} for any vertical vector field $V$ on $TQ$.  This assertion can be easily reversed, so we can write the following result.
\begin{lemm}
\label{lem-lag1}
Let $D$ be a second order vector field on $TQ$ with admissible regular Lagrangian $\cL$, and  $X^{(N)}$ be  a Newtonian vector field on $TQ$. If $X^{(N)}$ is $\oL$-Hamiltonian, with $i_{X^{(N)}}\oL=\dd\varphi$, then the function $u=i_{X^{(N)}}\tL-\varphi$ satisfies the conditions \eqref{Lag01}. If a function $u\in\mathcal F(TQ)$ satisfies the conditions \eqref{Lag01}    for \emph{any} vertical vector field $V$
and \emph{any} second order vector field $D$, then  $X^{(N)}$ is $\oL$-Hamiltonian with Hamiltonian function $\varphi=i_{X^{(N)}}\tL-u.$
\end{lemm}

\noindent We can now use this result to set a correspondence between a class of  infinitesimal symmetries for a dynamics and a class of constants of the motion within the Lagrangian formalism. Let us assume as in \eqref{eq32} that the dynamics is a second order vector field $D$ which  satisfies $i_D\oL=\dd \EL$, and  that $X^{(N)}$ is a Newtonian $\oL$-Hamiltonian  infinitesimal symmetry for it, i.e. $[X^{(N)},D]=0$ with a $\varphi\in\mathcal F(TQ)$ such that $i_{X^{(N)}}\oL=\dd\varphi$. It is clearly $$L_D\varphi=0,$$ i.e. $\varphi$ is invariant along the integral curves of $D$, and from the previous analysis, there exists a function $u\,=\,i_{X^{(N)}}\tL-\varphi\,\in\mathcal F(TQ)$ with $L_Vu=0$ for any vertical vector field $V$ such  that $L_D(i_{X^{(N)}}\tL)=L_Du$. From \eqref{eq34} and the first identity out of \eqref{eq35} this gives 
\beq
\label{eq36a}
L_{X^{(N)}}\cL=L_Du.
\eeq
 A one parameter group of point transformations on $TQ$, whose infinitesimal generator $X^{(N)}$ satisfies such relation for a given function $u\in\mathcal F(TQ)$ which is constant along any vertical vector field (and therefore can be identified with $u=\pi^*f$, for $f\in\mathcal F(Q)$) is usually referred to as a \emph{symmetry} for the Lagrangian $\cL$.
In this case we say the Lagrangian is quasi invariant, or that the Lagrangian changes by a total time derivative of a function of the position coordinates.

 \noindent Assume now that $X^{(N)}$ is a Newtonian vector field on $TQ$. From \eqref{eq26} and \eqref{eq27} we see that the commutator $[X^{(N)},D]$ is vertical, so that $i_{[X^{(N)},D]}\tL=0$. If $X^{(N)}$ generates  a symmetry for the Lagrangian, i.e. there exists a function $u=\pi^*f$ with $f\in\mathcal F(Q)$ such that 
$ L_{X^{(N)}}\cL=L_Du$, 
we immediately see from \eqref{eq34} that 
\beq
\label{impoL1}
\varphi=i_{X^{(N)}}\tL-u
\eeq is constant along $D$. From \eqref{eq35} one proves that 
\beq
\label{eq44}
i_{X^{(N)}}\oL=\dd\varphi.
\eeq
 Since both $D$ and $X^{(N)}$ are therefore  $\oL$-Hamiltonian, one has
$$
0\,=\,L_D\varphi\,=\,\{\varphi, \EL\}\,=\,-L_{X^{(N)}}\EL
$$ with respect to the Poisson structure associated to $\oL$, and this gives 
\beq
\label{eq45}
[X^{(N)}, D]=0,
\eeq
 i.e. the Newtonian $X^{(N)}$ vector field is an infinitesimal symmetry for the dynamics $D$, with the energy function $\EL$  invariant along $X^{(N)}$. 
This analysis gives the usual formulation of the Noether theorem within the Lagrangian setting, that we write as
\begin{prop}
\label{noeL1} 
    A Newtonian vector field $X^{(N)}$ gives an infinitesimal $\oL$-Hamiltonian symmetry for the (regular) Lagrangian dynamics $D$  (and therefore a function $\varphi\in\mathcal F(TQ)$, via the map $\tilde{\tau}$,  which is invariant along $D$) if and only if it generates a one parameter group of symmetry point transformations for the Lagrangian $\cL$.
    \end{prop}

\begin{example}
 \label{exemplum-lag1}   
Recalling the example \ref{I-exemplum1}[I], we consider the dynamics $\Gamma$ of a point particle moving in an external radial force field. The configuration space for such system is $Q=\R^2$ with global coordinates $(x,y)$, the corresponding tangent bundle is $TQ$ with fiber coordinates $(v_x,v_y)$,  the dynamics is Lagrangian with 
\beq
\label{exL1}
\cL\,=\, \frac{1}{2}(v_x^2+v_y^2)-V(r),
\eeq   
giving $$\theta_{\cL}=v_x\dd x+v_y\dd y.$$ It is immediate to check that the vector field $$X\,=\,y\del_x-x\del_y$$ provides an infinitesimal point transformation (infinitesimal rotation) given by $$X^{(N)}\,=\,y\del_x-x\del_y+v_y\del_{v_x}-v_x\del_{v_y}$$ such that $L_{X^{(N)}}\cL=0.$ The corresponding constant of the motion is (see \eqref{I-Lfun}[I])
\beq
\label{exL2}    
i_{X^{(N)}}\tL=yv_x-xv_y.
\eeq    
    
\end{example}    
 
    \noindent Although elementary, the example above shows that, when the dynamics has a Lagrangian formulation with $\cL=(1/2)g_{ab}v^av^b-V$ (where the kinetic energy term is given by a positive definite quadratic form $g_{ab}$, and $V=V(x)$ depends on the configuration space variables), then the constant of the motion provided by the Noether theorem and given by \eqref{impoL1} is (affine)  linear in the $v^a$ variables.  This is not an accident, as the following proposition  (whose  proof we omit, since it comes from the above discussion), clarifies, focussing  on the so called \emph{inverse} Noether theorem. 
\begin{prop}
\label{noeL1bis}
Let $D$ be a second order vector field on $TQ$ which has a Lagrangian formulation, i.e. $L_D\tL=\dd\cL$ for an admissible regular Lagrangian function $\cL$. 

\noindent If there exist a Newtonian vector field $X^{(N)}$ on $TQ$ and a function 
$u\in\mathcal F(TQ)$ with $L_Vu=0$ for any vertical vector field $V$, such that the function $\varphi=i_{X^{(N)}}\tL-u$ is invariant along $D$, that is $L_D\varphi=0$, then it is:
\begin{itemize}
\item $L_{X^{(N)}}\cL=L_Du$ (i.e. $X^{(N)}$ generates a point symmetry for $\cL$); 
\item $i_{X^{(N)}}\oL=\dd\varphi$ and $[X^{(N)},D]=0$ \\ (i.e. $X^{(N)}$ is an infinitesimal $\oL$-Hamiltonian symmetry for the dynamics);
\end{itemize}
If there exists a Newtonian vector field $X^{(N)}$ which is an infinitesimal symmetry for the dynamics, that is $[X^{(N)}, D]=0$, and generates a symmetry for the Lagrangian $\cL$, that is $L_Du=L_{X^{(N)}}\cL$ for a given $u$ such that $\,L_Vu=0$, then $\varphi=i_{X^{(N)}}\tL-u$ is invariant along the integral curves of $D$.

\end{prop}

\noindent We further notice   that, if  the (regular)  Lagrangian $\cL$ for the given second order dynamics $D$ on $TQ$   has a set of symmetries, that is there exist the $\oL$-Hamiltonian Newtonian vector fields $X^{(N)}_a$ and corresponding $u_a$ (with $a=1,\dots,k$) with $ L_{X_a^{(N)}}\cL=L_Du_a$, then one proves\footnote{The Poisson brackets are defined in terms of the symplectic form $\oL$.} 
\beq
\label{eq45p1}
i_{[X_a^{(N)},X_b^{(N)}]}\oL\,=\,\dd\{\varphi_b,\varphi_a\}
\eeq
with 
\beq
\label{eq45p2}
 \{\varphi_b,\varphi_a\}\,=\,i_{[X_a^{(N)},X_b^{(N)}]}\tL\,-\,u_{\{\varphi_b,\varphi_a\}}
 \eeq
where the function $$u_{\{\varphi_b,\varphi_a\}}\,=\,\{u_b,\varphi_a\}\,-\,\{u_a,\varphi_b\}$$  is easily seen to be constant along any vertical vector field on $TQ$. This shows that the set of Newtonian symmetries for a Lagrangian dynamics closes a Lie algebra. 

\begin{example}
\label{exemplum-lagoa}
An interesting example comes by considering the dynamics of three dimensional isotropic harmonic oscillator
$$
D\,=\,v^a\frac{\del}{\del q^a}\,-\,q^a\frac{\del}{\del v^a}, \qquad\qquad a=1,\dots,3.
$$
On $Q=\R^3$, with respect to global coordinates $(q^a,v^a)$, any function   
\beq
\label{oa1}
\cL=\frac{1}{2}h_{ab}(v^av^b-q^aq^b),
\eeq
with $h$ a symmetric non singular $(3\times 3)$ matrix, gives an admissible regular Lagrangian for $D$. The signature of $h$ characterises the classes of non equivalent Lagrangians, so we consider the following two examples.
\begin{itemize}
\item For $h_{ab}=\delta_{ab}$ we have that the vector fields
$$
\mathfrak X(Q)\,\ni\,X_j\,=\,\varepsilon_{jab}q^a\frac{\del}{\del q_b}
$$
give 
$[X_j,X_k]=\varepsilon_{jks}X_s$, i.e. they generate an action of the Lie algebra $\mathfrak g=\mathfrak{su}(2)$ on $Q$,  and also that  their  tangent lift $X^{(N)}_j$ give an action of the same Lie algebra  $[X^{(N)}_j,X^{(N)}_k]=\varepsilon_{jks}X^{(N)}_s$ in terms of Newtonian infinitesimal symmetries for $D$.  The Noether corresponding invariant functions are easily seen to be 
$$
f_{a}\,=\,\varepsilon_{abc}q^bv^c,
$$
i.e. the components of the angular momentum.
\item For $h={\rm diag}(1,1,-1)$ we have that the vector fields 
$$
J\,=\,q^1\del_2-q^2\del_1;\qquad K_1\,=\,q^3\del_1+q^1\del_3, \qquad K_2\,=\,q^3\del_2+q^2\del_3
$$ generate an action of the Lie algebra $\mathfrak g\,=\,\mathfrak{so(2,1)}$ on $Q$, their Newtonian lift generate an action of the same Lie algebra in terms of Newtonian symmetries for the dynamics. The corresponding constants of the motion coincide with the $f_a$ above.  
\end{itemize}
It is interesting to notice that the Lie algebras of symmetries we considered is given by   Killing vectors for the metrics given by $h$ on $Q$. 
\end{example}

\subsection{A (first) generalisation of Noether symmetries}
\label{subsub:ex}

 A generalisation of the Noether theorem within the Lagrangian formalism has been studied in \cite{pv0,cpv1,cpv2}. Consider a vector field $X\in\mathfrak X(TQ)$ which is not necessarily Newtonian, and assume that a function $u\in\mathcal F(TQ)$ exists, such that 
 \beq
 \label{Lag001}
 L_X\tL=\dd u.
 \eeq
  Assume further  that the function $u$ and the vector field $X$ satisfy the relations (notice that the first generalises the previous \eqref{eq36a}) 
 \begin{align}
 &L_X\cL=L_Du, \nn\\ &i_{[X,D]}\tL=0. \label{Lag02}
 \end{align}
  It is 
\beq
\label{eq43}
i_X\oL\,=\,-i_X\dd\tL\,=\,-L_X\tL\,+\,\dd(i_X\tL)\,=\,\dd\{i_X\tL\,-\,u\},
\eeq
which proves that $X$ is $\oL$-Hamiltonian, with respect to the Hamiltonian function 
\beq
\label{eq43poi}
\varphi=i_X\tL-u.
\eeq As in the previous section, it is then immediate to see that $L_D\varphi=0$ and that $[X,D]=0$, i.e. $X$, although not Newtonian, turns out to be an infinitesimal $\oL$-Hamiltonian symmetry for the Lagrangian dynamics $D$.  An example of such a symmetry is given in section 15.2 in \cite{mssv85}. 

\noindent Notice that such a result can be inverted. Let $\varphi\in\mathcal F(TQ)$ be a constant of the motion for the Lagrangian dynamics $D$ given by \eqref{eq32}. Define the corresponding Hamiltonian vector field $X$ on $TQ$ by $i_X\oL\,=\,\dd\varphi$, and the function $u\,=\,i_X\tL-\varphi$. It is immediate to prove that $[X,D]=0$ (i.e. $X$ is an infinitesimal, non necessarily Newtonian symmetry for the dynamics) and that the relations \eqref{Lag001} e \eqref{Lag02} are valid. We write 
\begin{prop}
\label{noeL2}
Consider a regular Lagrangian dynamics $D$ with a Lagrangian function $\cL$. If a function $u\in\mathcal F(TQ)$ exists, such that $L_X\cL=L_Du$  and the relations \eqref{Lag001} and \eqref{Lag02} are valid, then $X$ is $\oL$-Hamiltonian with corresponding Hamiltonian function $\varphi=i_X\tL-u$, and such function is invariant along the dynamics. 

\noindent Conversely, if an element  $f\in\mathcal F(TQ)$ such that $L_Df=0$ exists, then the Hamiltonian vector field defined by $i_X\oL=\dd f$ is an infinitesimal symmetry for the dynamics and for the function $u=i_X\tL-f$ the relations \eqref{Lag001}-\eqref{Lag02} hold.
\end{prop}
\noindent Before describing a more general setting for the Noether theorem within the Lagrangian formalism, it is interesting to notice that the correspondence between symmetries and constants of the motion for a given dynamics strongly depends on the Lagrangian function $\cL$ on $TQ$, as the following example shows.
\begin{example}
\label{exemplum-lag11}
 Consider (see \cite{other}) the dynamics of a two-dimensional harmonic oscillator, given by the second order vector field on $T\R^2$
\beq
\label{osar1}
D\,=\,v^a\frac{\del}{\del q^a}\,-\,q^a\frac{\del}{\del v^a}
\eeq
with $(a=1,2)$, which has a Lagrangian formulation with 
$$
\cL=\frac{1}{2}((v^1)^2+(v^2)^2-(q^1)^2-(q^2)^2),
$$
 giving $\oL=\dd q^a\wedge\dd v^a$. A point symmetry for this dynamics is given by the squeezing, i.e. a dilation along the $q^1$ coordinate and a contraction (by the same factor) along the $q^2$ coordinate. The infinitesimal generator of such a symmetry is the Newtonian vector field
$$
X_S\,=\,q^1\frac{\del}{\del q^1}+v^1\frac{\del}{\del v^1}-q^2\frac{\del}{\del q^2}-v^2\frac{\del}{\del v^2}.
$$
One directly proves that $[D,X_S]=0$, but the hypothesis of both claims of the proposition \ref{noeL2} do not hold.  From $$L_{X_S}\theta_{\cL}=2(v^1\dd q^1-v^2\dd q^2)$$ one sees that there is no element $u\in\mathcal F(TQ)$ such that $L_{X_S}\tL=\dd u$, and moreover $\dd(i_{X_S}\oL)\neq0$, which means that the squeezing is an infinitesimal symmetry which is not Hamiltonian with respect to $\oL$. One can also prove that $L_{X_S}\cL=L_Du$ with $$u=q^1v^1-q^2v^2,$$ and (see \eqref{eq43poi}) that $\varphi=i_{X_S}\theta_{\cL}-u=0$. We have then an example of an infinitesimal symmetry that does not provide, via the Noether theorem, a constant of the motion. 

\noindent It is nonetheless possible to describe the same dynamics (i.e. the vector field $D$ from \eqref{osar1}) within the Lagrangian formalism with
\beq
\label{osar2}
\cL'\,=\,v^1v^2-q^1q^2.
\eeq
It is $$\omega_{\cL'}=\dd q^1\wedge\dd v^2+\dd q^2\wedge\dd v^1.$$ Notice that, although admissible for the same dynamics, the Lagrangians $\cL,\cL'$ are not equivalent, i.e. although not singular they do not provide the same symplectic form. It is immediate to see that the Noether theorem applies, i.e. the hypothesis of the proposition  \ref{noeL2} 
hold, with $L_{X_S}\theta_{\cL'}=0$ and $L_{X_S}\cL'=0$. The infinitesimal symmetry given by $X_S$ corresponds to the function $$\varphi=i_{X_S}\theta_{\cL'}=v^2q^1-v^1q^2,$$ which is invariant along the dynamics. We conclude this example by referring to \cite{cara99}, where a more complete analysis on alternative Lagrangian description for the classical harmonic oscillator is performed. 
\end{example}

\subsection{A (further) generalisation: Newtonoid symmetries}
\label{subsub:N}
The analysis in section \ref{subsub:ex} above extends the class of constants of the motions and infinitesimal symmetry which are connected by a Noether theorem to a more general class then the  Newtonian vector fields analysed in \ref{subsubL2}, which are connected to invariant functions  \eqref{impoL1}  under the \eqref{Lag01} conditions. In order to characterise a suitable class of infinitesimal symmetry corresponding to the invariant functions \eqref{eq43poi}  we refer to   \cite{mms83, mssv85, mm86}.  The starting point is to consider that, given a Lagrangian dynamics $D$ on $TQ$, the set of interesting infinitesimal symmetries for it should be scrutinised within the set of those vector fields $X\in\mathfrak X(TQ)$ which map the second order $D$ into a second order field, \emph{not necessarily any} second order vector field into a second order vector field.  

\noindent Recalling the relation \eqref{eq26bis}, it is natural, given a second order vector field $D\in\mathfrak X(TQ)$, to define $X\in\mathfrak X(TQ)$ to be \emph{Newtonoid} with respect to $D$ if $S([X,D])=0$. A vector field $X$ is seen to be Newtonoid with respect to $D$ if and only if it can be written in a coordinate chart $(q,v)$ as 
\beq
\label{eq36}
X\,=\,X^a\,\frac{\del}{\del q^a}\,+\,(L_DX^a)\,\frac{\del}{\del v^a}
\eeq
with $X^a\in\mathcal F(TQ)$. Notice the analogy between \eqref{eq36} and \eqref{eq26}: a Newtonoid vector field reduces to a Newtonian vector field if $X^a=\pi^*\tilde{X}^a$ with $\tilde X^a\in\mathcal F(Q)$. The map
\beq
\label{eq37}
X\quad\mapsto\quad X^{(D)}\,=\,X\,+\,S[D,X]
\eeq
associates to any vector field $X\in\mathfrak X(TQ)$ a Newtonoid vector field $X^{(D)}$ with respect to a second order vector field $D$. Written in local coordinates, the action \eqref{eq37} coincides with \eqref{eq36}. It is immediate to see  that, if $V$ is a vertical vector field on $TQ$, then $$S([D,V])=-V$$ so it is  $V^{(D)}=0$. Moreover, since the set of second order vector fields on $TQ$ is an affine space modelled on $\mathfrak X^V(TQ)$, one computes that 
\beq
\label{eq40}
X^{(D+V)}\,=\,X^{(D)}+S([V,X^{(D)}])
\eeq
 for any $V\in\mathfrak X^V(TQ)$.

\noindent Let $X^{(D)}$ be a Newtonoid vector field for the Lagrangian dynamics $D$ such that a function $u\in\mathcal F(TQ)$ exists fullfilling
\beq
\label{eq38}
L_{X^{(D)}}\cL\,=\,L_Du.
\eeq
Given the identity 
\beq
\label{eq39}
i_A\tL\,=\,i_{S(A)}\dd\cL\,=\,L_{S(A)}\cL
\eeq
(valid for any $A\in\mathfrak X(TQ)$ and any $\cL\in\mathcal F(TQ)$), with $S([D,X^{(D)}])=0$ valid for the Newtonoid  $X^{(D)}$, it is immediate to see from \eqref{eq34} that  
\beq
\label{eq41}
L_D(i_{X^{(D)}}\tL-u)\,=\,i_{X^{(D)}}(\dd \EL-i_D\oL),
\eeq
i.e. $$\varphi=i_{X^{(D)}}\tL-u$$ is a constant\footnote{Since one also proves from \eqref{eq39} that $i_{X^{(D)}}\tL\,=\,i_X\tL$, it is $\varphi=i_X\tL-u$.} of the motion for the dynamics $D$ from \eqref{eq32}. 
This is the form of a \emph{direct} Noether type theorem for Lagrangian dynamics $D$: if the relation \eqref{eq38} is valid for a given $X\in\mathcal F(TQ)$, then $\varphi \in\mathcal F(TQ)$ is invariant along the dynamics. 

\noindent As in \cite{mms83} we can prove  the identities
\begin{align}
&\dd\varphi\,-\,i_{X^{(D)}}\oL\,=\,L_{X^{(D)}}\tL\,-\,\dd u, \nn\\
&L_{X^{(D)}}\EL\,=\,i_D(\dd\varphi\,-\,i_{X^{(D)}}\oL),\label{eq42.a}
\end{align}
with the second in  \eqref{eq42.a} being equivalent to  \eqref{eq41}. 
Further, one has 
\begin{align}
&i_{[X^{(D)},D]}\oL\,=\,L_D(\dd\varphi\,-\,i_{X^{(D)}}\oL)\,+\,L_{X^{(D)}}(\dd \EL\,-\,i_D\oL), \nn \\
&i_{S(A)}L_{X^{(D)}}(i_D\oL-\dd\EL)\,+\,i_A(\dd\varphi-i_{X^{(D)}}\oL)=\{i_{A+[D,S(A)]}-L_Di_{S(A)}\}(L_{X^{(D)}}\tL-\dd u),
\label{eq42}
\end{align}
for any $A\in\mathfrak X(TQ)$. It is clear from these relations that, contrary to what happens in the cases we considered in the sections \ref{subsubL2} e \ref{subsub:ex} above, if the Lagrangian describing a second order dynamics $D$ on $TQ$ has a Newtonoid symmetry $X^{(D)}$ as in \eqref{eq38}, it is not in general true that $X^{(D)}$ is $\oL$-Hamiltonian, nor that $X^{(D)}$ commutes with $D$, nor that the energy function $\EL$ is invariant along $X^{(D)}$. We notice that, in the last relation out of \eqref{eq42}, the vector fields $S(A)$ and $A+[D,S(A)]$ are vertical for any choice of $A$. Straightforward calculations prove that 
\beq
\label{eq46}
i_V(L_{X^{(D)}}\tL\,-\,\dd u)\,=\,L_{S([V,X^{(D)}])}\cL\,-\,L_V u
\eeq
for any vertical vector field $V$ on $TQ$.  Upon merging \eqref{eq46} with \eqref{eq40} we have 
\begin{align}
i_V(L_{X^{(D)}}\tL\,-\,\dd u)&=\,L_{X^{(D+V)}}\cL\,-\,L_{X^{(D)}}\cL\,-\,L_V u \nn \\&=L_{X^{(D+V)}}\cL\,-L_{V+D} u,
\label{eq47}
\end{align}
(where the last line comes from \eqref{eq38}) for any vertical vector field $V$ on $TQ$. If we indeed \emph{define}, following \cite{mms83, mssv85, mm86}, that a Newtonoid vector field $X^{(D)}$ is a symmetry for the Lagrangian dynamics $D$ on $TQ$ if a function $u\in\mathcal F(TQ)$ exists, such that the relation 
$L_{X^{(D)}}\cL\,=\,L_D u$
is valid \emph{for any} second order field $D$, then we have that the r.h.s. of the last line in \eqref{eq42} vanishes, for any $A\in\mathfrak X(TQ)$. Notice that this condition can be equivalently written as
\begin{align}
&L_{X^{(D)}}\cL\,=\,L_D u, \nn \\
&L_{S([V,X^{(D)}])}\cL\,=\,L_V u\quad\forall\, V\in\mathfrak X(TQ)\,:\,S(V)=0
\label{eq48}
\end{align}
Under this assumption, which is evidently stricter than the previous \eqref{eq38}, many interesting  consequences are valid. Again from the last line out of \eqref{eq42} it follows that $$\dd\varphi=i_{X^{(D)}}\oL,$$ i.e. $X^{(D)}$ is $\oL$-Hamiltonian with Hamiltonian function $\varphi=i_{X^{(D)}}\tL-u$ which, from \eqref{eq39}, does not depend on the vertical part of $X^{(D)}$. From the second and the third relation out of \eqref{eq42} it follows that $$[X^{(D)},D]=0,$$ i.e. the Newtonoid vector field $X^{(D)}$ is an infinitesimal symmetry for the dynamics, and that $\EL$ is invariant along $X^{(D)}$. What we have proven under the hypothesis \eqref{eq48} is another form of a direct Noether theorem, that is a procedure that allows to map a Newtonoid symmetry (satisfying the relation \eqref{eq48}) for a Lagrangian dynamics $D$ into a constant of the motion for such a dynamics. 

\noindent This theorem can be inverted. If $D$ is a Lagrangian dynamics on $TQ$, and $\varphi\in\mathcal F(TQ)$ is a constant of the motion for it, that is $L_D\varphi=0$, then one defines (the Lagrangian is indeed regular) the $\oL$-Hamiltonian vector field $X_{\varphi}$ by $$i_{X_{\varphi}}\oL=\dd\varphi$$ and the function $u=i_{X_{\varphi}}\tL-\varphi$. It is then  possible  to prove that one has
\begin{align}
&L_{X_\varphi}\cL=L_Du, \nn \\
&[X_\varphi,D]=0, \nn \\
&\dd u=L_{X_\varphi}\tL
\label{eq49}
\end{align}
Since $X_{\varphi}$ commutes with $D$, one has that $X_{\varphi}^{(D)}=X_{\varphi}$. From the third relation in \eqref{eq49} one sees  that $X_\varphi$ provides a Newtonoid vector field for $D$ which satisfies the conditions \eqref{eq48}. This result gives an inverse to the Noether theorem and characterises the class of vector fields which corresponds, via the symplectic 2-form $\oL$, to a generic invariant function for the second order (regular) Lagrangian dynamics. Such a class is given by the set of  Newtonoid vector fields on $TQ$.  We write
\begin{prop}
\label{noeL3}
If a Lagrangian dynamics $D$ has an infinitesimal Newtonoid symmetry (i.e. a vector field $X^{(D)}$ which satisfies the conditions \eqref{eq38}, \eqref{eq48} with a given $u\in\mathcal F(TQ)$), then the function $\varphi=i_{X^{(D)}}\tL-u$ is invariant along the integral curves of $D$ and $X^{(D)}$ is $\oL$-Hamiltonian corresponding to $\varphi$. 

\noindent Conversely,  if $f\in\mathcal F(TQ)$ satisfies $L_Df=0$, then the corresponding $\oL$-Hamiltonian vector field $X_f$ is a Newtonoid symmetry for the dynamics.
\end{prop}

\begin{example}
\label{exemplum-Lag2}
Consider, on $TQ=\R^4$, the dynamics of the two dimensional harmonic oscillator (with frequency $\omega=1$) as in the example \ref{exemplum-lag1}, 
\beq
\label{exL3}
D=v^1\del_{q^1}+v^2\del_{q^2}-q^1\del_{v^1}-q^2\del_{v^2},
\eeq
which has a Lagrangian formulation with 
\beq
\label{exL4}
\cL=\frac{1}{2}((v^1)^2+(v^2)-(q^1)^2-(q^2)^2).
\eeq
It is easy to check that the function 
\beq
\label{exL5}
\varphi=v^1v^2+q^1q^2
\eeq
is invariant along $D$, i.e. $L_D\varphi=0$, and the corresponding $\oL$-Hamiltonian vector field, i.e. 
\beq
\label{exL6}
X\,=\,v^2\del_{q^1}+v^1\del_{q^2}-q^2\del_{v^1}-q^1\del_{v^2}
\eeq
with $$\oL=\dd q^1\wedge\dd v^1+\dd q^2\wedge\dd v^2$$ is $D$-Newtonoid without being Newtonian.  It is nonetheless immediate to see that the function 
\beq
\label{exL5-1}
\varphi'\,=\,\frac{1}{2}((v^1)^2-(v^2)^2-(q^1)^2+(q^2)^2)
\eeq
is invariant along $D$ and provides another example\footnote{We notice that the invariant functions $\varphi, \varphi'$ are mapped by the Legendre transform under $\cL$ into the functions $u_1,u_3$ in \eqref{I-eqsy8}[I] analysed in the example \ref{I-exemplum-mom1}[I].} of a $D$-Newtonoid symmetry which is not Newtonian.

\noindent This is specific example for the two dimensional case. For a $n$-dimensional harmonic oscillator dynamics $$D=v^a\del_{q^a}-q^a\del_{v^a}$$ with $$\cL=(1/2)\delta_{ab}(v^av^b-q^aq^b)$$ it is known \cite{inv} that the components $$Q^{ij}=v^iv^j+q^iq^j$$ of the quadrupole tensor are invariant along $D$. The corresponding $\oL$-Hamiltonian vector fields $X_{Q^{ij}}$ are easily seen to be $D$-Newtonoid infinitesimal symmetries for the dynamics. 

\noindent Analogously, for the Kepler problem on $Q=\R^3\backslash\{0\}$, with 
$$
D\,=\,v^a\frac{\del}{\del{q^a}}-c\frac{q^a}{r^3}\frac{\del}{\del v^a}
$$
with $c\in\R$ and $r^2=\delta_{ab}q^aq^b$, the functions
$$
R^s\,=\,(\delta_{ab}v^av^b-c/r)q^s-(\delta_{ab}v^aq^b)v^s
$$
(which comes from the Runge-Lenz vector) are invariant along $D$. Their $\oL$-Hamiltonian vector fields are seen to be $D$-Newtonoid.

\end{example}

\noindent As in the case of  Newtonian symmetries,  it is possible to prove that,  if  the (regular)  Lagrangian $\cL$ for the given second order dynamics $D$ on $TQ$   has a set (labelled by $a=1,\dots,k$) of Newtonoid symmetries \eqref{eq49},  then the relations \eqref{eq45p1} and \eqref{eq45p2} are valid, i.e. 
 \beq
\label{eq49p1}
i_{[X_a^{(D)},X_b^{(D)}]}\oL\,=\,\dd\{\varphi_b,\varphi_a\}
\eeq
with 
\beq
\label{eq49p2}
 \{\varphi_b,\varphi_a\}\,=\,i_{[X_a^{(D)},X_b^{(D)}]}\tL\,-\,u_{\{\varphi_b,\varphi_a\}}
 \eeq
upon defining  $u_{\{\varphi_b,\varphi_a\}}\,=\,\{u_b,\varphi_a\}\,-\,\{u_a,\varphi_b\}$.  This shows that also Newtonoid symmetries close a Lie algebra. For a more careful analysis of this topic we refer also to \cite{mukunda1}.


\subsection{Reduction within the Lagrangian formalism}
\label{sss:red}
Following our analysis on the Noether theorem for a second order (regular) Lagrangian dynamics $D$ on $TQ$, we consider, without dwelling upon a complete theory, examples of  reduction procedures driven by the presence of the Noether constants of the motions within the Lagrangian formalism. 

\noindent We start by considering again the  example  \ref{exemplum-lag1} of a particle moving in an external radial force field.
\begin{example}
\label{exemplum-Lag3}
We consider only motions such that for the invariant function (the angular momentum, see \eqref{exL2}) we have $$L=q^2v^1-q^1v^2\neq0,$$ so that we can consider $Q_0=\R^2_0$ and the radial $(r,\theta)$ coordinates. We write  
\beq
\label{exL7}
\cL\,=\,\frac{1}{2}(v_r^2+\frac{L^2}{r^2})-V(r)
\eeq 
for the Lagrangian, with
\beq
\label{exL8}
\oL\,=\,\dd r\wedge\dd v_r+\dd\theta\wedge\dd L
\eeq
for the symplectic form, and 
\begin{align*}
&L=r^2v_{\theta},\\ &\dd L=2rv_{\theta}\dd r+r^2\dd v_{\theta}.
\end{align*}
 The equations of motions (see \eqref{I-sODE1}[I]) are written as 
 \begin{align}
\dot\theta=v_{\theta},&\qquad \dot L=0, \nn \\
\dot r=v_r,&\qquad \dot v_{r}\,=\,\frac{L^2}{r^3}-\del_rV.
 \label{exL9}
 \end{align}
Upon fixing $L=l\neq0$, it is immediate to see that radial motion is given by the second order vector field 
$$
D_l\,=\,v_r\frac{\del}{\del r}+(\frac{l^2}{r^3}-\del_rV)\frac{\del}{\del {v_r}}
$$
which satisfies the Lagrangian condition $$L_{D_l}\theta_{\mathcal L'}\,=\,\dd\mathcal L'$$
with 
\beq
\label{exL10}
\mathcal L'\,=\,\frac{1}{2}v_r^2-(\frac{l^2}{2r^2}+V)
\eeq
on $T\R_0^+$ with $r\in \R_0^+$.
Fixing a value $L=l\neq0$ amounts to consider the submanifold $$i_l\,:\,N_l\hookrightarrow TQ_{0}$$ with $$N_l=\{m\in TQ_0:L(m)=l\}.$$ As we already pointed out in the example \ref{I-exemplum1}[I], the manifold $T\R^+_0$ comes as the quotient of $N_l$ upon identifying the point along the integral curves of the vector field $\del_\theta$. Describing such quotient by the map $\pi\,:\,N_{l}\,\to\,T\R^+_0$, we see that for the Lagrangian function $\mathcal L'\in\mathcal F(T\R^+_0)$ one has that
$$
\cL_{\mid N_{l}}\neq\pi^*\mathcal L',
$$
the reason being the difference in the sign of the term depending on the angular momentum $L$.
This shows that the relation \eqref{I-riduH}[I] -- which is valid for a reduction driven by a constant of the motion within the Hamiltonian formalism -- does not hold for a reduction within the Lagrangian formalism. 
\end{example}

Several interesting remarks arise in this example. 
It shows that the reduction procedure we have examined does not produce directly a new Lagrangian dynamics: one first has to carry on a reduction within the symplectic setting, and afterwards searches for a Lagrangian description for the reduced dynamics. 
As an intermediate step, one may also need to analyse whether the reduced carrier space for  the dynamics has a tangent bundle structure.  

We recall that, if $M=T^*Q$ with $Q$ a smooth manifold and $X_Q=A^j(q)\del_{q^j}$ is a vector field on $Q$, then its cotangent (canonical) lift \eqref{I-califf}[I] is the vector field $$ X=A^j\del_{q^j}-p_s(\del_{q^j}A^s)\del_{p_j}$$ on $T^*Q$ which is Hamiltonian with respect to the canonical symplectic form $\omega_Q$ on $M$ with
$$
i_{X}\omega_Q\,=\,\dd f_A,
$$ 
where $f_A=p_jA^j$. If $X_Q$ has no fixed points, it is possible to prove that the codimension one submanifold given by $N_{\alpha}=f_A^{-1}(\alpha)$ for $\alpha$ a regular value of $f_A$ is transversal to the fibers of $T^*Q$ and can be further reduced by the flow of the vector field $ X$, giving a submanifold $$N_{\alpha}/X\simeq T^*(Q/X_Q),$$ where $Q/X_Q$ denotes the quotient of the base manifold $Q$ by the flow generated by $X_Q$.  More generally, if $G$ is a Lie group acting on $Q$ whose canonical lift is Hamiltonian and provides an associated momentum map (see section \ref{I-subsub:liesym}[I]) $\mu:T^*Q\to\mathfrak g^*$, then the reduced phase space is diffeomorphic to $T^*(Q/G)$. If the Hamiltonian dynamics $X_H$ is invariant under the action of $X$, then the reduced dynamics $\tilde X_{H}$ turns to be Hamiltonian on $T^*(Q/X_Q)$ (see \eqref{I-eqsy2}[I]). 

One can prove that the  reduction in the previous example \ref{exemplum-Lag3} can be generalised. If $X$ is a vector field on $Q$ whose tangent lift $X^{(N)}$ generates a symmetry for the regular  Lagrangian $\cL$ of a given second order dynamics $D$, such that $i_{X^{(N)}}\oL=\dd\varphi$, then the submanifold $N_{\alpha}/X^{(N)}$ (with $N_{\alpha}=\varphi^{-1}(\alpha)$ for a regular value $\alpha$ of the constant of the motion $\varphi$ on $TQ$, see the proposition \ref{noeL1bis}) is diffeomorphic to $T(Q/X)$. 

That the reduced dynamics $\tilde D$ on such a reduced tangent bundle has a Lagrangian formulation is not in general true, as the following example shows. 

\begin{example}
\label{exe:mm}
Consider the dynamics of a charged point particle in a magnetic monopole field. The equations of motions are given, in the Euclidean space $\R^3_0$, by the second order system
\begin{align}
&\dot x^a\,=\,v^a, \nn \\
&\dot v^a\,=\,\frac{\lambda}{r^3}\epsilon^a_{\,\,\,bc}x^bv^c
\label{monopeq1}
\end{align}
with $r^2=\delta_{ab}x^ax^b$ and $\lambda\,\in\,\R$ a suitable coupling constant. Recalling the analysis in \cite{bala-mono, mr87, ferrara88} we know that  the vector field 
\beq
\label{fe4}
D=v^a\frac{\del}{\del x^a}\,+\,\frac{\lambda}{r^3}\varepsilon_{abc}x^av^b\frac{\del}{\del v^c}
\eeq
is Hamiltonian with respect to the symplectic structure given by 
\beq
\label{fe1}
\omega\,=\,\dd x^a\wedge\dd v_a\,+\,\frac{\lambda}{2r^3}\varepsilon_{abc}x^a\dd x^b\wedge\dd x^b
\eeq
on $M=T(\R^3_0)$, with Hamiltonian function $$H=\frac{1}{2}v^av_a, $$
that is $i_D\omega=\dd H$. Since the vector fields 
$$
R_a\,=\,\varepsilon_{abc}(x^b\frac{\del}{\del x^c}+v^b\frac{\del}{\del v^c})
$$
are Hamiltonian and commute with $D$, the Noether constant of the motion corresponding to such infinitesimal symmetries within the symplectic formalism are given by 
$$
J_a\,=\,\varepsilon_{abc}x^bv^c+\lambda\frac{x^a}{r}
$$
with $i_{R_a}\omega=\dd J_a$. Notice that such $J_a$ functions are not the components of the \emph{orbital} (say) angular momentum, since they contain a $\lambda$-dependent \emph{elicity} term.

It is immediate to prove that $\dd\omega=0$, but $\omega$ is not exact. While it is obviously  $\dd x^a\wedge\dd v_a=\dd(x^a\dd v_a)$, the term depending on $\lambda$ in \eqref{fe1} gives indeed a multiple of the solid angle 2-form on the sphere ${\rm S}^2$, whose integral on the boundary of any regular connected domain in $\R^3_0$ around the origin is not zero. This means that there is no function $\cL$ on $T(\R^3_0)$ such\footnote{Notice also that, on any open and connected $U\subset\R^3_0$ which is homotopic to a point, it is $\tiny{\frac{1}{2r^3}}\varepsilon_{abc}x^a\dd x^b\wedge\dd x^c=\dd A=A_s\dd x^s$ with $A$ a 1-form on $U$, so that the restriction of $D$ to $TU$ has a (regular) Lagrangian formulation with Lagrangian  function $\cL=\tiny{\frac{1}{2}}v^av_a+\lambda A_sv^s$.}  that $\omega=-\dd\theta_{\cL}$. That this is one of the steps shaping a path  to prove that the second order dynamics $D$ has no (global) Lagrangian formulation is described  in \cite{mr87}. 

It is nonetheless possible to extend the carrier space of such a dynamics to a larger tangent bundle and to exhibit a global Lagrangian dynamics whose reduction to $T(\R^3_0)$ gives $D$. Consider the configuration space $$Q=\R^4_0\simeq \R^+_0\times {\rm S}^3$$ as the total space of a Hopf ${\rm U}(1)$-fibration, i.e. $$\pi\,:\,Q\,\simeq\,\R^+_0\times {\rm SU}(2)\,\stackrel{\rm U(1)}{\longrightarrow}\,\R^3_0\,\simeq\, \R^+_0\times \mathrm S^2.$$
 Upon parametrising $Q$ via  the radial coordinate $r>0$ and $ g\in{\rm SU}(2)\simeq{\rm S}^3$ via 
 $$
 g\,=\,\begin{pmatrix}
 u & -\bar v \\ v & \bar u\end{pmatrix}
 $$
with $\bar{u}u+\bar{v}v=1$, the projection map $\pi$ can be written as $(r,g)\,\mapsto\,x^a$ where the $x^a$ coordinates are implicitly given by the relation 
$$
x^a\sigma_a\,=\,rg\sigma_3g^{-1}
$$
with respect to the Pauli matrices $\sigma_a=\sigma_a^\dagger$. The right action of the gauge group ${\rm U}(1)$, which is a Lie subgroup of ${\rm SU}(2)$, can be written as 
$$
 (g,z)\quad\mapsto\quad g\,z\,=\,\begin{pmatrix}
 u & -\bar v \\ v & \bar u\end{pmatrix}\begin{pmatrix} z & 0 \\ 0 & z^*\end{pmatrix}
 $$
with $z\in\C$ such that  $zz^*=1$ representing an element in ${\rm U}(1)$. The function (see \cite{gfd})
\beq
\label{fe2}
\cL\,=\,\frac{1}{2}{\rm Tr}(\frac{\dd}{\dd t}(rg\sigma_3g^{-1}))^2-\lambda({\rm Tr}(\sigma_3g^{-1}\dot g))^2
\eeq
gives a regular Lagrangian function on $TQ$ which is evidently invariant under the left action of ${\rm SU}(2)$ on itself and under the right action of the gauge group\footnote{Such a Lagrangian is an example of a \emph{Kaluza-Klein} system}. In order to describe the corresponding equations of the motions we notice that ${\rm S}^3\simeq{\rm SU}(2)$ is parallelisable (as it is a Lie group manifold). There exists a global basis of left and right invariant vector fields on it, that we denote by $L_a$ and $R_b$, and dual globally defined bases for the set of 1-forms given by left $\theta^a$ and right $\phi^b$ forms, with $$i_{L_a}\theta^b=i_{R_a}\phi^b=\delta^b_a.$$ Since the vector fields $L_a$ close the Lie algebra commutation relations of $\mathfrak g=\mathfrak{su}(2)$, they are not associated to a holonomic coordinate chart on ${\rm S}^3$: with this proviso, the relations 
$$
\dot\theta^a=i_\Gamma\theta^a
$$
for any second order vector field $\Gamma$ on $T{\rm S}^3$ give a set of functions on  $T{\rm S}^3$ which are linear in the fiber variables and are functionally independent. The Lagrangian \eqref{fe2} is regular, and reads
\beq
\label{fe3}
\cL\,=\,\frac{1}{2}v_r^2+\frac{1}{4}r^2((\dot{\theta}^1)^2+(\dot{\theta}^2)^2)+\lambda(\dot{\theta}^3)^2,
\eeq
giving a globally defined second order dynamics $\mathfrak D_{\cL}$ on $TQ$. 
The function $\dot\theta^3$ is a constant of the motion for $\mathfrak D_\cL$, and corresponds via the Noether theorem to the Newtonian vector field $\tilde X_3$ on $T{\rm S}^3$ given by the tangent lift of the vector field $X_3$ (which generates the action of the gauge group and therefore is the vertical vector field of the Hopf fibration). If we denote by $N_{\alpha}=\{\dot\theta^3=\alpha\}$ the codimension one submanifold given by the level set of $\dot\theta^3$ for a regular value $\alpha$, it is possible to prove \cite{gfd} that for the quotient it is 
$$
N_{\alpha}/\tilde X_3\,\simeq\,T(\R_0^3)
$$
and that the dynamics $\mathfrak D_{\cL}$ reduces to $D$ \eqref{fe4}. This  shows an example of a Lagrangian dynamics (of the  Kaluza-Klein type)  whose reduction is not Lagrangian, but just symplectic. This dynamics will also be studied in the example \ref{exemplum-sL2} via a singular Lagrangian. 

\end{example}
A different analysis of the reduction process for a regular Lagrangian system on a tangent bundle $TQ$ driven by a symmetry for the Lagrangian is in \cite{CMPR03, lrbs99, MaSc93}. If the Lagrangian $\mathcal L$ is invariant under a point transformation generated by  the action of a symmetry group $G$ on $Q$, so that $\mathit{l}\in\mathcal F(TQ/G)$ is the reduced Lagrangian,  then the resulting equations of the motion for the reduced system are proven to come as stationary points for an action functional $\int_{t_1}^{t_2}\dd t\,\mathit{l}([q,\dot q]_G)$ with respect to a suitable family of variations which have an interesting geometric interpretation when  written (they are also known as Lagrange-Poincar\'e variational principles) 
upon considering the bundle isomorphism $TQ/G\simeq T(Q/G)\oplus\mathfrak g$ with $\mathfrak g$ the natural fibre of the adjoint bundle. 

\noindent We close this section by noticing that the problem of studying symmetries for a regular Lagrangian dynamics $D$
is related to the (inverse) problem of determining other admissible Lagrangians for the same dynamics. It can be proven (see \cite{inv}) that, if $D$ is a second order Lagrangian vector field on $TQ$ with admissible Lagrangian $\cL$, and  $X^{(N)}\in\mathfrak X(TQ)$ is the Newtonian lift of a given vector field $X$ on $Q$, then the function $$\cL'=L_{X^{(N)}}\cL$$ is an admissible Lagrangian for $D$ if and only if $$[X^{(N)}, D]=0.$$ Nothing prevents $\cL'$  to be equivalent to $\cL$, or to be a pure gauge function: if this happens, then $X^{(N)}$ generates a Noether symmetry for $D$. This means that, in order to find an admissible Lagrangian for $D$ using such a theorem, one has to inspect the possibility of having \emph{non} Noether symmetries for $D$. For this problem, which goes beyond our analysis,  we refer to \cite{nNesp, crampinN}.

\section{Implicit  equations from singular Lagrangian dynamics}
\label{sec:vincoli}

Within the geometric setting described in the previous pages, our attention will be mainly focussed on a specific class of implicit differential equations, namely those corresponding to singular Lagrangian functions on a tangent bundle manifold which we shall analyse in terms of the presymplectic formalism that we discussed in section \ref{I-sec:pre}[I]. Such class  can be unified under the name of \emph{generalised Hamiltonian systems}, which we now describe.

\subsection{Generalised Hamiltonian systems}
\label{sub:ghs}
Assume that $(M,\omega)$ is a symplectic manifold of dimension $n$. It is immediate to see  that  $$(TM, \omega_N=\dd_N\omega)$$ is a symplectic manifold, with $$\omega_N=-\dd\theta_N$$ where $$\theta_N=-i_N\omega.$$ 
Given the local coordinate system  $\{x^a,v^a\}_{a=1,\dots,\dim \,M=n}$, with $\omega=\omega_{ab}\dd x^a\wedge\dd x^b$, one has
\beq
\label{tabuo}
\omega_N\,=\,2\{v^a(\del_{x^c}\omega_{ab})\dd x^c\wedge\dd x^b\,+\,\omega_{ab}\,\dd v^a\wedge\dd x^b\}
\eeq
and 
\begin{align}
& \{x^i,x^j\}_N\,=\,0, \nn \\
& \{x^i,v^j\}_N\,=\,-\frac{1}{2}(\omega^{-1})^{ij}, \nn \\
& \{v^i,v^j\}_N\,=\,\frac{1}{2}(v^a\del_{x^b}\omega_{ac})((\omega^{-1})^{bj}(\omega^{-1})^{ci}-(\omega^{-1})^{bi}(\omega^{-1})^{cj})
\label{tabuo1}
\end{align}
for the corresponding Poisson structure.
From the relations \eqref{imp06} one proves that 
$$
L_{X^{(N)}}\omega_N\,=\,0\qquad\Leftrightarrow\qquad L_{X}\omega\,=\,0;
$$
moreover, if $L_X\omega=0$ (i.e. $X$ is a locally Hamiltonian vector field on $M$), then 
$$
i_{X^{(N)}}\omega_N\,=\,\dd(i_Ni_X\omega),
$$
which means that its tangent lift on $TM$ is globally Hamiltonian. In particular, if  $f\in\mathcal F(M)$ and $X_f$ is the corresponding Hamiltonian vector field on $M$, then $$i_{X_f^{(N)}}\omega_N\,=\,\dd F$$ with  $F\,=\,i_{X_f^{(N)}}\theta_N\,=\,\dd_Nf$. 

A first order differential equation on the symplectic manifold $(M,\omega)$, i.e. a submanifold $\mathfrak Z\subset TM$, is called a \emph{generalised Hamiltonian system} with respect to $(TM,\omega_N)$ if $\mathfrak Z$ is a Lagrangian submanifold, namely if its tangent distribution $\D_{\mathfrak Z}$ satisfies the condition $$\D_{\mathfrak Z}^{\perp}=\D_{\mathfrak Z}$$ with respect (see the definition \eqref{I-orthosymp}[I] in section \ref{I-sub:dirac}[I]) to $\omega_N$. 

Both Hamiltonian and regular Lagrangian systems are generalised Hamiltonian systems, as the following \eqref{e28.1} and \eqref{e28.2} show.  
\begin{enumerate}[(a)]

\item
\label{e28.1}
If $(M,\omega)$ is a symplectic manifold,  with a local coordinate system given by $\{x^a\}_{a=1,\dots,n}$, and $\pi:TM\to M$ denotes  the tangent bundle fibration, with $z\in TM$ and $\pi(z)=m$, we write again a local coordinate system for $TM$ as $z\simeq(x^a,v^a)$.

\noindent Let $\alpha\in\Lambda^1(M)$ be a 1-form on $M$. The set 
\beq
\label{impli01}
\mathfrak Z\,=\,\{z\,\in\,TM\,:\,\omega(v,v')=\alpha(v') \,\,\forall\,\, \,v' \,\in\, T_{m}M\}
\eeq
is  a Lagrangian submanifold in $(TM,\omega_N)$. Indeed we can write $\mathfrak Z$ as the zero set level
\beq
\label{impli011}
\mathfrak Z\,=\,\{z\,\in\, TM\,:\,\psi_a\,=\,\omega_{ba}v^b\,-\,\alpha_a\,=\,0,\qquad a=1,\dots,\dim \,M\}
\eeq
with $\omega=\omega_{ab}\dd x^a\wedge\dd x^b$ and $\alpha=\alpha_b\dd x^b$. It is possible to prove, using \eqref{tabuo1},  that 
$$
\{\psi_i,\psi_j\}_N\,=\,\frac{1}{2}\,(\del_{x^j}\alpha_i-\del_{x^i}\alpha_j )
$$
i.e. the submanifold $\mathfrak Z$ is generated by a set of $n$ first class constraints in the $2n$-dimensonal symplectic manifold $TM$ if $\alpha$ is closed, that is if $\dd\alpha=0$.
If the 1-form $\alpha$ is exact, with $\alpha=\dd H$, then it is straightforward to prove that the set $\mathfrak Z$ defined above can be written as 
$$
\mathfrak Z\,=\,\{z\,\simeq\,(x^a,v^a)\,: \quad v^a\,=\,\{x^a,H\}\}
$$
with respect to the Poisson bracket on $M$ corresponding to the symplectic 2-form $\omega$.  This shows that the range of the section on $TM$ corresponding to (ordinary) Hamiltonian vector fields on $(M,\omega)$  provides a Lagrangian  submanifold in $(TM,\omega_N)$.

\noindent Notice that the submanifold $\mathfrak Z$ defined in \eqref{impli011} can be written, given the invertibility of the matrix $\omega_{ab}$, as 
$$
v^k=\Lambda^{ks}\alpha_s
$$
where $\Lambda^{ks}$ is  the matrix representing the Poisson tensor corresponding to the symplectic tensor $\omega$. The condition for a function $f$ on $M$ to be a constant of the motion for the equation \eqref{impli011} can be written as
$$
(\dd_Nf)_{\mid\mathfrak Z}\,=\,(v^a\del_{x^a}f)_{\mid\mathfrak Z}\,=\,(\Lambda^{ab}\alpha_b\del_{x^a}f)_{\mid M}\,=\,0,
$$
which is the well known condition on a (locally) Hamiltonian vector field.

\noindent A natural example comes upon  recalling that the isomorphism between the dual and the bidual of a finite dimensional vector space allows to prove that, for any smooth manifold $Q$, there exists a canonical diffeomorphism \cite{ccss89} $\mathfrak a\,:\,T(T^*Q)\,\to\,T^*(TQ)$ which  in local coordinates can be written as (see section  \ref{I-sec:hamilton}[I])
\beq
\label{mappaa}
\mathfrak a\,:\,(q^a,p_a, v^a, w_a)\,\mapsto\,(q^a, v^a, w_a,p_a).
\eeq
Given the canonical symplectic form $\omega_Q=-\dd\theta_Q$ on $T^*Q$ (see \eqref{I-symfo2}[I]), it is easy to see that 
\beq
\label{symfo2q}
\omega_N\,=\,\dd_N\omega_Q\,=\,\dd v^a\wedge\dd p_a\,+\,\dd q^a\wedge\dd w_a
\eeq 
is a symplectic form on $TT^*Q$, that is also $$\omega_N=\mathfrak a^*(\omega_{TQ}),$$ i.e. it comes from the canonical symplectic form $$\omega_{TQ}\,=\,\dd q^a\wedge\dd p_a\,+\,\dd v^a\wedge\dd w_a$$ on $T^*TQ$, which is a cotangent bundle.
If we consider the cotangent bundle $(M=T^*Q, \omega_Q)$, a Hamiltonian vector field $X_H\in\mathfrak X(M)$  gives the f.o.d.e. $\mathfrak Z$ represented by the zero level set 
\begin{align}
&v^a\,=\,\frac{\del H}{\del q^a}, \nn \\ &w_a\,=\,-\frac{\del H}{\del p_a}
\label{imp04}
\end{align}
on $TM=T(T^*Q)$, which  is Lagrangian with respect to $\omega_N$.  

\medskip
\item
\label{e28.2}
Let $C\hookrightarrow M$ be a $(N-k)$-dimensional submanifold, with $H\in\mathcal F(C)$. The set 
\beq
\label{impli02}
\mathfrak Z\,=\,\{z\,\in\,TM\,\,:\,\omega(v,v')\,=\,\dd H(v')\quad\forall\,v'\,\in\,T_mC\,:\,m\in \,C\}
\eeq
is proven to be a Lagrangian submanifold in $(TM,\omega_N)$. Such equations are called \emph{Dirac systems}, since their analysis generalises  the analysis by Dirac and Bergmann for the class of Euler-Lagrange equations associated to a singular Lagrangian.  If the submanifold $C$ is defined by the embedding given by 
\beq
\label{C01}
C\,=\,\{m\,\in\,M\,:\,\phi_a(x)\,=\,0\}
\eeq
with $\{\phi_a\}_{a=1,\dots,k}$ independent elements in $\mathcal F(M)$, then it is possible to prove that the submanifold $\mathfrak Z$ defined above can be written as
\beq
\label{impli03}
\mathfrak Z\,=\,\{z\,\simeq\,(x^a,v^a)\,:\,\phi_a(x)=0, \quad \omega_{ab}v^b\,=\,\del_a(H+\lambda _s\phi_s)\}
\eeq
where $\omega=\omega_{ab}\dd x^a\wedge\dd x^b$ and $\{\lambda_s\}_{s=1,\dots k}$ is a set of Lagrange multipliers. The integrability of a Dirac system has been studied in \cite{mmt95,mmt97,me-tu78}: it turns out that a Dirac system is integrable if and only if the manifold $C$ is coisotropic with respect to the symplectic structure $\omega$ on $M$, that is (recalling the analysis on submanifolds of symplectic manifolds in section \ref{I-sub:dirac}[I]) if and only if the set $\phi_a$ \eqref{C01} is a set of first class constraints. 

\end{enumerate}

Let  $\cL$ be a function on $TQ$. The range of  its differential $\dd\cL$ gives a section   of the cotangent bundle $T^*(TQ)$  which, via the (inverse of the) map $\mathfrak a$ (see \eqref{mappaa}), provides a set $\mathfrak Z_{\cL}$ of the tangent bundle $T(T^*Q)$, given as follows
$$
(q^a,v^a)\,\in\,TQ\qquad\mapsto\qquad (q^a, \frac{\del\cL}{\del v^a},v^a,\frac{\del \cL}{\del q^a})\,\in\,T(T^*Q)
$$
in the natural coordinate chart. Such  $\mathfrak Z_{\cL}$ is the set whose points $(q^a, p_a, v^a, w_a)$  in $T(T^*Q)$ satisfy the conditions 
\begin{align}
w_a&=\,\frac{\del \cL}{\del q^a} \nn \\ 
p_a&=\, \frac{\del\cL}{\del v^a}, \label{ds01}
\end{align}
i.e. $\mathfrak Z_{\cL}$ gives the Euler-Lagrange equations for the Lagrangian $\cL$ on $TQ$.  Solutions to this equation are those curves $(q^a(t), p_a(t))$ in $T^*Q$ whose first order prolongation $$(q^a(t), p_a(t), v^a(t)=\dot q^a, w_a(t)=\dot p_a)$$ gives points in $\mathfrak Z_{\cL}$ (see section \ref{ss:2}). 
It is easy to see that, for any $\cL=\cL(q,v)$, the set $\mathfrak Z_{\cL}$ is a Lagrangian submanifold embedded in $T(T^*Q)$ with respect to the symplectic form $\omega_N$ in \eqref{symfo2q}. 

\noindent When $\cL$ is regular, it is possible to invert the last relations out of \eqref{ds01} and to write $v^a=v^a(q,p)$, so that the $\mathfrak Z_\cL$ turns to be the graph of a Hamiltonian vector field on $T^*Q$ as in \eqref{imp04}, with Hamiltonian function given by the Lagrangian Energy $E_\cL=v^ap_a-\cL$ in coordinate form. 

\noindent The study of the subset $\mathfrak Z_{\cL}$ for singular Lagrangians is what we address our attention to in the following section. Here we limit ourselves to mention that the example provided by 
\beq
\label{impli06}
\cL\,=\,-m(-g_{\mu\nu}\dot q^\mu\dot q^\nu)^{1/2}
\eeq
i.e. a Lagrangian for  a relativistic free particle in the flat Minkowski spacetime $Q=\R^4$, with metric tensor  $g_{\mu\nu}={\rm diag}(-1,+1,+1,+1)$ is interesting. The equation $\mathfrak Z_{\cL}$ is given as the submanifold 
\begin{align}
&\psi_\alpha\,=\,p_\alpha-\frac{m\dot q_\alpha}{(-g_{\mu\nu}\dot q^\mu\dot q^\nu)^{1/2}}\,=\,0  \nn \\
&\psi_{\alpha+4}\,=\,\dot p_\alpha\,=\,0 \label{impli08}
\end{align}
which cannot be represented as a vector field on $T^*Q$, since the tangent bundle projection maps $\mathfrak Z_{\cL}$ onto the submanifold $C$ given by
\beq
\label{impli07}
g_{\mu\nu}p^\mu p^\nu+m^2=0
\eeq
which does not coincide with the whole $T^*Q$.

We conclude such introduction to  generalised Hamiltonian systems by  briefly describing systems formulated in terms of Morse functions \cite{bg1}. 
Let $\tilde \pi\,:\,\tilde M\,\to\,M$ be a smooth fibration, with the tangent bundle fibrations $\pi_{\tilde M}\,:\,T\tilde M\to\tilde M$ and $\pi_M\,:\,TM\,\to\,M$, and let $G\in\mathcal F(\tilde M)$ be a smooth function. The set
\beq
\label{impli04}
\mathfrak Z\,=\,\{z\,\in\,TM\,:\,\omega(T\tilde\pi(\tilde v),v)\,=\,\dd G(\tilde v) \quad\,\forall \,\tilde v\,\in\,T_{\tilde m}\tilde M\,:\,\tilde \pi(\tilde m)\,=\,m\}
\eeq
gives a Lagrangian submanifold in $(TM, \omega_N)$ provided the function $G$ satisfies a suitable condition.   
Via the vertical subbundle of the  fibration $\tilde\pi$ define the \emph{critical set} 
\beq
\label{crS}
\tilde M\,\supset\,S\,=\,\{\tilde m\,\in\,\tilde M\,:\,\dd G(u)\,=\,0 \quad\forall \,u\,:\,T\tilde\pi(u)=0\},
\eeq
then, 
with $(x^a,\lambda^s)$ a local coordinate system for $\tilde M$, consider the matrix
$$
W\,=\,\begin{pmatrix} \frac{\del^2G}{\del \lambda^s\del \lambda^k} & \frac{\del^2G}{\del \lambda^s\del x^a}\end{pmatrix}. $$ The function $G$ gives a Lagrangian submanifold \eqref{impli04} if the rank of the matrix $W$ is maximal at each point $y\in S$. In such a case, the function $G$ is called a \emph{Morse function}, the f.o.d.e. is called a \emph{generalised Hamiltonian system}. If $(x^a,v^a)$ denote a local coordinate chart on $TM$, the f.o.d.e. $\mathfrak Z$ \eqref{impli04} is 
\beq
\label{crS1}
\dot{x}^a\,=\,\{x^a, G\}, \qquad\quad \del_{\lambda^s}G=0
\eeq
in terms of the Poisson structure on $M$ corresponding to $\omega$, while the condition $\del_{\lambda^s}G=0$ represents the 
critical set \eqref{crS}.

\noindent For interesting examples we refer to \cite{mmt95,mmt97}. We limit ourselves here to notice that if we consider $Q\,=\,\R^4$ and $(M=T^*Q, \omega_Q=\dd q^a\wedge\dd p_a)$, with $$\tilde{\pi}\,:\,\tilde M=\R\times T^*Q\,\to\,T^*Q,$$ the function 
\beq
\label{impli05}
G\,=\,\lambda((-g_{\mu\nu}p^{\mu}p^{\nu})^{1/2}+m)
\eeq
turns to be a Morse function, for the equations of motion of a free relativistic particle. The  critical set is 
\beq
\label{mshe}
S\,=\,\{(-g_{\mu\nu}p^{\mu}p^{\nu})^{1/2}+m=0\},
\eeq
 the projected submanifold $\tilde \pi(S)$ defines the \emph{mass shell} on $T^*Q$  of the relativistic dynamics we are considering.  The advantage of such a formulation with respect to the one above in terms of a Dirac system is twofold. The first  is that, as a generalised Hamiltonian system, the integrability of such a dynamics can be  studied avoiding the reference to  a Hamiltonian function which would be zero on the mass shell since the Lagrangian \eqref{impli06} is linear homogeneous in the velocity variables. The second is that, since the mass shell \eqref{mshe} is different from the set defined in \eqref{impli07}, the difference between particles and antiparticles is preserved.

\subsubsection{\blu{A Noether theorem for generalised Hamiltonian systems}}\label{ss:28.b}
We refer to \cite{mmt95,mmt97} for an analysis of the integrability conditions for Dirac and generalised systems. 
Within this setting, a Noether theorem can be proven \cite{mmt92}. It is the analogue of the proposition \ref{I-noesymp2}[I] given in section \ref{I-sec:hamilton}[I]. 
\begin{prop}
\label{gh1}
Let $(M,\omega)$ be a symplectic manifold and $\mathfrak Z\subset TM$ a generalised Hamiltonian first order differential equation with respect to the symplectic manifold $(TM,\omega_N=\dd_N\omega)$. 

If $f\in\mathcal F(M)$ is a constant of the motion for $\mathfrak Z$, then the Hamiltonian vector field $X_f$ is a canonical infinitesimal symmetry for $\mathfrak Z$, i.e. $X_f^{(N)}\in\mathfrak X(\mathfrak Z)$.

If $X\in\mathfrak X(M)$ is an infinitesimal canonical symmetry for generalised Hamiltonian first order equation $\mathfrak Z$, then the Hamiltonian function $f$ for $X$, i.e. $X=X_f$, satisfies $\dd_Nf_{\mid \mathfrak Z}=k$ for a constant $k\in\R$. 
\end{prop}

 We refer to \cite{mmt92} for the details of the calculations  that provide a set of constants of the motion associated to the action of the Poincar\'e group  for the dynamics given by  \eqref{impli08}. Limiting our attention to the infinitesimal generators of the spatial rotations $$\mathfrak X(Q)\ni\,J_a=\varepsilon_{abc}q^b\del_c,$$ we see that they  are lifted to $J^{(N)}_a\in\mathfrak X(T^*Q)$,  with  $$L_{J_a^{(N)}}\psi_{\alpha}=0,\qquad L_{J_a^{(N)}}\psi_{\alpha+4}=0.$$ This means they are infinitesimal symmetries for the the dynamics, corresponding to the invariant functions $$f_a\,=\,m\,\frac{\varepsilon_{abc}q^b\dot{q}^c}{(-g_{\mu\nu}p^{\mu}p^{\nu})^{1/2}}.$$


\subsection{Singular Lagrangian dynamics }
\label{ss:vincoli}

An analysis of the implicit differential equations $\mathfrak Z_\cL$ defined in \eqref{ds01}  when $\cL$ is singular was first performed  in \cite{dirac50,dirac-lec}: in these works  Dirac studied the integrability of $\mathfrak Z_{\cL}$ on $T^*Q$, that is he was interested in considering solutions given by curves $(q(t), p(t))$ on subsets of $T^*Q$. He showed that, when the rank of the Hessian \eqref{H0} $$H_{ab}=\frac{\del^2\cL}{\del v^a\del v^b}$$ is $\rho<N$ with $N=\dim \,Q$, and does not vary over $TQ$, then $(N-\rho)$ conditions in \eqref{ds01} are (primary) constraints which define a submanifold $M\hookrightarrow T^*Q$. The condition that Cauchy data on $M$ are evolved by the equations $\mathfrak Z_{\cL}$ without leaving $M$ may result in further constraints, in a sequence that may converge\footnote{One sees that this comes if the higher order generations of constraints give a constant number of functionally independent conditions.}   to a submanifold $M'\hookrightarrow M$. When this happens, the implicit equation $\mathfrak Z_{\cL}$ results integrable as a tangent distribution to $M'$, where the rank of the distribution is given by the number of independent first class primary constraints.  This path has clear analogies with the path described in section \ref{I-sec:pre}[I] for pre-symplectic systems. For this reason, we shall not review the original Dirac's analysis of the problem\footnote{We suggest the analysis in \cite{sm-book}.}: in order to  present the analysis of the Lagrangian equations of the motions for singular $\cL$ both on $TQ$ and on $T^*Q$ we shall describe his results within the setting of pre-symplectic geometry.

The Euler-Lagrange equations $\mathfrak Z_{\cL}$ given in \eqref{ds01}, that we write as 
\begin{align}
\frac{\dd q^a}{\dd t}&=\,v^a, \nn \\
\frac{\dd}{\dd t}(\frac{\del \cL}{\del v^a})&=\,\frac{\del\cL}{\del q^a}
\label{ancor1}
\end{align}
 can indeed be studied on $TQ$. As we have seen in section \ref{ss:L}, these equations give the  stationary conditions for  the action functional corresponding to a Lagrangian function $\cL\in\mathcal F(TQ)$. They can also be written as  \eqref{eq32}, i.e.  as   
\beq
\label{eiL}
i_D\oL=\dd\EL,
\eeq
with $D$ a second order vector  field on $TQ$, i.e. $D\,=\,v^k\del_{q^k}+a^k\del_{v^k}$. If we write \eqref{ancor1} as
 \begin{align}
\label{sl1} 
&H_{ks}a^k\,=\,\left(\frac{\del \cL}{\del q^s}\,-\,\frac{\del^2\cL}{\del v^s\del q^k}\,v^k\right),
\end{align}
we see that, 
when the Hessian $H_{ks}$  has rank $N$ (i.e. the Lagrangian is  regular), then the equations of motions \eqref{sl1} can be cast in explicit form, i.e. the dynamics is described by  a unique second order  vector field $D$ on $TQ$. If $\cL$ is singular,  one can multiply both sides of \eqref{sl1} by the elements in the kernel of the Hessian matrix (i.e. the eigenvectors corresponding to the zero eigenvalue): with the l.h.s. identically vanishing, the r.h.s provides constraints for the possible Cauchy data, since it does not contain accelerations. Such constraints may\footnote{hopefully! This is indeed the stage of the analysis when the assumptions that the rank of the Hessian $H_{ab}$ does not change is crucial.} provide a submanifold in $TQ$, and one is therefore left to solve the equations of the motions on such a submanifold, on which the Lagrangian 2-form and the Energy function can be pulled back. It may happen  that null eigenvectors for the Hessian arise, so that the submanifold of the allowed Cauchy data is further restricted. When this procedure stabilizes, one has to solve the dynamics  on such a submanifold. It is clear that also
this problem presents close analogies to the theory of presymplectic systems we considered in section \ref{I-sec:pre}[I].  

In order to show such analogies (see \cite{mms83, hgn78,sm-book}), we discuss the problem in intrinsic form. 
We recall that the map
\beq
\label{eq50}
\Phi_{\cL}\,:\,(q^a,v^a)\,\in\, T_qQ\,\mapsto\,(q^a,p_a=\frac{\del \cL}{\del v^a})\,\in\,T_q^*Q
\eeq
is the Legendre transform (i.e. the fiber derivative) corresponding to a given Lagrangian $\cL\in\mathcal F(TQ)$ with $\dim Q=N$. 
 When $\cL$ is regular, the map $\Phi_{\cL}$  is a local  diffeomorphism\footnote{When the corresponding Legendre transform $\Phi_{\cL}$ is a global diffeomorphism  $TQ\leftrightarrow T^*Q$, the Lagrangian $\cL$   is usually called \emph{hyperregular}.} with $$\FL^*\omega_Q=\oL,$$ where $$\omega_Q=\dd q^s\wedge\dd p_s$$  is the canonical symplectic 2-form (see \eqref{I-symfo2}[I]) on $T^*Q$ described in section \ref{I-sec:hamilton}[I]. If one then defines $H\in\mathcal F(T^*Q)$ via 
\beq
\label{mise01}
\FL^*H=\EL,
\eeq
the integral curves on $TQ$ of the Hamiltonian vector field $D$ with $i_{D}\oL=\dd\EL$ are bijectively mapped under $\FL$ into the  integral curves on $T^*Q$ of the Hamiltonian vector field $X_H$ defined by  
\beq
\label{mise02}
i_{X_H}\omega_Q=\dd H.
\eeq  
When the rank  of the matrix $H_{ab}$ is not maximal, i.e. ${\rm rk}\,H_{ab}=\rho<N$ (notice that we assume here, as we did in section \ref{I-sec:pre}[I], the standard hypothesis that such rank is constant on $TQ$), the vertical subspace $$\ker^V\oL\,=\,\{\ker\oL\cap\mathfrak X^{(V)}(TQ)\}\,\subseteq\ker\oL$$ is spanned by  $(N-\rho)$ vertical vector fields $K^V_{(\sigma)}\,=\,A_{(\sigma)}^j\del_{v^j}$ with $A^j_{(\sigma)}\in\mathcal F(TQ)$ and $\sigma=1,\dots,N-\rho$, where the $A^j_{(\sigma)}$ can be chosen in such a way that   
\beq
\label{nucleoH}
A^j_{(\sigma)}H_{js}=0
\eeq
 for any $s=1,\dots,N$.  Upon assuming the (usual) hypothesis that the distribution given by $\ker^V\oL$ gives a regular foliation (notice that the closedness of $\oL$ allows to prove that not only  is the distribution  $\ker\oL$ integrable, as already stressed in [I], but also the distribution $\ker^V\oL$ is)  with connected leaves so that one has a manifold structure for the quotient $$P=TQ/\ker^V\oL,$$ (with  the projection $\pi_{\cL}\,:\,TQ\,\to\, P$ being  a regular submersion\footnote{Notice that such a condition selects, as in \cite{GoNe2}, those  presymplectic systems which are called \emph{admissible}.}) one sees that 
the range of the Legendre transform 
\beq
\label{mise03} M=\FL(TQ)\subset T^*Q
\eeq is a submanifold which is diffeomorphic to $P$. 
It is also possible to prove that  a local coordinate chart $(q^a, p_b, p_{\sigma})$ with $a=1,\dots,N, \,b=1,\dots,\rho, \,\sigma=\rho+1,\dots, N$ exists on $T^*Q$ and a set of functions $f_{\sigma}(q,p_b)$ exists such that one can define 
\beq
\label{eq51}
\mathcal F(T^*Q)\,\ni\,\varphi_{\sigma}^{(0)}(q,p)\,=\,p_{\sigma}\,-\,f_{\sigma}(q,p_b)
\eeq
and characterise the embedding $$i_M\,:\,M\,\hookrightarrow\,T^*Q$$ via the conditions $$\varphi_{\sigma}^{(0)}=0$$ in $T^*Q$. Such functions $\varphi_{\sigma}^{(0)}$ are called  (adopting the definition introduced by Bergmann) the \emph{primary} constraints of the system  with the given Lagrangian $\cL$.  The set $(q^a, p_b)$ (with $a=1,\dots,N,\,b=1,\dots,\rho$) gives a local coordinate chart on $M$. Moreover, and referring to the local coordinate chart on $TQ$ introduced above, a   basis for $\ker^V\oL$ is given by the vertical vector fields 
$$
K_{(\sigma)}^V=\frac{\del}{\del v^{\sigma}}+\FL^*(\{q^a, \varphi_{\sigma}^{(0)}\})\frac{\del}{\del v^a},
$$
 with respect to the canonical Poisson bracket on $T^*Q$. 
 
 Before analysing the problem following Dirac's approach, we present an example describing a limiting case, which is also related to the problem of reduction within the Hamiltonian formalism (see also \cite{mukunda1}).

\begin{example}
\label{exluc}
Let $\Gamma$ be a Hamiltonian vector field on the exact symplectic manifold $(M,\omega=-\dd\theta)$, with 
\beq
\label{lu1}
i_\Gamma\omega=\dd H.
\eeq
Upon writing $X\,=\,X^k\del_k$ and $\theta=\theta_j\dd x^j$ with respect to a local coordinate system $\{x^i\}_{i=1,\dots,2N}$ for $2N=\dim \,M$, the previous relation \eqref{lu1} reads
\beq
\label{lu2}
(\del_a\theta_k-\del_k\theta_a)X^k\,=\,\del_aH,
\eeq
with $\omega=\omega_{ab}\dd x^a\wedge\dd x^b$ and $\omega_{ab}=\del_b\theta_a-\del_a\theta_b$. As already analysed in [I], the components of the vector field $\Gamma=X_H$ can be written in an explicit form in terms of the Poisson structure corresponding to $\omega$, which is symplectic and then invertible. The question we wonder is: although not of  second order, can the set of ordinary differential equations associated to $\Gamma$, i.e. 
\beq
\label{lu4}
\dot x^k\,=\,X^k
\eeq
be described as a suitable system of implicit differential equations of Lagrangian type? The answer is in the affirmative. Consider the tangent bundle manifold $TM$ and the function 
\beq
\label{27m3}
\cL\,=\,v^a\theta_a-H(x),
\eeq
 where by $\{v^i\}_{i=1,\dots,2N}$ we denote the velocity (i.e. the fiber) coordinates on $TM$. We see that for the corresponding Lagrangian 2-form it is $$\oL\,=\,\pi^*(\omega)$$ (given  $\pi\,:\,TM\,\to\,M$ the tangent bundle projection) with $\ker^V\oL=\ker\oL$ span by any bases of vertical vector fields for the tangent bundle fibration and clearly $M=TM/\ker^V\oL$. 
The associated Euler-Lagrange equations \eqref{eq29} read
\begin{align}
&(\del_a\theta_k-\del_k\theta_a)v^k\,=\,\del_aH \label{lu3}\\ 
&\dot x^k\,=\,v^k. \nn 
\end{align}
Given the non degeneracy of $\omega$ on $M$, a comparison between \eqref{lu2} and \eqref{lu3} shows that the set of ordinary differential equations in \eqref{lu4} comes as the reduction to $M$ of the implicit Euler-Lagrange equations on $TM$. 

Interesting examples of such a reduction come from the dynamics of a finite level quantum system. With $\cH=\C^N$ a finite dimensional Hilbert space, the quantum infinitesimal evolution of any observable (described by  a linear Hermitian operator $A\in\mathbb{B}(\cH)$, i.e. $A=A^{\dagger}$)  is given by the Heisenberg equations of the motions
\beq
\label{lu5}
\dot A\,=\,i[H,A]
\eeq
where $H=H^{\dagger}\in\mathbb{B}(\cH)$ and the commutator is defined upon antisymmetrising the associative product in $\mathbb{B}(\cH)$ as $[A,B]=AB-BA$. The set $\mathbb{B}(\cH)$ is itself  indeed a Hilbert space with respect to the Hermitian scalar product 
\beq
\label{lu6}
{\rm h}(A,B)\,=\,\frac{1}{2}{\rm Tr}(A^{\dagger}B).
\eeq
The dynamics \eqref{lu5} is linear on the real vector space $\A\simeq\R^{N^2}$ of Hermitian linear operators on $\cH=\C^N$. 

Being $\A$ in general not even dimensional, one first explores the possibility of describing such a dynamics within the Poisson formalism. The map $A\,\mapsto\,\tilde{A}=iA$ is a real vector space isomorphism between the set $\A$ of Hermitian and those $\tilde{\A}$ of anti-Hermitian linear operators on $\cH$, which is isomorphic to the Lie algebra $\mathfrak{u}_N$ of unitary automorphisms\footnote{Notice that under the name of unitary automorphisms \emph{in} (i.e. not \emph{on}) $\mathbb{B}(\C^N)$ we mean those elements $U\in\,\mathcal U\subset\mathbb B(\C^N)$ such that $U^{\dagger}U=\mathbb{I}$.} in $\mathbb{B}(\C^N)$. Under such isomorphism, the equations of the motions \eqref{lu5} can be written as
\beq
\label{lu5p1}
\frac{\dd{\tilde A}}{\dd t}=[\tilde{H},\tilde{A}]
\eeq
If $\{\tau_j\}_{j=1,\dots,N^2}$ denotes an orthonormal basis for $\tilde{\A}$ (with respect to the Hermitian scalar product \eqref{lu6}, that is ${\rm Tr}(\tau_j^\dagger\tau_k)=2\delta_{jk}$) and along such a basis we can write $\tilde{A}=y_j\tau_j$ with $y_j={\rm Tr}(\tilde{A}^\dagger\tau_j)$ a system of global coordinates, then from the Lie algebra structure $[\tau_j,\tau_k]=c_{jk}^{\,\,\,\,s}\tau_s$ on $\mathfrak{u}_N$ we can define the Poisson tensor (which is equivalent to the one \eqref{I-podu}[I] on the dual space to $\mathfrak{u}_N$) which can be written as 
\beq
\label{lu5p2}
\Lambda\,=\,c_{jk}^{\,\,\,\,s}y_s\frac{\del}{\del y_j}\wedge\frac{\del}{\del y_k}
\eeq
or  as 
$$
\{y_j, y_k\}\,=\,c_{jk}^{\,\,\,\,s}y_s
$$
and prove directly that the Heisenberg  dynamics \eqref{lu5} can be written as
\beq
\label{lu5p3}
\dot{y}_j\,=\,\{y_j,H\}\,=\,\Lambda(\dd y_j, \dd H)
\eeq
with $H=h_sy_s$ where $\tilde H=h_s\tau_s$.  

In order to have a Hamiltonian description for the Heisenberg dynamics, we consider the $\C$-vector space isomorphisms $\mathbb{B}(\C^N)\simeq\C^{N^2}$ and the $\R$-vector space isomorphism 
\beq
\label{lu5p3p1}
\C^{N^2}\simeq\R^{2N^2}\simeq\A\oplus_{\R}\tilde\A.
\eeq
An orthonormal (with respect to \eqref{lu6}) basis for $\C^{N^2}$ is given by $\{\sigma_{j}\}_{j=1,\dots,N^2}$, with (adopting the same notation we used above) $\tilde{\sigma}_j=\tau_j$ giving an orthonormal basis for $\tilde\A$: one can then write  
$$
A\,=\,(x_j+iy_j)\sigma_j
$$
with 
$$
{\rm h}(\sigma_j, A)\,=\,x_j+iy_j.
$$
The set $\{x_j, y_j\}_{j=1,\dots,N}$ gives a global (real) coordinate system for $\R^{2N^2}=\cV$.   
Upon identifying the tangent space to $\cV$ with $\cV$ itself, the Hermitian product induces a Euclidean metric $\gh$ and a symplectic structure $\omega_h$ as
\beq
\label{lu8}
{\rm h}\,=\,\gh\,+\,i\omega_h\,=\,(\dd x_{j}\otimes\dd x_{j}\,+\,\dd y_{j}\otimes\dd y_{j})\,+\,i(\dd x_{j}\otimes\dd y_{j}\,-\,\dd y_{j}\otimes\dd x_{j}).
\eeq
The tensor $${\rm J}_h\,=\,\del_{y_j}\otimes\dd x_{j}-\del_{x_{j}}\otimes\dd y_{j},$$ with ${\rm J}_h^2=-1$ gives the complex structure compatible with both $\gh$ and $\omega_h$, since 
\begin{align*}
&\gh({\rm J_h}u,v)\,=\,\omega_h(u,v), \\
&\gh({\rm J_h}u,{\rm J_h}v)\,=\,\gh(u,v), \\ &\omega_h({\rm J}_hu, {\rm J}_hv)\,=\,\omega_h(u,v)
\end{align*}
for any $u,v\,\in\,\cV$. The linear evolution given in \eqref{lu5} is Hermitian, that is  
it preserves both the Euclidean $\gh$ and the symplectic $\omega_h$ tensors. If the Hamiltonian is written as $H=h^a\sigma_a$, the evolution reads 
\begin{align}
&\dot x^a=c_{bs}^{\,\,\,a}x^bh^s, \nn \\
&\dot y^a=c_{bs}^{\,\,\,a}y^bh^s
\label{27m1}
\end{align}
on $T\mathcal V$. The vector field $\Gamma$ describing such a dynamics is linear, with $i_{\Gamma}\omega_h=\dd f_H$ for a quadratic Hamiltonian function 
\beq
\label{27m2}
f_H=c_{sba}h^sy^bx^a
\eeq
where the indices of the structure constants have been lowered via $\gh$.
We can compare such description with the theory outlined in section \ref{I-susec:linPoisson}[I]. 
The dynamics $\Gamma$ on $\cV$ is a linear vector field $\Gamma=\,M_{rs}z_r\del_s$, where we have collectively denoted by $z=(x,y)$ the coordinates on $\cV$ with $s,r=1,\dots,2N^2$.  Upon such ordering for the coordinate system, we see from \eqref{27m1} that  the matrix $M$ has the block form  
$$
M\,=\,\begin{pmatrix} \alpha & 0 \\  0 & \alpha \end{pmatrix}
$$ with $\alpha=-\alpha^T$ (it is $\alpha_{ab}=c_{sab}h^s$)  so that $H=-\omega_hM=H^T$ and the condition \eqref{I-eq23bi}[I] is satisfied. The vector field $\Gamma$ is Hamiltonian with respect to the symplectic form $\omega_h$, and the corresponding Hamiltonian function \eqref{27m2} can be written, in a matrix form, as
$$
f_H\,=\,\frac{1}{2}\begin{pmatrix} x & y \end{pmatrix} \begin{pmatrix} 0 & -\alpha \\ \alpha & 0 \end{pmatrix} \begin{pmatrix} x \\ y\end{pmatrix}.
$$
We notice that $\A$ and $\tilde\A$ are Lagrangian submanifold of $\cV$ with respect to $\omega_h$. The equations \eqref{27m1} show that  the dynamics $\Gamma$ projects onto both submanifolds. If $\pi_{\A}\,:\,\cV\,\to\,\A$ is the natural projection associated to the direct sum decomposition in \eqref{lu5p3p1}, it is clear  that  the vector field $\pi_{\A*}(\Gamma)$ exists on $\A$ and describes the Heisenberg dynamics \eqref{lu5} as a reduction of the Hamiltonian dynamics $\Gamma$ on the unfolding symplectic space $(\cV,\omega_h)$. 

The dynamics $\Gamma$ on $\cV$ has a Lagrangian formulation on $T\cV$, with (see \eqref{27m3}) the singular Lagrangian
\beq
\label{27m4}
\cL\,=\,\frac{1}{2}(y^a\dot x^a-x^a\dot y^a)-c_{sba}h^sx^by^a.
\eeq
If we parametrise the real manifold $\cV\simeq\mathbb B(\C^N)$ in terms of matrices $A,A^{\dagger}$, such a Lagrangian can be written as
\beq
\label{27m5}
\cL\,=\,-\frac{i}{8}{\rm Tr}(\dot A^{\dagger}A-A^{\dagger}\dot A)-\frac{1}{4}{\rm Tr}(A[H,A^{\dagger}]).
\eeq
This is the Lagrangian written in \cite{modpl20}. The corresponding equations of the motions can be reduced to $\cV$ and coincide with the  Heisenberg equations \eqref{lu5}. 

Why is the possibility of describing a Hamiltonian dynamics $\Gamma$ on the exact symplectic manifold $(M,\omega=-\dd\theta)$ in terms of a (albeit singular) Lagrangian $\cL$ on $TM$ interesting? If $i_S\,:\,S\hookrightarrow M$ denotes the embedding of a submanifold $S$ into $M$, such that $S$ does not inherit a symplectic structure (i.e. $i_S^*\omega$ is degenerate) or such that $\Gamma$ is not tangent to $S$ (i.e. $\Gamma$ can not be reduced to $S$), then the tangent lift  $Ti_S\,:\,\mathcal F(TM)\to\mathcal F(TS)$ gives  a well defined function $Ti_S^*\cL$ on $TS$ which  defines a (singular) Lagrangian  dynamics on $TS$ that might be reduced to $S$.

This procedure provides an interesting result since it allows for a Lagrangian description of the 
  Landau - von Neumann equations of the motions for a finite level quantum system. If $\rho$ is a quantum state (a density matrix, i.e. a Hermitian positive element with ${\rm Tr}\rho=1$) the quantum dynamics is given by 
\beq
\label{28m1}
\dot\rho=i[\rho, H],
\eeq
which appears as a dualization of the Heisenberg dynamics \eqref{lu5}. The evolution of any state $\rho$ can be written as 
$$
\rho(t)=e^{-itH}\rho e^{itH}
$$
with $U(t)=e^{-itH}\in\mathcal U$ the operator exponential. The orbits of such a dynamics are then given by homogeneous spaces  $O_{\rho}\simeq \mathcal U/\mathcal U_{\rho}$, where $\mathcal U_{\rho}$ is the isotropy subgroup of the action $(U,\rho)\mapsto U^\dagger\rho U$; the orbit $O_{\rho}$ is therefore  identified by the spectrum of the diagonal and positive element $\rho_0$ such that $\rho=U^\dagger \rho_0 U$, whose existence is given by the spectral properties of Hermitian positive operators. The map 
$$
i_{\rho_0}\,:\,U\,\mapsto\,\sqrt{\rho_0}\,U
$$
defines an embedding of the 
orbits $O_{\rho_0}$ of the  unitary group $\mathcal U$ into  $\mathcal B(\C^N)$. It is possible to prove (see \cite{modpl20}) that, if $\cL$ is the   Lagrangian in \eqref{27m5}, then  
$Ti_{\rho_0}^*\cL$ gives a singular Lagrangian on $TU_{\rho_0}$. The corresponding dynamics can be reduced to $O_{\rho_0}$, where it coincides with the  Landau - von Neumann dynamics \eqref{28m1} 

\end{example}

\subsection{Singular Lagrangian dynamics on the cotangent bundle}
\label{subsec:H}
 Since the 2-form $\oL$ is closed, and $i_{K^V}\oL=0$ for any element $K^V\in\ker^{V}\oL$, there exists a closed 2-form $\omega$ with constant rank on $M$ such that $$\FL^*\omega=\oL,$$ which is proven to come also as  $$\omega=i_M^*\omega_Q.$$ A direct calculation shows that 
 \beq
 \label{eqpe1}
 L_{K^V}\EL=0,
 \eeq
  so there  exists an element $H_0\in\mathcal F(M)$ such that $\FL^*H_0=\EL$. The triple 
 $$
 (M, \,\omega,\,\dd H_0)
 $$
 is, recalling what we described in section \ref{I-sec:pre}[I], a pre-symplectic system associated to the singular Lagrangian $\cL$ on $TQ$, with $\dim \,M\,=\,N+\rho$.  Our aim is now to analyse such a system also by using the Dirac's theory of constraints introduced in section \ref{I-sub:dirac}[I]. So 
 consider the matrix 
$$
M_{ab}=\{\varphi_{\sigma_a}^{(0)},\varphi_{\sigma_b}^{(0)}\}
$$
 on $M$. Assume it has constant rank $L-\rho={\rm rk}\,M_{ab}$ with $L\leq N$. A $(N-L)$-dimensional basis $\tilde\varphi_{\mu}^{(0)}$ for the kernel of $M_{ab}$ spans the set of what are called the \emph{first class} primary constraints for the system\footnote{We remark that the definition of the manifold $M\hookrightarrow T^*Q$ in \eqref{mise03} comes in terms of $N-\rho$ primary constraints functions corresponding to the vertical kernel $\ker^{(V)}\oL$  of the 2-form $\oL$, whose  dimension cannot exceed $N$.  Along such a line, the rank of the matrix $M_{ab}$  is assumed to be $L-\rho$ with $L\leq N$: so we have $N-L$ first class constraints.  A comparison with the description of general submanifolds of a symplectic manifold in terms of the Dirac's theory of constraints as given in section \ref{I-sub:dirac}[I] comes with  $N-\rho=2N-\delta$ and $L-\rho=2N-\delta-k$, which means $\delta=N+\rho$ and $k=N-L$: first class constraints do not exceed $N$. }. The Hamiltonian vector fields $X_{\tilde{\varphi}_\mu^{(0)}}$ generated by first class primary constraints turn to be  tangent to $M$ at each point $m\in M$ (see \eqref{I-dco4}[I]). In analogy to what we described about the symplectic reduction,  one proves that the intersection \eqref{I-eqsy1}[I]-\eqref{I-eqsy1poi}[I] is given by $$X\in\ker\omega\,\Leftrightarrow\,X=X_{\tilde\varphi^{(0)}}$$ for a first class primary constraint $\tilde\varphi^{(0)}$. 
The \emph{second class} primary constraints $\varphi_r^{(0)}$ (with $r=\rho+1, \dots,L$ upon a suitable reshuffling of the labels) are given by quotienting the set of primary constraints by the subset span by $\tilde\varphi_{\mu}^{(0)}$, with $\mu=L+1,\dots,N$ (notice that the second class constraints are defined up to the addition of first class ones). Upon considering only the second class constraints on $M$, the $(L-\rho)$-dimensional matrix $M_{ij}=\{\varphi_{r_i}^{(0)}, \varphi_{r_j}^{(0)}\}$ with $r_i,r_j=\rho+1,\dots,L$ turns out to be invertible, so one  can indeed  define the elements $C^{js}$ on $M$ by  
\beq
\label{matDi01}
\{\varphi^{(0)}_{r_i},\varphi^{(0)}_{r_j}\}C^{js}=\delta_i^s.
\eeq    Once the kernel of the closed 2-form $\omega$ is given, the analysis developped in section  \ref{I-sec:pre}[I] for pre-symplectic systems can be mimicked. 
We start upon assuming that 
\beq
\label{matD01}
(i_{X_{\tilde\varphi^{(0)}}}\dd H_0)_{\mid m\in M}\,=\,\{H_0, \tilde\varphi^{(0)}\}_{\mid m\in M}=0.
\eeq
 This means that $(M,\omega,\dd H_0)$ gives a \emph{global} pre-symplectic system: at each point $m\in M$ there exists a vector field $Y$ which solves the problem 
\beq
\label{eq53.e}
i_{Y}\omega=\dd H_0.
\eeq
 Such a vector field is not uniquely determined, but it is defined up to an element in $\ker\omega$. 
As already noticed, since $\dd\omega=0$, one has the quotient  $$\tilde\pi_M\,:\,M\stackrel{\ker\omega}{\rightarrow}\tilde{M}$$ given by the involutive distribution $\ker\omega$, so that $(\tilde M, \tilde\omega)$ is a symplectic manifold with $\omega=\tilde\pi^*_M\tilde\omega$. From \eqref{matD01} it is  $L_{X_{\tilde\varphi^{(0)}}}H_0=0$, so that the equation \eqref{eq53.e} on $\tilde{M}$ is written as  $$i_{\tilde Y}\tilde\omega=\dd\tilde H_0$$ and has a unique solution, with $H_0=\tilde\pi^*_M\tilde H_0$. The corresponding set of solutions  $Y$ can be lifted to $T^*Q$: under this expression we mean that  the action of the vector field  
\beq
\label{eq54}
\mathfrak X(T^*Q)\,\ni\,X_{Y}\,=\,\{\varphi^{(0)}_i, H\}C^{ij}X_{\varphi^{(0)}_j}\,+\,X_H\,+\,u^{\mu}X_{\tilde\varphi^{(0)}_{\mu}}, 
\eeq
-- where $H\in\mathcal F(T^*Q)$ extends $H_0$, i.e. $H_0=i_{M}^*H$ (notice that it is determined up to the addition of a primary constraint) and $u^{\mu}$ are $(N-L)$ free multipliers --  satisfies the relation
\beq
\label{eq541}
i_M^*(X_{Y}f)\,=\,Y(i_M^* f)
\eeq
for any $f\in\mathcal F(T^*Q)$. 
The freedom in the choice of $H$ and of the second class $\varphi^{(0)}_j$ constraints is easily seen to be accomodated by suitable changes in $u^{\mu}$.  If  $$\ker\oL=\ker^V\oL$$ there are no primary first class constraints, so $(M, \omega)$ is a symplectic manifold and the solution $Y$ to \eqref{eq53.e} is unique on $M$. 
\begin{rema}
\label{reD}
The set $\mathfrak Z_{\cL}$ defined in \eqref{ds01} is a Lagrangian submanifold in $T(T^*Q)$. Our analysis on the integrability of such equation shows that, if a singular Lagrangian gives only functionally independent primary constraints $\varphi_{\sigma}^{(0)}=0$ defining a submanifold $M$ in $T^*Q$  such that the matrix $M_{ab}=\{\varphi_a^{(0)}, \varphi_b^{(0)}\}$ has a constant rank on $M$, then the Euler-Lagrange equations are integrable on $M$  as a distribution $\mathfrak Z'_{\cL}\hookrightarrow TM$. 
At each point $m$ in $M$ the vectors in \eqref{eq54} give the elements $X_m\in T_mM$ such that one can locally write 
$$
\mathfrak Z'_{\cL}\,\ni\,z\,\simeq\,(m, X_m). 
$$  
One immediately sees that $\dim\,\mathfrak Z'_{\cL}=2N+\rho-L$. If the Lagrangian gives only first class (primary) constraints it is $\rho=L$, so that $\mathfrak Z_{\cL}\simeq\mathfrak Z'_{\cL}$ is globally integrable. If the Lagrangian gives only second class constraints, then $\mathfrak Z'_{\cL}$ results in the graph of a vector field on $M$.

Notice that $\mathfrak Z_\cL$ is \emph{not} a Dirac system, since the condition \eqref{eq53.e} can not be written as  \eqref{impli02} on the manifold $M$. An analysis on the integrability of a more general class of implicit differential equations which can be written as constrained Hamiltonian equations is in \cite{mmt95}.

\end{rema}
When the pre-symplectic system $(M,\omega, \dd H_0)$ is not global, i.e. the relations 
\beq
\label{secD1}
\chi\,=\,\{H_0,\tilde\varphi^{(0)}\}_{\mid m\in M}\,=\,0
\eeq
 are not identically satisfied on $M$, the procedure outlined in section \ref{I-sec:pre}[I] can be applied. Assume that the sequence introduced in \eqref{I-ps1.1}[I] and \eqref{I-ps1.2}[I] gives the nested submanifolds  $M_{s+1}\hookrightarrow M_{s}$ (with $M_0=M$)  and \emph{has} a fixed point, which is a submanifold in $M$ that we denote by $M'$. The embedding $$i_{M'}\,:\,M'\hookrightarrow M$$ can be described in terms of a set of relations that we collectively denote by $\chi=0$ (the secondary $\varphi^{(1)}$, tertiary $\varphi^{(2)}$, and higher order constraints) with $\chi\in\mathcal F(T^*Q)$. 
Via such embedding we define, following what  we described in section  \ref{I-ss:genps}[I], (with $i_{M'}^*H_0=H_0'$ and $\omega'=i_{M'}^*\omega$) the pre-symplectic system 
\beq
\label{gldy}
(M', \omega', \dd H_0') 
\eeq
with a global dynamics.  Along the path outlined above it is possible to see that ${\rm ker}\,\omega'$ is spanned by the Hamiltonian vector fields $X_{\tilde\varphi},X_{\tilde\chi}$, where $\tilde\varphi$ and $\tilde\chi$ respectively denote the  first class constraints with respect to the whole set $(\varphi^{(0)}, \chi)$ (see \eqref{secD1}) of functions giving the constraints that define $M'$ in $T^*Q$. 
The solutions to the equation \eqref{I-ps3}[I], that we now write as 
\beq
\label{gldy1}
i_{Y'}\omega'=\dd H_0',
\eeq
 can  again be  lifted to $T^*Q$, reading  
\beq
\label{eq55}
\mathfrak X(T^*Q)\,\ni\,X_{Y'}\,=\,\{\theta_i, H\}C^{ij}X_{\theta_j}\,+\,X_H\,+\,u^{\mu}X_{\tilde\varphi^{(0)}_{\mu}}\,+\,v^{\rho}X_{\tilde\chi_{\rho}}, 
\eeq
where the elements $\theta_i$ (with $C^{ij}\{\theta_j,\theta_k\}=\delta^i_k$ on $M'$) give the set of second class constraints out of the whole family of constraints $(\varphi^{(0)}, \chi)$, while $u^{\mu},v^{\rho}$ are free multipliers, the function $H\in\mathcal F(T^*Q)$ satisfies the condition $H_0'=i^*_{M'}(i^*_{M}H)$. The set of vector fields $X_{Y'}$  in \eqref{eq55} give the lift to $T^*Q$ of the elements  $Y'\in\mathfrak X^{\omega'}(M')$ introduced in section  \ref{I-ss:genps}[I]. It is then possible to prove that uniquely the elements 
\beq
\label{eq56}
\mathfrak X(T^*Q)\,\ni\,X_{Y}\,=\,\{\theta_i, H\}C^{ij}X_{\theta_j}\,+\,X_H\,+\,u^{\mu}X_{\tilde\varphi^{(0)}_{\mu}}, 
\eeq
i.e.  those elements from $\mathfrak X^{\omega'}(M')$  depending on free multipliers which corresponds \emph{only} to primary first class constraints with respect to the whole set of constraints defining the manifold $M'$, are elements whose action lifts to $T^*Q$ the action of the vector fields   $Y\in\mathfrak X^{\omega}(M')$ which gives the vector fields solving  \eqref{I-ps2}[I]. Notice that this gives an interesting example for the inclusion written in \eqref{I-ciocio}[I].


\subsection{Singular Lagrangian dynamics on the tangent bundle}
\label{subsec:L}
We now focus on the problem  of the integrability of the Euler-Lagrange equations $\mathfrak Z_{\cL}$ \eqref{ancor1} on $T(TQ)$. The path to analyse it starts by considering, for a given Lagrangian $\cL$ on $TQ$, a more general version of the equation \eqref{eiL}, namely we study the implicit differential equation 
\beq
\label{a2} 
i_D\oL=\dd\EL
\eeq
with $D\in\mathfrak X(TQ)$, and afterwards the conditions that lead to a \emph{second order} solution for this problem. If we label points in $T(TQ)$ by the coordinate system $(q^j,v^j,u^j,a^j)_{j=1,\dots,\dim Q}$ we see that \eqref{a2} can be written as
\begin{align}
\frac{\del^2\cL}{\del v^k\del v^j}(u^j-v^j)&=0, \nn \\
-\frac{\del^2\cL}{\del v^k\del q^j}u^j-\frac{\del^2\cL}{\del v^k\del v^j}a^j-\frac{\del^2\cL}{\del q^k\del v^j}(v^j-u^j)&=-\frac{\del\cL}{\del q^k}
\label{impf1}
\end{align}
or equivalently, in matrix form
\begin{align}
\begin{pmatrix}
\frac{\del^2\cL}{\del q^k\del v^j}-\frac{\del^2\cL}{\del v^k\del q^j} & -\frac{\del^2\cL}{\del v^k\del v^j}  \\ \frac{\del^2\cL}{\del v^k\del v^j} & 0 \end{pmatrix}\,\begin{pmatrix} u^j \\ a^j\end{pmatrix}\,=\,\begin{pmatrix} \frac{\del^2\cL}{\del q^k\del v^j}v^j-\frac{\del\cL}{\del q^k} \\ \frac{\del^2\cL}{\del v^k\del v^j}v^j\end{pmatrix}
\label{impf2}
\end{align}
We see immediately from the first relations out of \eqref{impf1} that, if $\cL$ is regular, then $u^j=v^j$, that is any solution $D$ of \eqref{a2} is a second order vector field, and moreover that $a^j$ are uniquely determined. It is in general possible to prove that (see \cite{pepin90})
\beq
\label{eq57}
S(D)-\Delta_Q\,\in\,\ker^V\oL.
\eeq
The equation \eqref{a2} gives a presymplectic system (given that the rank of the Hessian  $H_{jk}$ does not vary on $TQ$) 
$$
(TQ, \oL, \dd\EL)
$$
An interesting case is given if it is $L_K\EL=0$ for any $K\in\ker\oL$, so that there are, within the Lagrangian setting on $TQ$, no first generation  constraints\footnote{The relation \eqref{eqpe1} shows that, for vertical elements $K=K^V\in\ker\oL$, the condition $L_K\EL=0$ is an identity which gives no constraints. Within the Lagrangian setting on $TQ$, we denote as \emph{first, second, $\dots$ l-ary generation} of constraints what we have defined as primary, secondary and so on constraints on $T^*Q$.}. This means that the given presymplectic system has a global dynamics, that is  there exists a set $\mathfrak X^{\oL}(TQ)$ of vector fields on $TQ$ which is an affine space modelled on $\ker\oL$. Let $D\in\mathfrak X^{\oL}(TQ)$: the element $$D'=D+K$$ (with $K\in\ker\oL$) satisfies the second order condition (i.e. $S(D')=\Delta_Q$) if and only if 
$$
S(K)=\Delta_Q-S(D);
$$ 
from \eqref{eq57} we see  that the set of elements $K\in\ker\oL$ which satisfy such condition is an affine space modelled on $\ker^V\oL$.  Examples  of singular Lagrangians for which $$S(\ker \oL)=\ker^V\oL$$ (i.e. the restriction of $S$ to $\ker\oL$ is surjective onto $\ker^V\oL$) indeed exist: they are referred to as type II (see \cite{belgio-spagna}, and see also \cite{pepin-alberto85} for an analysis of this condition for a Lagrangian with secondary constraints). For this class of singular Lagrangians, it is then possible to exhibit  a set of second order vector fields on all $TQ$ satisfying \eqref{eiL}. A type II singular Lagrangian dynamics is given in the example \ref{exemplum-sL2}. 

Another interesting case is given by  singular Lagrangians for which $\ker\oL=\ker^V\oL$. Along the previous lines, one has that no first generation constraints exist and, since $S^2=0$, every solution $D\in\mathfrak X^{\oL}(TQ)$ satisfies the second order condition. Moreover,  the comparison with the problem \eqref{eq53.e} given in the Hamiltonian setting shows that any solution $D\in\mathfrak X^{\oL}(TQ)$ is projectable   under the Legendre map on the unique $Y_D\in\mathfrak X(M)$, i.e. 
\beq
\label{eqa0}
\Phi_{\cL}^*(Y_D f)\,=\,D(\Phi^*_{\cL} f)
\eeq
for any $f\in\mathcal F(M)$.

In general, if we apply  to the presymplectic system $(TQ, \oL, \dd\EL)$ the iterative procedure outlined in section \ref{I-sec:pre}[I] we have a sequence of nested submanifolds $P_{s+1}\hookrightarrow P_s$ (with $P_0=TQ$) which has a fixed point given by the submanifold $$i_{P'}\,:\,P'\,\hookrightarrow\,TQ.$$ 
We denote as $\psi^{(s)}=0$ the $s$-ary generation of constraints that give the embedding $P_s\hookrightarrow P_{s-1}$.
We write the equation \eqref{eiL} on $P'$ as in  \eqref{I-ps2}[I], that is
\beq
\label{eqa1}
\oL(D,Y)\,=\,L_Y\EL 
\eeq
for $D\in\mathfrak X(P')$ and any $Y\in\mathfrak X(TQ)$, and denote under $\mathfrak X^{\oL}(P')$ the set of its solutions.   
It is possible to prove (see \cite{batlle86,batlle88,dLdD96,GoNe1,GoNe2,pons88}) that, under the admissibility  assumptions we already considered  on $\cL$,  the restriction of  Legendre map $\Phi_{\cL}:P_s\to M_s$ is a surjective submersion providing a suitable fibration at each step $s$ (notice that these results generalise what we already  studied in more detail for $s=0$), with $\Phi_{\cL}:P'\to M'$.

An interesting generalisation of the projectability analysis  \eqref{eqa0}  above is given in \cite{GoNe1,mms83}. If $\mD$ is an element in $\mathfrak X^{\oL}(P')$ which is projectable onto a vector field $Y_{\mD}$ on $\mathfrak X(M')$, then such $Y_{\mD}$ is an element in $\mathfrak X^{\omega}(M')$, i.e. if $\mD$ is a solution of the equation \eqref{eqa1}, then $Y_{\mD}$ is a solution of the equation \eqref{I-ps2}[I] 
\beq
\label{eqa2}
i_{Y_\mD}\omega=\dd H_0
\eeq
on $M'$\footnote{We recall that a vector field $D$ on $P'$ is projectable under $\Phi_{\cL}$ into $\mD$ on $M'$ if the relation $L_D(\Phi_{\cL}^*f)=\Phi^*_{\cL}(L_{\mD}f)$ is valid for any $f\in\mathcal F(M')$ (see \eqref{eq541} and \eqref{eqa0}).}. On the contrary, it is also possible to prove that if $Y\in\mathfrak X^{\omega}(M')$ and $\mD_Y$ is any vector field on $P'$ which is projectable onto $Y$, then $\mD_Y\in\mathfrak X^{\oL}(P')$, i.e. $\mD_Y$ is a solution for \eqref{eqa1}. These results are usually considered as providing an equivalence between the formulation of the problem \eqref{eiL} within the  Lagrangian setting (that is on $TQ$) and that within the Hamiltonian setting (that is on $T^*Q$).   

We notice that one of the motivations to study the second order problem for the vector field $\mD\in\mathfrak X^{\oL}(P')$ corresponding to \eqref{eiL} is that the integral curves of \eqref{eiL} are proven to follow from a variational principle if and only if the differential equations  satisfy the second order condition, which we write $S(\mD)=\Delta_Q$ as before (see \cite{crampin81,crampinteam94, inv}).  

Let $Y\in\mathfrak X^{\omega}(M')$ and let $\mD_Y\in\mathfrak X(P')$ be any vector field on $P'$ which is  projectable (under the fibration given by the Legendre transform $\Phi_{\cL}:P'\to M'$) on $Y$. There exists (see \cite{GoNe2, dLdD96}) a (differentiable) section $\sigma:M'\to P'$ such that the restriction of the  vector field ${\mD}_Y$ on $\sigma(M')\subset TQ$ satisfies the second order condition.  We denote $\Sigma=\sigma(M')$. Since ${\mD}_Y$ is in general not tangent to $\Sigma$, one proves that the vector field $\tilde D\in\mathfrak X(\Sigma)$ given as the lift of $Y$ on $M'$ associated to the section $\sigma$ solves the equation 
\beq
\label{eqa3}
i_{\tilde{D}}\oL=\dd\EL
\eeq
on $\Sigma$ and 
satisfies the second order condition. Given a solution $\mD_Y\in\mathfrak X^{\oL}(P')$ which is projectable onto a $Y$ in $\mathfrak X^{\omega}(M')$, one has that the set of possible sections $\sigma:M'\to P'$  providing a second order solution for the pre-symplectic problem \eqref{eqa3} is parametrised by the intersection $\ker^V\oL\cap \mathfrak X(P')$.   This analysis provides a complete answer to the existence problem of a second order  solution for the implicit system given in \eqref{eiL}, that is to the integrability problem for the Euler-Lagrange (second) order equations on $TQ$.
\begin{example}
\label{exemplum-sL1}
In order to clarify what we are describing, we consider $Q=\R^2$ and the dynamics given by the singular Lagrangian 
$$
\cL\,=\,\frac{1}{2}av_1^2
$$
where $(q=(q_1,q_2), v=(v_1,v_2))$ denote a global coordinate system on $TQ=\R^4$ and $a\in\mathcal F(Q)$. 
We have 
\begin{align}
&\tL\,=\,a(q)v_1\dd q_1, \nn \\
&\oL\,=\,a\,\dd q_1\wedge\dd v_1\,+\,v_1(\del_2a)\dd q_1\wedge \dd q_2, \nn \\
&E_{\cL}\,=\cL
\label{esL01}
\end{align}
so we assume $a(q)\neq 0$ and $\del a/\del q_2=\del_{q_2}a\neq0$ in order for the 2-form $\oL$ to have constant rank. 
The vector field on $TQ$ given by
$$D\,=\,\alpha\frac{\del}{\del q_1}+\beta\frac{\del}{\del q_2}+\gamma\frac{\del}{\del v_1}+\delta\frac{\del}{\del v_2}$$
satisfies $i_D\oL=\dd E_{\cL}$ if and only if the relations
\begin{align*}
&a(\alpha-v_1)=0, \\
&(\del_{q_2}a)v_1(\alpha-v_1/2)=0, \\
&a\gamma+\beta v_1(\del_{q_2}a)+(\del_{q_2}a)v_1^2/2=0
\end{align*}
hold. Such relations give a vector field $D$ which is not necessarily a second order one; moreover they can  be satisfied only on the submanifold given by $v_1=0$ in $TQ$.  In order to develop the geometric analysis along what we described, we notice that the kernel of $\oL$ is the left module $$\ker\oL=\mathcal F(TQ)<K^V,K>$$ spanned by the vector fields  
\beq
\label{esL02}
K^V\,=\,\frac{\del}{\del v_2},\qquad\qquad K\,=\,-a\frac{\del}{\del q_2}\,+\,v_1(\del_{q_2}a)\frac{\del}{\del v_1},
\eeq
with the  vertical part $\ker^V\oL$ being   generated by $K^V$. 

We consider first the problem on the cotangent bundle $T^*Q$. The Legendre transform is the map
$$
\Phi_{\cL}\,:\,(q_a,v_a)\quad\mapsto\quad(q_a, \,p_1=av_1, \,p_2=0),
$$
the relation 
$$
\varphi\,=\,p_2\,=\,0
$$
gives the primary constraint within the Hamiltonian formulation of the dynamics, the 3-dimensional submanifold 
$$
M\,=\,\{m\in\,T^*Q\,:\,p_2=0\}
$$ 
gives the range of the Legendre transform, with $i_M:M\hookrightarrow T^*Q$ the corresponding embedding. From 
$$\omega_M=i_M^*\omega_Q=\dd q_1\wedge\dd p_1$$ we see that the only secondary constraint is $$\chi\,=\,p_1\,=\,0,$$ and that there are no tertiary constraints,  so the  
final constrained manifold is given by
$$
M'\,=\,\{m\,\in\,M\,:\,p_1=p_2=0\}\,\simeq\,Q .  
$$
From $\mathcal F(M)\,\ni\,H_0=p_1^2/2a$ it is  easy to compute that
\beq
\label{esL03}
\mathfrak X^{\omega}(M')\,=\,B\,\frac{\del}{\del q_2}
\eeq
with $B\in\mathcal F(Q)$ is the set of solutions for \eqref{eq53.e}, while, 
since $M'$ is a Lagrangian submanifold with respect to the canonical symplectic form $\omega_Q$ on $T^*Q$, 
the set 
$$
\mathfrak X^{\omega'}(M')\,=\,A\frac{\del}{\del q_1}\,+\,B\,\frac{\del}{\del q_2}
$$
with $A,B\,\in\mathcal F(Q)$ gives the solutions for the global dynamics in \eqref{gldy1}. 

Within the tangent bundle geometry, we identify $P=TQ$ and then, from $L_{K^V}E_{\cL}=0$ we have a single first generation constraint 
$$
\psi^{(0)}\,=\,L_KE_{\cL}\,=\,0\quad\Leftrightarrow\quad v_1=0
$$
and easily compute that there are no further higher generation constraints. This means that we can identify 
$$
P'\,=\,\{x\,\in\,TQ\,:\,v_1=0\}
$$
and then prove that 
\beq
\label{esL04}
\mathfrak X^{\oL}(P')\,=\,\beta\frac{\del}{\del q_2}\,+\,\delta\frac{\del}{\del v_2}
\eeq
for any $\beta, \delta\in\mathcal F(P')$ is the set of solution for \eqref{eqa1}. 

We notice that the Legendre transform is a fibration $\Phi_{\cL}:P\to M$ and $\Phi_{\cL}:P'\to M'$ with vertical vector field given by $K^V$. 

An element $\mD\in\mathfrak X^{\oL}(P')$ is projectable under the Legendre map into a vector field $Y_\mD\in\mathfrak X(M')$ if and only if $$\mD=\beta\del_{q_2}+\delta\del_{v_2}$$ with $L_{K^V}\beta=0$, i.e. $\beta=\beta(q_1,q_2)$. In such a case, one has that the projected vector field  is $$Y_\mD=\beta\del_{q_2}\in\mathfrak X^{\omega}(M').$$  

On the contrary, if $Y\in\mathfrak X^{\omega}(M')$, a vector field $\mD_Y\in\mathfrak X(P')$ is projectable onto $Y$ if and only if $$\mD_Y=\beta\del_{q_2}+\delta\del_{v_2}$$ with $L_{K^V}\beta=0$. In such a case, it is $\mD_Y\in\mathfrak X^{\oL}(P')$. 

We can now address the second order problem for the class of elements $\mD\in\mathfrak X^{\oL}(P')$ which are projectable onto $Y_{\mD}\in\mathfrak X^{\omega}(M')$. Given  
$$
\mD\,=\,\beta\,\frac{\del}{\del q_2}+\delta\frac{\del}{\del v_2}
$$
with $\beta\in\mathcal F(Q)$, we define the section $\sigma:M'\to P'$ via
\beq
\label{esL05}
\sigma:(q_1,q_2)\quad\mapsto\quad(q_1,q_2,v_2=\beta(q_1,q_2)).
\eeq
It is clear that the restriction of $\mD$ to $\Sigma=\sigma(M')$ is not tangent to $\Sigma$. The lift via the section $\sigma$ of the vector field $Y_{\mD}$ gives the vector field
\beq
\label{esL06}
\tilde D\,=\,\beta\frac{\del}{\del q_2}-\beta\frac{\del\beta}{\del q_2}\frac{\del}{\del v_2}\,
\eeq
which, considered  the condition $v_1=0$ on $P'$, is a second order vector field on $\Sigma$ which solves \eqref{eqa3}. We can conclude this example by noticing that, as we mentioned at length, the problem of characterising the vector field $D$ on $TQ$ satisfying the condition $i_D\oL=\dd E_{\cL}$ for a singular Lagrangian has in general not a unique solution, and such solutions are defined on a suitable submanifold of $TQ$.

\end{example}

Interesting relations between the set of constraints on the $T^*Q$ bundle (Hamiltonian, say) and the set of constraints on the $TQ$ bundle (Lagrangian, say) can be indeed described (see \cite{batlle86,batlle88,pons88}). These relations give a different approach to the analysis of the problem \eqref{eiL} on $TQ$.  One can introduce the operator $\mathfrak K:\mathcal F(T^*Q)\to\mathcal F(TQ)$ whose action is 
\beq
\label{opeK}
\mathfrak K(f)\,=\,(\Phi_{\cL}^*\{f,p_j\})v^j\,+\,(\Phi^*_{\cL}\{q^j,f\})\frac{\del L}{\del q^j}.
\eeq 
It is proven that all the Lagrangian constraints $\psi\in\mathcal F(TQ)$ can be written in the form 
\beq
\label{LcH}
\mathfrak K(\varphi^{(s)})\,=\,0,
\eeq
with $\varphi^{(s)}$ a $s$-ary Hamiltonian constraint. In particular, the first generated Lagrangian constraints are given by 
$$
\psi_{\sigma}^{(1)}\,=\,\Phi_{\cL}^*\varphi^{(0)}_{\sigma}\,=\,\tilde A^s_{(\sigma)}\left(\frac{\del \cL}{\del q^s}\,-\,\frac{\del^2\cL}{\del v^s\del q^k}\,v^k\right)\,=\,0
$$
for a given suitable basis of $\ker^V\oL$. From the relations
\begin{align}
\Phi_{\cL}^*\{\varphi_{\sigma}^{(0)},H_0\}&=\,\psi_{\sigma}^{(1)}, \nn \\
\Phi_{\cL}^*\{\varphi_{\mu}^{(0)}, \varphi_{\sigma}^{(0)}\}&=\,L_{K^V_{(\sigma)}}\psi^{(1)}_{\mu}
\label{pond88}
\end{align}
one sees that the Hamiltonian splitting of the primary constraints $\varphi^{(0)}$ into first and second class reads a splitting of the first generated Lagrangian constraints $\psi^{(1)}$. First class primary constraints $\tilde\varphi^{(0)}$ on $T^*Q$ bijectively correspond to Lagrangian constraints $\tilde\psi^{(1)}$ for which $$L_{K^V_{(\sigma)}}\tilde\psi^{(1)}=0,$$ so that there exist functions $\gamma\in\mathcal F(M)$ such that $$\Phi_{\cL}^*\gamma=\tilde\psi^{(1)},$$ while second class primary constraints are mapped under \eqref{LcH} into elements $\psi^{(1)}$ which cannot be written as the pullback via the Legendre map $\Phi_{\cL}$ of any function on $M$. In particular, one can prove that the conditions $\tilde\psi^{(1)}=0$ reduce the manifold $TQ=P_0$ to $P_1$ (such conditions are usually referred to as \emph{dynamical constraints}) while the conditions $\psi^{(1)}=0$ reduce $P_1$ to the submanifold where the second order condition for $D$ is satisfied. The requirement that the dynamics respects the constraints provides a second generation of constraints, which again can be split into dynamical or second order conditions.  This procedure can be iterated, until a stable set (say $S_f$) is reached. As proven in \cite{batlle86}, this formalism is equivalent to the Hamiltonian formalism described above, although the dimensions of $M'$ can differ from the dimension of $S_f$, since (see \cite{coppia-dirac}) the dynamics \eqref{eq56} on $M'$ is proven to have the same number of free Lagrangian multipliers as the second order dynamics on $S_f$.

\subsection{A Noether theorem for singular Lagrangian systems}
\label{ss:cL}
We can now generalise the notion of Newtonoid symmetry to the dynamics described in terms of a degenerate Lagrangian. The previous analysis provided, for a suitably regular pre-symplectic system $(TQ,\oL, \dd\EL)$, a final submanifold $S_f$ embedded in $TQ$ and a class of second order vector fields $D\in\mathfrak X(S_f)$ which solve, as we described, the equation $i_D\oL=\dd\EL$ on $S_f$. We denote by $\tilde{\mathfrak X}^{\oL}(S_f)$ such a set. 

A natural generalisation and a merging of what we described in section \ref{ss:lag} and in section \ref{I-ss:genps}[I] is to say that a vector field $X\in\mathfrak X(TQ)$ gives a (generalised, i.e. Newtonoid) infinitesimal symmetry for the (pre-symplectic) dynamics given by $\tilde{\mathfrak X}^{\oL}(S_f)$  if the Newtonoid (see \eqref{eq37}) vector field $X^{(D)}\in\mathfrak X(S_f)$ (i.e. it is tangent to the final constraint manifold $S_f$) and if $[X^{(D)},D]\in\ker^V\oL\cap \mathfrak X(S_f)$. Notice that one can prove that this intersection  is spanned by
$$
\mathfrak X(TQ)\,\ni\,K_{\mu}\,=\,(\Phi_{\cL}^*\{q^j,\tilde\varphi_{\mu}^{(0)}\})\frac{\del}{\del v^j}
$$
for the (see \eqref{eq56}) primary constraints which are first class with respect to all the Hamiltonian constraints of the system. Analogously, we say that a vector field $X\in\mathfrak X(TQ)$ is an infinitesimal symmetry for the Lagrangian $\cL$ if a function $u\in\mathcal F(TQ)$ exists such that 
\beq
\label{nps1}
L_{X^{(\Gamma)}}\cL=L_{\Gamma}u
\eeq
 for any second order vector field $\Gamma$ on $TQ$. 

It is possible to prove (see \cite{mms83, CaRa88, arianna88, arianna90, arianna92} for systems without tertiary constraints) that if $X\in\mathfrak X(TQ)$ is an infinitesimal symmetry for the singular Lagrangian $\cL$, that is the relation \eqref{nps1} is valid, then the function $F=i_{X^{(\Gamma)}}\theta_{\cL}-u$ satisfies, for any $D\in\tilde{\mathfrak X}^{\oL}(S_f)$ the relations
\begin{align}
&L_{K_{\mu}}F=0; \nn \\
&i_{X^{(D)}}\oL\,=\,\dd F; \nn \\
&L_{X^{(D)}}\EL\,=\,0; \nn \\
&L_{X^{(D)}}F\,=\,0
\label{ntps1}
\end{align}
on $S_f$. Moreover, one also proves that 
the Newtonoid vector field $X^{(D)}$ is an element in $\mathfrak X(S_f)$ for any $D\in\tilde{\mathfrak X}^{\oL}(S_f)$, and it is an infinitesimal symmetry for the dynamics. This theorem can be inverted. If $F\in\mathcal F(TQ)$ satisfies 
$$
L_{D}F=0
$$
on $\Sigma$ for any $D\in\tilde{\mathfrak X}^{\oL}(S_f)$, then there exists a vector field $X\in\mathfrak X(TQ)$
such that $X$ is an infinitesimal symmetry for the dynamics, and also a symmetry for the Lagrangian. This is what we consider as the pre-symplectic generalisation of the Noether theorem and its converse. 
\begin{example}
\label{exemplum-sL2}
As in the example \ref{exe:mm}, and adopting the same notations,  we consider again the  dynamics of a charged point particle in a magnetic monopole field, i.e. the second order vector field 
\beq
\label{mon1}
D\,=\,v^a\frac{\del}{\del x^a}+\frac{\lambda}{r^3}\epsilon^a_{\,\,\,bc}x^bv^c\frac{\del}{\del v^a}
\eeq
on $T(\R_0^3)$. As in \cite{bala-mono, mr87, ferrara88}, with respect to the Hopf fibration $$\pi\,:\,Q\,\simeq\,\R^+_0\times {\rm SU}(2)\,\stackrel{\rm U(1)}{\longrightarrow}\,\R^3_0\,\simeq\, \R^+_0\times \mathrm S^2$$ 
we consider the Lagrangian 
\beq
\label{mon2}
\cL'\,=\,\frac{1}{2}{\rm Tr}(\frac{\dd}{\dd t}(rg\sigma_3g^{-1}))^2+i\lambda{\rm Tr}(\sigma_3g^{-1}\dot g)
\eeq
on $TQ$. Notice that the difference between $\cL'$ in \eqref{mon2} and the Lagrangian $\cL$ in \eqref{fe2} is in  the exponent of the interaction term. If we parametrise  again 
 $Q$ via  the radial coordinate $r>0$ and $ g\in{\rm SU}(2)\simeq{\rm S}^3$ via 
 $$
 g\,=\,\begin{pmatrix}
 u & -\bar v \\ v & \bar u\end{pmatrix}
 $$
with $\bar{u}u+\bar{v}v=1$ as 
\begin{align*}
&u\,=\,(\cos\theta/2)\,e^{i(\phi+\psi)/2}, \\ &v=(\sin\theta/2)\,e^{i(\psi-\phi)/2}
\end{align*}
in terms of the Euler angles $\phi\in[0,2\pi), \,\theta\in[0,\pi),\,\psi\in[0,2\pi)$, it is  
\beq
\label{mon3}
\cL'\,=\,\frac{1}{2}(v_r^2\,+\,r^2v_{\theta}^2\,+\,r^2(\sin^2\theta)v_{\phi}^2)\,+\,\lambda\,(v_{\psi}+v_{\phi}\cos\theta).
\eeq
The interaction term
$$
(v_{\psi}+v_{\phi}\cos\theta)\,=\,i_D\xi
$$
(with $D$ any second order vector field on $TQ$) comes from the monopole connection for the Hopf bundle, i.e.
$$
\xi\,=\,i{\rm Tr}(\sigma_3 g^{-1}\dd g) 
$$
where $g^{-1}\dd g$ is the left invariant Maurer-Cartan form and $\sigma_3$ is the Pauli matrix associated to the generator of the ${\rm U}(1)$ action.  Some straightforward calculations read
\begin{align}
&\theta_{\cL'}\,=\,v_r\dd r\,+\,r^2v_\theta\dd\theta\,+\,\lambda\dd\psi\,+\,(r^2v_\phi\sin^2\theta +\lambda\cos\theta)\dd\phi, \nn \\
&E_{\cL'}\,=\,\frac{1}{2}(v_r^2+r^2(v_\theta^2+(\sin^2\theta) v_\phi^2)):
\label{mon4}
\end{align}
notice that, since the interaction term in $\cL'$ is linear in the velocity variables, the corresponding energy function $E_{\cL}$ has only the kinetic term. 

The Lagrangian $\cL'$ is singular. Omitting the explicit expression of the 2-form $\omega_{\cL'}=-\dd\theta_{\cL'}$ (see \cite{arianna88, arianna92}), it is easy to see that 
\beq
\label{mon5}
\ker\,\omega_{\cL'}\,=\,\{\alpha\,\del_\psi\,+\,\beta\,\del_{v_{\psi}}\}
\eeq
with $\alpha, \beta\,\in\,\mathcal F(TQ)$. Comparing \eqref{mon5} with \eqref{mon4} makes it immediate to see that $L_K\E_{\cL'}=0$ for any $K\in\ker\,\omega_{\cL'}$,  and that moreover the restriction of the soldering endomorphism $S$
is surjective onto the vertical $\ker^V\omega_{\cL'}$, so we have that the Lagrangian \eqref{mon3} is type II. The vector fields $\mathfrak D'$  defined by $L_{\mathfrak D'}\theta_{\cL'}=\dd\cL'$ are given by 
\begin{align}
\mathfrak D'\,=&v_r\frac{\del}{\del r}+v_\theta\frac{\del}{\del \theta}+v_\phi\frac{\del}{\del \phi} \nn \\ &\qquad+(r\,v_\theta^2+rv_\phi^2\sin^2\theta)\frac{\del}{\del v_r}+(v_\phi^2\sin\theta\,\cos\theta-2v_\theta-\frac{\lambda}{r^2}v_\phi\sin\theta)\frac{\del}{\del v_{\theta}} \nn \\ &\qquad\qquad +(\frac{\lambda}{r^2\sin\theta}v_{\theta}-\frac{2\sin\theta}{r}v_r-\frac{2\cos\theta}{\sin\theta}v_{\theta})\frac{\del}{\del v_\phi}+\alpha\frac{\del}{\del\psi}+\beta\frac{\del}{\del v_\psi}
\label{mon6}
\end{align}
on all $TQ$. They provide $\mathfrak X^{\omega_{\cL'}}(TQ)$. It is evident that, for any such a $\mathfrak D'$, the sum $$\tilde{\mathfrak D}'={\mathfrak D}'+(v_{\psi}-\alpha)\del_\psi$$ gives a second order vector field on all $TQ$.

We close\footnote{Notice that the paper \cite{CaRa88} focuses on the Noether theorem for type II Lagrangians.  } this example by noticing that the vector field $$X\,=\,f\del_\psi+(L_Df)\del_{v_\psi}$$ with $f\in\mathcal F(TQ)$ is a Newtonoid infinitesimal symmetry for $D$. Moreover, it is easy to see that the vector fields in $\mathfrak X^{\omega_{\cL'}}(TQ)$ can all be projected to vector fields on $T(\R^3_0)$, where they give the equations of motions \eqref{mon1}. 

This example shows then that the  dynamics of a charged particle in a monopole magnetic field can be given as a reduction of a global (on a larger carrier space) singular Lagrangian.

\end{example}

\section{Symmetries and conservation laws for the Hamilton-Jacobi theory}
\label{sec:HJ}

What is referred to as the Hamilton-Jacobi (HJ) formalism of classical dynamics is very elegant and economical with respect to both the Hamiltonian and the Lagrangian formalisms. It provides an important physical example of the deep relations existing between a class of first order non linear \emph{partial} differential equations (p.d.e.'s)  on a configuration manifold $Q$ and systems of Hamiltonian \emph{ordinary} differential equations on the cotangent bundle manifold $T^*Q$.  Moreover, it comes as a natural topic within the general approach we have taken in this paper, as it emerges as a semiclassical approximation of one of the two equations one derives from the Schr\"odinger equation when written on $Q$ or in terms of Gaussian coordinates. 

Consider a (smooth and orientable) configuration space $Q$ equipped with a non degenerate metric tensor $g=g_{ab}\dd q^a\otimes\dd q^b$ along a coordinate system $\{q^a\}_{a=1, \dots, N}$. We recall that, for $f\in\mathcal F(Q)$, the gradient $\nabla f$ is the vector field implicitly defined by the condition 
$$
g^{-1}(\dd f,\alpha)=i_{\nabla f}\alpha
$$
for any 1-form $\alpha$ on $Q$, while the Laplace-Beltrami operator is 
$$
\Delta\phi={\rm div}(\nabla\phi),
$$
where the  divergence of a vector field $X$ on $Q$ is defined by 
\beq
\label{2807211}
L_X\tau\,=\,({\rm div}X)\tau 
\eeq
with respect to any\footnote{Although the relation \eqref{2807211} defines the divergence of a vector field on an orientable smooth manifold $M$ with respect to \emph{any} volume form $\tau$, it is customary to use, when $M$ is equipped with a non degenerate metric tensor $g=g_{ab}\dd q^a\otimes\dd q^b$, the so called \emph{metric} volume form, that is $\tau=\sqrt{|\det[g_{ab}]|}\dd q^1\wedge\dots\dd q^N$.} volume form $\tau$ on $Q$. 
The Schr\"odinger equation describing the evolution of the wave function $\psi=\psi(q,t)$  for a point particle of mass $m$ moving in such configuration space   under an external potential $V=V(q)$ is\footnote{Here $\hbar=h/2\pi$ is the reduced Planck's constant.}
$$
i\hbar\frac{\del\psi}{\del t}\,=\,-\frac{\hbar^2}{2m}\Delta\psi+V(q)\psi,
$$
where  $\psi(t)\in L^2(Q, \tau)$ (i.e. a square integrable function on $Q$). 
Since $\psi$ is a $\C$-valued function, we can write it in terms of an amplitude and a phase, i.e.  $$\psi(q,t)=A(q,t)e^{iS(q,t)/\hbar}$$ where $A=A(t,q),\,S=S(t,q)$ are both real valued functions, with $A$ square integrable with respect to the position variables. The Schr\"odinger equation reads
\begin{align}
&\frac{\del A}{\del t}+\frac{1}{m}\nabla A\cdot\nabla S+ \frac{1}{2m}A\Delta S 
=0 \label{17.1} \\ 
& A(\frac{\del S}{\del t}+\frac{1}{2m}\nabla S\cdot\nabla S+V(q))=\frac{\hbar^2}{2m}\Delta A, 
\label{17.2}
\end{align}
where we have written 
$$
\nabla A\cdot\nabla S\,=\,g(\nabla A, \nabla S)\,=\,g^{-1}(\dd A, \dd S).
$$
The equation \eqref{17.1} represents a continuity relation for the density $\rho=A^2$ and the current  vector $\nabla S/m$, since it gives
\beq
\label{17.3}
\frac{\del \rho}{\del t}+\frac{1}{m}{\rm div}(\rho \nabla S)=0,
\eeq
while on the domains defined by $A\neq0$ the equation \eqref{17.2} can be written as\footnote{For the analysis of the conditions under which the solutions of this equations on different disconnected domains defined by $A\neq0$ can be glued, we refer the reader to the analysis of the so called JWKB approximation. See \cite{ems, emms}.}
\beq
\label{17.4}
\frac{\del S}{\del t}+\frac{1}{2m}\nabla S\cdot\nabla S+V(q)\,=\,\frac{\hbar^2}{2m}\frac{\Delta A}{A}.
\eeq
If the term depending on $\hbar^2$ is neglected, such equation is the well known  HJ equation for a Hamiltonian $H=\delta^{ab}p_ap_b/2m+V(q)$ on the phase space $T^*Q$, as we shall more geometrically describe in the following sections. 

The Schr\"odinger equation, which is a linear p.d.e. (partial differential equation) for a $\C$-valued wave function $\psi$ on a configuration space, reads two non-linear p.d.e.'s for the $\R$-valued amplitude and phase $\psi=Ae^{iS/\hbar}$. One of these equations gives a continuity relation, one gives a quantum correction to the classical HJ equation for the eikonal function $S$, which will be recovered as a principal function. The two equations are decoupled in the semiclassical approximation given by neglecting (at the lowest order of approximation) the terms in $\hbar$.  The HJ equation appears then in quantum mechanics as a semi-classical  approximation of the Schr\"odinger equation describing the quantum evolution. 

\begin{rema}
\label{uub}
A Hamilton-Jacobi equation is given as an approximation of the Schr\"odinger also when it is written in terms of the Hodge Laplacian\footnote{That one can define inequivalent Laplacian operators on a smooth orientable manifold $Q$ equipped with a metric tensor $g$ reading a non flat curvature is well known, and for that we refer to \cite{jjost}. For an explicit analysis of the cases of low dimensional spheres, see \cite{dz1,dz2}.  }, where  such operator is defined by 
$$
\tilde{\Delta}\,\alpha\,=\,(-1)^{kN}(\star\dd\star\dd\alpha+(-1)^N\dd\star\dd\star\alpha)
$$
for any exterior form $\alpha$ on $Q$ with respect to the Hodge duality $\star\,:\,\Lambda^{k}(Q)\to\Lambda^{N-k}(Q)$ defined via the metric volume tensor. If we consider 
$$
i\hbar\frac{\del\psi}{\del t}\,=\,-\frac{\hbar^2}{2m}\tilde{\Delta}\,\psi+V(q)\psi,
$$
with $\psi\in L^2(Q,\tau=\star(1))$ and set $\psi=Ae^{iS/\hbar}$, we indeed have
\begin{align}
&\frac{\del A}{\del t}+\frac{1}{m}\star(\dd S\wedge(\star\dd A))+\frac{1}{2m}A\,\tilde{\Delta} S=0, \label{ub2} \\
&A(\frac{\del S}{\del t}\,+\,\frac{1}{2m}\star(\dd S\wedge\star(\dd S))+V(q))\,=\,\frac{\hbar^2}{2m}\tilde{\Delta} A. \label{ub1} 
\end{align}
Recalling that 
$$
\star(\dd\star \alpha)={\rm div}(\alpha)
$$
defines the divergence of a $k$ form $\alpha$ on $Q$, while 
$$
g^{-1}(\nabla f, \nabla \tilde f)=\star(\dd f\wedge\star(\dd \tilde f))
$$
provides a definition for the scalar product of (exact, here) 1-forms, one sees that \eqref{ub2} gives a continuity relation, while neglecting the term depending on $\hbar^2$ in \eqref{ub1} where $A\neq0$ gives a HJ equation. 

This example can be generalised. 
Using the notion of principal symbol of a linear differential operator acting upon a set  $\mathcal F(Q)$ of functions  
on a configuration space, it is  possible to show that  higher (homogeneous) order linear p.d.e.'s  on $Q$ read, in a suitable limiting process,   HJ-type (non linear) equations associated to Hamiltonian functions $H(q,p)$ which are higher (homogeneous) order  polynomials in the momentum variables. When physical problems require to deal with non homogeneous differential operators, it is possible to see that, upon adding an auxiliary variable  (that is, upon suitably extending the configuration space of the system) non homogeneous differential operators can be made homogeneous at the highest degree. Along such path,   it is indeed possible to associate HJ-type (non linear) equations to non homogeneous linear differential operators (for details, see \cite{vinovteam}). 

We conclude this remark by noticing that quantizing a classical (Hamiltonian) dynamics may amount to suitably linearise the non-linear classical HJ equation to a p.d.e. which encodes also a continuity condition for a probability current. For a more precise analysis of this topic  we refer to \cite{q-HJ, hjs}.  
\end{rema}

It is time for us to describe the relations between the HJ p.d.e. on $Q$ and the Hamiltonian formulation of a classical dynamics starting from the symplectic geometry of the cotangent bundle $T^*Q$.

We start by recalling that, if $(M,\omega)$ is a $2N$-dimensional symplectic manifold, then a diffeomorphism $\phi:M\to M$ is canonical if and only (i.e. $\phi^*\omega=\omega$) if and only if its graph $\Sigma_\phi\subset M\times M$ given by $(m,\phi(m))$ for $m\in M$ is a Lagrangian submanifold in $M\times M$ with respect to the symplectic tensor 
\beq
\label{8giu2}
\omega_{M\times M}\,=\,\omega\,\ominus\,\omega 
\eeq
on the cartesian product, that is the second $M$ factor is equipped with the symplectic structure $-\omega$. When $(M=T^*Q, \omega=\omega_Q)$ one writes the diffeomorphism $\phi\,:\,(q^a,p_a)\,\mapsto\,(x^a, k_a)$ so that the symplectic structure on $T^*Q\times T^*Q$ is 
$$
\dd q^a\wedge\dd p_a-\dd x^a\wedge\dd k_a.
$$
From the equivalence $T^*Q\times T^*Q\simeq T^*(Q\times Q)$ one can consider those Lagrangian submanifolds which can be written as the graph
$$
\dd S\,:\,Q\times Q\,\to\, T^*(Q\times Q)
$$
for a so called generating function $S\,:\,Q\times Q\,\to\,\R$. If $H(q,p,t)$ is the Hamiltonian of a classical dynamics on  the extended phase space $T^*Q\times\R$, such a canonical transformation\footnote{Notice that the equivalence $T^*Q\times T^*Q\simeq T^*(Q\times Q)$ (where $T^*Q\times T^*Q$ can be equipped with any linear combination of two copies of $\omega_Q$, slightly generalising \eqref{8giu2}) is not canonical with respect to the symplectic structures $\omega_{Q\times Q}$ on $T^*(Q\times Q)$. It is possible to identify the basis of the bundle $T^*Q\times T^*Q$ with different Lagrangian submanifolds and the graph $\Sigma_{\phi}$ as coming from a generating function $S$ which may depend on suitable choices of both position and momenta variables. This results in what is usually referrred to as \emph{types} of generating functions, and consequently different HJ equations.} results in the relation 
$$
p_a\dd q^a-H\dd t=k_a\dd x^a-H'\dd t+\dd S,
$$
with $S=S(q,x,t)$ and transformed Hamiltonian $H'$ on $T^*Q\times\R$ so that the Hamiltonian vector field $X_H$ is transformed into the Hamiltonian vector field $X_{H'}$. The canonical transformation $\phi$  is given by 
\begin{align}
&p_a\,=\,\frac{\del S}{\del q^a},\nn \\
\label{hj1}
& k_a\,=\,-\frac{\del S}{\del x^a}.
\end{align}
under  the condition that  the matrix
\beq
\label{hj2}
S_{ab}\,=\,\frac{\del^2 S}{\del q^a\del x^b}
\eeq
is invertible for any value $t$ of the time parameter, and 
\beq
\label{hj3}
H'\,=\,H\,+\,\frac{\del S}{\del t}.
\eeq
If a function $S$ exists, such that $H'$ is constant, or depends only on the $k_a$ variables (i.e. the momenta), then the dynamics $X_{H'}$ is completely integrable and then integrable by quadratures. Finding such a generating function $S$ is usually referred to as solving the Hamilton - Jacobi (HJ) partial differential equation for a given Hamiltonian $H$, which is 
\beq
\label{hj4}
H\left(q, \frac{\del S}{\del q}, t\right)\,+\,\frac{\del S}{\del t}\,=\,0.
\eeq

Our presentation will consider only time independent Hamiltonian systems, which give, under the ansatz $$S\,=\,W-Et$$ (with $E$ a constant which can be identified in the mechanical case with the energy and $W$ not depending on $t$), the time independent HJ equation
\beq
\label{hj5}
H\left(q, \frac{\del W}{\del q}\right)\,=\,E.
\eeq
Jacobi's motivation to investigate the (nowadays) called HJ p.d.e. was to elaborate a method for integrating the f.o.d.e. of a dynamical system in the Hamiltonian form. The strategy introduced in  \cite{jac} is to split up such integration problem in two steps.  If $W=W(q,x)$ solves the HJ equation, then the integration of the Hamilton's equations of the motion is given by solving first a set of f.o.d.e.'s on $Q$, namely 
\beq
\label{2m.1}
\frac{\dd q^a}{\dd t}=\frac{\del H}{\del p_a}\mid_{p_a=\frac{\del W}{\del q^a}}, 
\eeq
then  the relations
\beq
\label{2m.2}
p_a\,=\,\frac{\del W}{\del q^a}\mid_{q^a=q^a(t)}
\eeq
with $q^a(t)$ the solution to \eqref{2m.1}, 
 give a full set of integral curves for the Hamiltonian dynamics  $X_H$ with initial conditions $$(q^a(0), p_a(0)=\frac{\del W}{\del q^a} (q(0),x)).$$ 
 A complete solution of the  HJ p.d.e. gives a  \emph{family} of f.o.d.e. on $Q$ (labelled by the values of the variables $x$ in \eqref{2m.1}) whose solutions are sufficient, via \eqref{2m.2}, to give solutions for the Hamilton's f.o.d.e. on $T^*Q$. This family raised the interest of Dirac, who wrote in \cite{Df}:
 
 \emph{Such family does not have any importance from the point of view of Newtonian's mechanics;  but it is a family which corresponds to one state of motions in the quantum theory, so presumably the family has  some deep significance in nature, not yet properly understood.}

One further interesting point is to be described. As clearly elucidated in many textbooks (see for example \cite{lanczos, sm-book}) when a dynamics is Hamiltonian on a symplectic manifold $(M,\omega)$, the time evolution $\Phi_t$  itself is the time unfolding of a one parameter family of canonical transformations on $M$ that we write, on  $(M=T^*Q, \omega=\omega_Q)$, as
$$
\Phi_t\,:\,(q(0), p(0))\,\mapsto\,(q(t),p(t)).
$$
It is possible to prove that, when the Hamiltonian $H(q,p,t)$ for such a dynamics is given in terms of the Legendre transform of a regular Lagrangian $\cL(q,v,t)$, then the function
\beq
\label{15.1}
S=S(q(0),q,t)\,=\,\int_0^t\dd t\,\cL(q(t), v(t),t))
\eeq
given by the action functional when  the integral is computed \emph{along the actual  solutions} of the Cauchy problem given by the Hamilton equations of the motions gives a solution to \eqref{hj4} upon identifying $x=q(0)$.

\subsection{A geometric setting for the Hamilton-Jacobi theory}
\label{ss:o}  
When it comes to analyse a notion of solution and  of symmetry for the Hamilton-Jacobi equation \eqref{hj4} associated to a Hamiltonian vector field $X_H$ on the symplectic phase space $(T^*Q, \omega_Q)$, the deep difference between this picture and the others we have described (namely the Poisson, the symplectic, the Lagrangian pictures of classical mechanics) becomes evident. The HJ equation is  a first order (non linear) p.d.e., and its natural setting is within the geometry of jet spaces.  

If $\tilde Q=Q\times \R$, the submersion $\tau\,:\,\R\times\tilde Q\,\to\,\tilde Q$ which we locally write as $(S, q^j, t)\mapsto(q^j,t)$ gives a vector bundle whose sections are identified by functions $S=S(q^j, t)$. The quotient of such set of sections via the equivalence relation (compare it with \eqref{25.6})
\beq
\label{18.1}
S\sim S'\quad\Leftrightarrow\quad \frac{\del S}{\del q^j}=\frac{\del S'}{\del q^j}, \quad  \frac{\del S}{\del t}=\frac{\del S'}{\del t}
\eeq
is proven to be a smooth manifold. It is denoted by $J^1\tau$, and its elements are the equivalence classes $[S]$ defined by any  element $S\in\mathcal F(\tilde Q)$. It is easy to prove that (compare it with \eqref{25.7})
\beq
\label{18.2}
J^1\tau\simeq \R\times T^*\tilde Q
\eeq
is a smooth manifold with local coordinates given by $(\sigma, q^j,t,p_j,p_0)_{j=1,\dots,N}$: any element $[S]$ in $J^1\tau$ can be identified, as the relation \eqref{18.1} suggests,  by the 1-form $\dd S$. 
Moreover, if $S\in\mathcal F(\tilde Q)$, its first order prolongation is defined to be the section $[S]$ which is locally written as the section $\dd S\,:\,\tilde Q\,\to\,T^*\tilde Q$. 

In analogy to what we have described in section \ref{ss:2} about first order ordinary differential equations, a first order partial differential equation can be defined as a subset (usually  assumed to be a submanifold) $\mathfrak P\subset J^1\tau$. The HJ p.d.e. is given by the submanifold  $\mathfrak P_H$ implicitly defined by \eqref{hj4}, i.e.
$$
H(q^j,p_0,t)+p_0=0.
$$
A \emph{solution} of such equation is given by a function $S$ on $\tilde Q$  whose first order prolongation is in $\mathfrak P_H$. This condition amounts to 
$$
(j^1S)^*(H(q^j,p_0,t)+p_0)\,=\,H(q, \frac{\del S}{\del t}, t)+\frac{\del S}{\del t}=0,
$$
 A \emph{symmetry} for this equation will be given by a map $Y\,:\,\tilde Q\,\to\,\tilde Q$ whose first order prolongation (which turns to be expressible in terms of the pull back
$Y^*(\dd S)$) maps solutions into solutions.  
\begin{rema}
\label{re18}
The equivalence \eqref{18.2} allows to define a first order partial differential equation on $\tilde Q$ as a submanifold (we assume) $\mathfrak P\subset\R\times T^*\tilde Q$. If $\mathfrak P$ is defined (as it is the case for the HJ equation) as the zero level set 
\beq
\label{18.3}
G(\sigma,q^j,t,p_0, p_j)=0
\eeq
(where $p_0$ is the fiber coordinate to $t$) 
together with the constraint
$$
\alpha=\dd \sigma-p_0\dd t-p_j\dd q^j=0
$$
on each cotangent space, then a symmetry for $\mathfrak P$ can be defined to be a map $Y\,:\,\R\times T^*\tilde Q\to\R\times  T^*\tilde Q$ which maps $\mathfrak P$ into itself. Its infinitesimal counterpart can be written in analogy to the relation \eqref{new01} already described for first order ode's. A vector field $X$ on $\R\times T^*\tilde Q$ is an infinitesimal symmetry for $\mathfrak P$ given by \eqref{18.3} if there exist functions $\lambda, \mu, \nu, \rho$ on $\R\times T^*\tilde Q$ such that 
\begin{align}
& L_XG=\lambda G, \nn \\
& L_X\alpha=\mu \alpha+\nu \dd G+G\dd \rho.
\nn 
\end{align}

\end{rema}
It is not the aim of the present paper\footnote{We refer the reader to \cite{boyer, shov} and references therein.} to analyse such a general theory for the notions of solutions and  of symmetries for the HJ equation. Coherently with the path we have already travelled, we focus our attention to describe solutions and symmetries which come from the fact that the HJ equation encodes the properties related to Lagrangian submanifolds embedded into symplectic manifolds. As we described in the introduction to the first part [I] of this paper, it will be the symplectic structure on the phase space which will give a Noether-type theorem.

\subsection{Lagrangian embeddings and solutions of the Hamilton-Jacobi equation}
\label{ss:HJ1} 
In order to cast this problem within a  geometric formalism that allows to describe a Noether theorem for it, we start by recalling\footnote{We refer the reader to \cite{sw-hj, mmm-hj} for a more general theory.} that, given a $2N$-dimensional symplectic manifold $(M,\omega)$, a map $$\varphi:M'\to M,$$ with $M'$ a $N$-dimensional  manifold, is called a \emph{L-embedding} if  it is an embedding\footnote{that is, the range $\varphi(M')$ is a submanifold of $M$ such that $M'$ and $\varphi(M')$ are diffeomorphic, see the appendix \ref{I-app2}[I].} and if the tangent distribution to $\varphi(M')$ is Lagrangian with respect to $\omega$ in $M$.  
Analogously, a map $$\varphi:M\to N'$$ (with $N'$ a manifold) is called a \emph{L-submersion} if  it is a submersion\footnote{that is, if its differential has constant rank equal to the the dimension of $N'$, see the appendix \ref{I-app2}[I].} such that the set $\varphi^{-1}(n')$ is a Lagrangian submanifold in $M$ for any $n'\in N'$. In that case, there is a family of L-submanifolds in $M$, with each $m\in M$ belonging to exactly one of them. 

An interesting example of a L-submersion comes upon considering $N$ globally defined and independent functions $f^a\in\mathcal F(M)$ which are in involution, i.e. which satisfy the conditions
\begin{align}
&\dd f^1\wedge\dots\wedge\dd f^N\,\neq\,0, \nn \\
&\{f^a,f^b\}\,=\,0;
\label{hj6}
\end{align}
 the map $\varphi:M\to N'=\R^N$ given by $$\varphi(m)\,=\,f^1(m), \dots, f^N(m)$$ turns out to be a L-submersion, and $M$ is foliated by the level sets of $\varphi$, with each leaf given by $\varphi^{-1}(x)$ for any $x\in\R^N$.  Notice that a leaf $\varphi^{-1}(x)$ is an example of a constrained submanifold in $M$ defined by a set of $N$ first class constraints, following Dirac's formulation\footnote{We recall here from section \ref{I-sub:dirac}[I]:  a submanifold $\Gamma\hookrightarrow M$ embedded in a $2N$-dimensional symplectic manifold $(M,\omega)$ is Lagrangian if and only if one can write  $\Gamma\,=\,\{m\in M\,:\,f_a(m)=0, \,\,\,a=1,\dots,N\}$ with $f_a\in\mathcal F(M)$ such that $\{f_a,f_b\}_{\mid \Gamma}=0$ and $\dd f_1\wedge\dots \wedge \dd f_N\neq0$. }.

With $M$  the total space of the fiber bundle $\pi:M\to B$, we say that $(M,\omega,\pi,B)$ is a \emph{L-bundle} if, for any $b\in B$, the fiber $\pi^{-1}(b)$ is a L-submanifold in $M$. A familiar example is given by considering the cotangent bundle $(T^*Q, \omega_Q, \pi_Q, Q)$ since each fiber is  the cotangent space $T^*_qQ$, which is a L-submanifold in $T^*Q$. If, with respect to the example above \eqref{hj6}, the set $\varphi^{-1}(x)$ for any $x\in\R^N$ can be given a manifold structure, then $(M,\omega, \pi=\varphi, \R^N)$ is another example of a L-bundle\footnote{We refer the reader to \cite{dade-87} for a more complete analysis on Lagrangian fiberings within the context of action-angles variables, and to \cite{dui74} within the context of geometric quantization.}.

If $\varphi:M'\to M$ is a L-embedding and $(M,\omega, \pi, B)$ is a L-bundle, then the composition $$\psi=\pi\circ\varphi:M'\to B$$ is called the  \emph{L-map}
associated to a given L-embedding and a given L-bundle. It is clear that, if the L-embedding is the inclusion map $i_{M'}:M'\hookrightarrow M$, then the corresponding L-map $\psi=\pi\circ i_{M'}$ is the restriction of the projection $\pi$ to the submanifold $M'$ in $M$. 

In general, the L-embedding $\varphi:M'\to M$ is called \emph{transversal} with respect to the L-bundle $(M,\omega, \pi, B)$ if, at every $m=\varphi(m')$, it is $$\varphi_*(T_{m'}M')\oplus\ker(\pi_*(m))\,=\,T_mM.$$
This means that an L-embedding $\varphi:M'\to M$ is transversal with respect to a given L-bundle if the map $\varphi_*$ transforms the tangent space to $M'$ into a distribution on $M$ with a trivial intersection to the vertical distribution of the bundle.

Transversality is not a property intrinsic to $\varphi(M')$ as a Lagrangian submanifold in $M$, but relates it  to the L-bundle structure given on $M$ by the projection $\pi$. The \emph{caustic} (or \emph{catastrophe}) set of the L-map $$\psi\simeq(M'\stackrel{\varphi}{\to}M\stackrel{\pi}{\to}B)$$ is the set of the critical values\footnote{see \cite{AbRo67} for details.} for $\psi$, i.e. the set of elements $b\in\psi(M')$ such that the differential $\dd\psi$ fails to have rank $N$. 
It is then possible to prove that a L-embedding $\varphi:M'\to M$ is transversal with respect to the L-bundle $(M,\omega,\pi, B)$ if and only if the caustic set of the corresponding L-map $\psi=\pi\circ\varphi:M'\to B$ is empty.

Within such a general setting, consider the L-bundle given by the cotangent bundle $$(T^*Q, \omega_Q=\dd q^a\wedge\dd p_a, \pi_Q, Q)$$ for a given smooth $N$-dimensional manifold $Q$.
A smooth 1-form $\alpha\in\Lambda^1(Q)$, whose local expression is $\alpha=\alpha_a\dd q^a$, defines a section $\sigma^{(\alpha)}:Q\to T^*Q$ locally represented by  
\beq
\label{hj7}
q^a\mapsto(q^a,p_a=\alpha_a).
\eeq
 It is immediate to see that $\sigma^{(\alpha)}$ is a L-embedding if and only if $\dd\alpha=0$. The L-map associated to the cotangent bundle L-structure reads  $\psi=\pi_Q\circ\sigma^{(\alpha)}={\rm id}_Q$, so  the L-embedding is transversal (see \cite{wlodek-L1, wlodek-L2}).

Following the conventional approach, a \emph{complete integral} of the HJ p.d.e.  \eqref{hj5} for a Hamiltonian $H$ defined on a $2N$ dimensional cotangent bundle $T^*Q$ is a function $W\,\in\,\mathcal F(Q\times U)$ with $U\subseteq\R^N$ such that the condition (analogue to \eqref{hj2}) $$\det(\frac{\del^2W}{\del q^a\del u^b})\neq0$$ (where  $\{u^a\}_{a=1,\dots N}$ is a local coordinate chart on $U$) is fulfilled. For any fixed $u\in U$, the 1-form $\dd W$ on $Q$ defines as in \eqref{hj7} a section $\sigma^{(\dd W)}:Q\,\to\,T^*Q$ given by 
\beq
\label{hj8}
\sigma^{(\dd W)}\,:\,q\,\mapsto (q, p=\frac{\del W}{\del q}).
\eeq
The comparison with \eqref{hj7} shows that such a section provides, for any $u\in U$, a transversal L-embedding: upon defining $$\mathcal F(T^*Q)\ni f_j^{(W)}=p_j-\del_{q^j}W,$$ such embedded L-submanifold  -- denote it by $\Gamma^{(W)}$ --  can be described,  for any $u\in U$, as given by the constraint relations
\beq
\label{hj9}
f_j^{(W)}=0
\eeq
which are immediately seen to satisfy the relations \eqref{hj6}. Upon noticing that the time independent HJ equation can be written, using 
\eqref{hj8}, as 
\beq
\label{hj11}
\sigma^{(\dd W)*}(H-E)=0,
\eeq it is possible to prove that 
\beq
\label{hj10}
\{f_j^{(W)},H\}_{\mid \Gamma^{(W)}}=0,
\eeq
that is the Hamiltonian vector field $X_H$ is tangent, for any $u\in U$, to the L-submanifold $\Gamma^{(W)}$, and from the relation $\{f_a^{(W)}, f_{b}^{(W)}\}=0$ 
that the Hamiltonian vector fields $X_{f_{j}^{(W)}}$ span the tangent space to $\Gamma^{(W)}$ for any $u\in U$. This means also that $\Gamma^{(W)}$ is mapped into itself under the time evolution driven by the Hamiltonian $H$, and so the time evolution does not develop caustics. 

If we denote by $\Sigma_E$ the submanifold (we do not consider the critical points of the Hamiltonian function $H$) in $T^*Q$ defined by $H(q,p)=E$, then it is also easy to see that $\Gamma^{(W)}\subset \Sigma_E$, so $E$ is included in the space of parameters $U$: for fixed $E$,  the submanifolds $\Gamma^{(W)}$ give a $(N-1)$-parameter foliation of $\Sigma_E$ by L-submanifolds, i.e. $\Sigma_E=\cup_u \Gamma^{(W)}$, where the union ranges over $u\in U$ provided $E$ is fixed\footnote{When the HJ equation has a \emph{solution} $W\in\mathcal F(Q\times U)$ with $U\subseteq \R^k$ with $k<N$, then $E$ can still be considered as one of the parameters in $U$, but the L-submanifold $\Gamma^{(W)}$ will depend on a $k$-dimensional space of parameters. We refer to \cite{mmm-hj} for more details.}. 

When $W$ is a complete solution of the HJ equation, the function 
\beq
\label{2m.3}
\dd W:Q\times U\to T^*Q
\eeq is a diffeomorphism and provides, via  \eqref{2m.2} (like  \eqref{hj9}) a $N$-dimensional foliation of $T^*Q$, whose leaves are Lagrangian and transverse to the bundle structure in $T^*Q$. Since $X_H$ is tangent  to each leaf, its restriction projects to a vector field on $Q$ whose integral curves are given by \eqref{2m.1}. In this sense, one says  that $X_H$ can be replaced by a family of vector fields on $Q$.

\subsection{A Noether theorem for the Hamilton-Jacobi theory}
\label{ss:HJ2}

The notion of complete integral for the HJ equation \eqref{hj5} has been formulated within a geometric setting in \eqref{hj11} via \eqref{hj8}. In order to study a coherent notion of symmetry for this problem we start by noticing that, with $(M,\omega)$ a symplectic $2N$ dimensional manifold and $\phi:M\to M$ a canonical diffeomorphism (see \eqref{I-ds00}[I]),  the map $$\varphi'=\phi\circ\varphi:M'\to M$$ is a L-embedding for a given  L-embedding 
$\varphi:M'\to M$, while the map $$\varphi'=\varphi\circ\phi^{-1}: M\to N'$$ is a L-submersion provided $\varphi:M\to N'$ is a L-submersion. 
 It is also easy to see that, if $(M, \omega, \pi, B)$ is a L-bundle, then the projection $$\pi\circ\phi:M\to B$$ defines a \emph{new} L-bundle structure for  $(M,\omega)$ onto the same basis manifold $B$. Analogously, the transversality of the map  $\psi\simeq(M'\stackrel{\varphi}{\to}M\stackrel{\pi}{\to}B)$ is not necessarily preserved under a generic canonical transformation $\phi$ on $M$: the map $\psi'=\pi\circ\phi\circ\varphi$ can indeed have caustics. 

The class of fiber preserving canonical transformations for the L-bundle $(M,\omega, \pi, B)$ is defined as the set of canonical diffeomorphisms $\phi$ on $M$ such that there exists a diffeomorphism $\phi_0:B\to B$ satisfying  
\beq
\label{un1}
\pi\circ\phi\,=\,\phi_0\circ\pi:
\eeq
the L-fiber $\Gamma_b=\pi^{-1}(b)$ for $b\in B$ is mapped under the given fiber preserving canonical map into the L-fiber $\Gamma_{\phi_0(b)}$.
It is immediate to prove that, if the L-map $\psi\simeq(M'\stackrel{\varphi}{\to}M\stackrel{\pi}{\to}B)$ is transversal and $\phi$ is a fiber preserving canonical map on $M$,  then  $$\psi'\,=\,\pi\circ\phi\circ\varphi=\phi_0\circ\psi:M'\to B$$ is a transversal L-map. 

When the L-bundle is given by the cotangent bundle $(T^*Q, \omega_Q,\pi_Q,Q)$, and $\phi$ is a fiber preserving canonical diffeomorphism with associated base diffeomorphism $\phi_0$ on $Q$ as in \eqref{un1}, one can write 
\beq
\label{hj16}
\phi\,=\,(\phi\circ T^*\phi_0^{-1})\circ T^*\phi_0.
\eeq
This proves that 
any fiber preserving canonical diffeomorphism $\phi$ on $T^*Q$  can be \emph{uniquely} decomposed as the product, in that order, of a fiber preserving base invariant canonical diffeomorphism $\beta=\phi\circ T^*\phi_0^{-1}$ on $T^*Q$ (i.e. such that $\pi_Q\circ\beta=\pi_Q$) and a canonical lift\footnote{We recall that, within the tangent bundle formalism, a map $\phi:TQ\to TQ$ is Newtonian if there exists a diffeomorhism  $\phi_0$ on $Q$ such that $\phi=T\phi_0$ (see \eqref{eq25}). Canonical lifts within the cotangent bundle setting are the analogue of the point transformations in the tangent bundle setting. They are given by \eqref{I-calif}[I].  } $T^*\phi_0$. It is moreover possible to prove that, given a  fiber preserving canonical map $\phi$ on $T^*Q$, there exists a  closed 1-form $\gamma\in\Lambda^1(Q)$ such that (see  \eqref{I-symfo2}[I]) 
\beq
\label{hj12}
\phi^*\theta_Q\,=\,\theta_Q\,+\,\pi_Q^*\gamma
\eeq

Under such a  fiber preserving canonical map $\phi$ on $T^*Q$ it is possible to prove that the transversal L-submanifold  (see \eqref{hj7}) given by the graph of the section $\sigma^{(\alpha)}$ for the  closed 1-form $\alpha\in\Lambda^1(Q)$ is transformed into the graph of the section  $\sigma^{(\alpha')}$ with 
\beq
\label{hj13}
\alpha'=\phi_0^{-1*}(\alpha+\gamma).
\eeq This result means that fiber preserving canonical diffeomorphisms on $T^*Q$ map graphs of closed 1-forms on $Q$ into graphs of closed 1-forms on $Q$.  
This allows us to  write that, if $W$ is a complete integral of the HJ equation, then the transversal L-submanifold given by the  graph of the section $\sigma^{(\dd W)}$ is transformed, under a fiber preserving canonical diffeomorphism $\phi$ on $T^*Q$, into the graph of $\sigma^{(\alpha')}$ with 
\beq
\label{hj14}
\alpha'=\phi_0^{-1*}(\dd W+\gamma).
\eeq
 This shows that the HJ equation is form covariant under a fiber preserving canonical diffeomorhism $\phi$ on $T^*Q$ such that the corresponding transformation \eqref{hj12} comes with an exact 1-form $\gamma=\dd f$ with $f\in\mathcal F(Q)$, so that  one can define $\alpha'=\dd W'$ with 
 \beq
 \label{hj15}
 W'=\phi_0^{-1*}(W+f).
 \eeq
Denote this group of canonical diffeomorphisms by $G$. If $\phi_s$ is a one parameter group of elements in $G$, the decomposition \eqref{hj16} allows to prove that its infinitesimal generator $X\in\mathfrak X (T^*Q)$ can be written as 
\beq
\label{hj19}
X\,=\,\tilde{X}_0\,+\,X^v,
\eeq
where $\tilde{X}_0$ 
is the canonical lift\footnote{We recall from section \ref{I-sub:exasy}[I]  that, given $X\in\mathfrak X(Q)$, its \emph{canonical lift} on $T^*Q$ is the unique Hamiltonian vector field $\tilde{X}\in\mathfrak X(T^*Q)$ which projects to $X$, i.e. such that $\pi_{Q*}(\tilde{X})=X$ and leaves the canonical 1-form $\theta_Q$ invariant, i.e. such that $L_{\tilde{X}}\theta_Q=0$. If, with respect to the natural cotangent bundle coordinate system it is $X\,=\,X^{a}\del_{q^a}$ with $X^a\in\mathcal F(Q)$, then (see \eqref{I-califf}[I]) one has $\tilde{X}\,=\,-p_b(\del_{q^a}X^b)\del_{p_a}+X^a\del_{q^a}$. } of the infinitesimal generator $X_0\in\mathfrak X(Q)$ of the one parameter group $(\phi_0)_s$ of diffeomorphisms on $Q$ induced by $\phi_s$, and $X^v$ is the infinitesimal generator of the given one parameter group of base invariant canonical diffeomorphisms  which we can denote by $\beta_s=(\phi_s\circ T^*(\phi_0)_s^{-1})$. Notice that $\pi_{Q*}(X^v)=0$, i.e. $X^v$ is vertical.  Moreover, it is possible to see that 
\begin{align}
&i_{\tilde{X}_0}\omega_Q=\dd(i_{\tilde{X}_0}\theta_Q), \nn \\
&i_{X^v}\omega_Q=\dd(\pi_Q^*f) \label{hj17}.
\end{align}
This proves that  the infinitesimal generator  of a one parameter group  $\phi_s$ of elements in $G$ is Hamiltonian, with $X=X_F$ and Hamiltonian function given by 
\beq
\label{hj18}
F=i_{\tilde{X}_0}\theta_Q+\pi^*_Qf.
\eeq
Among the elements in $G$,  it is natural to define a symmetry for the HJ equation those\footnote{It is easy to prove that such symmetry transformations form a group under composition.} satisfying the condition 
\beq
\label{18.10}
\phi^*H=H.
\eeq
 Denote such a group by $G_H$: from \eqref{hj19} and \eqref{hj18} we see that the infinitesimal generator of a one parameter group of elements in $G_H$  is given by those Hamiltonian vector fields $X_F$ such that $\{F,H\}=0$. The function $F$ is the corresponding  constant of the motion.  
We have sketched the proof of the following result. 
\begin{prop}
\label{prop-hj1}
Given a Hamiltonian dynamics $X_H$ on a cotangent bundle manifold $T^*Q$, with corresponding  time independent Hamilton-Jacobi equation as in \eqref{hj5}, if $\phi_s$ is a one parameter group of canonical fiber preserving diffeomorphisms providing a symmetry for the HJ equation  with infinitesimal generator $X\in\mathfrak X(T^*Q)$, then there exists a function $f\in\mathcal F(Q)$ such that $F=i_X\theta_Q+\pi^*_Qf$ is invariant along the integral curves of the dynamics.  
\end{prop}
It is immediate to see that, when expressed in terms of a coordinate chart $(q^a,p_a)$ on $T^*Q$,  the constants of the motion $F$ given in \eqref{hj18} are linear in the $p_a$ variables. It is then possible to prove the following result, which give a suitable converse to the previous one.
\begin{prop}
\label{prop-hj2}
Given a Hamiltonian dynamics $X_H$ on a cotangent bundle $T^*Q$ manifold, if there exists a function $F\in\mathcal F(T^*Q)$ which is invariant along $X_H$, i.e. $\{F,H\}=0$, and can be written as the sum $F=\pi^*_Qf+i_{\tilde{X}_0}\theta_Q$, with $f\in\mathcal F(Q)$ and $\tilde{X}_0$ on $T^*Q$ the  canonical lift of a vector field $X_0$ on $Q$,  then the vector field $X=X_f+\tilde{X}_0$ generates a one parameter group of symmetries for the time independent Hamilton-Jacobi equation corresponding to the Hamiltonian $H$.
\end{prop}
The analogy between such two propositions and the proposition \ref{noeL1bis} within the Lagrangian formalism for a dynamical system is evident. 
It is then natural to look for a generalisation of the HJ time independent formalism which allows to generalise so much  the notion of symmetry within the HJ formalism that any canonical transformation on $T^*Q$ fall in that set, in analogy to  the generalisation given by the notion of Newtonoid symmetry, in order to have a parallel with the notion of symmetry for a dynamics within the symplectic formalism on any symplectic manifold
\begin{example}
\label{exa-hj1}
Consider the Hamiltonian dynamics $X_H$ of the one dimensional harmonic oscillator, with $H=1/2(q^2+p^2)$ on $T^*\R\simeq \R^2$. Although $\Sigma_E=\{(q,p)\,\in\,T^*\R\,:\,H=E\}$ is, for any $E\neq0$, a (maximal) L-submanifold in $T^*\R$, it is not possible to represent it as the graph of an exact 1-form $\dd W$ on the configuration manifold $\R$, since such $W$ would not be defined on the whole $\R$ and, for $p=0$ would $\Sigma_E$ be not transversal (i.e. read caustics).  The solution usually found in textbook, i.e. 
$$
W\,=\,\pm\{E\sin^{-1}(\frac{q}{\sqrt{2E}})\,+\,q\sqrt{2E-q^2}\}
$$
is strictly local and does not allow for a more general analysis of the well known symmetries of the problem.
\end{example}

\subsection{A generalised Hamilton-Jacobi problem}\label{ss:HJ3}
Given a symplectic manifold $(M,\omega)$ and a Hamiltonian dynamics $X_H$ with $H\in\mathcal F(M)$, we define a solution for the \emph{generalised HJ problem} for the Hamiltonian $H$ and the  regular value $E$ of $H$ to be  \emph{any} L-submanifold $\Gamma$ embedded in the energy submanifold given by $\Sigma_E=\{m\in M\,:\,H(m)=E\}$. A maximal L-submanifold $\Gamma\hookrightarrow \Sigma_E$ gives a \emph{global} solution, a foliation of $M$ by global solutions for definite values of $E$ gives a \emph{complete} solution to the HJ generalised problem. 

Following \cite{mmm-hj}, it is possible to prove that the covariance group of the generalised HJ problem consists of the group of canonical transformations on $M$, and also that any function $F\in\mathcal F(M)$ which satisfies the condition $\{F,H\}=0$ generates a vector field $X_F$ which is the infinitesimal generator of a group of transformations mapping generalised solutions of the corresponding HJ problem into generalised  solution of the same problem.  
\begin{example}\label{exa-hj2}
Consider the Hamiltonian dynamics on $\R^4\simeq T^*\R^2$ given by the homogeneous harmonic oscillator with 
$$
H\,=\,\frac{1}{2}(q_1^2+q_2^2+p_1^2+p_2^2).
$$
The Energy surface 
$$
\Sigma_E\,=\,\{m\in \R^4\,:\,H=E\}
$$
is clearly diffeomorphic to the Euclidean $\mathrm S^3$ in $\R^4$ with radius $r=\sqrt{2E}$. 

In order to construct solutions to the generalised HJ equation corresponding to the Hamiltonian $H$, that is in order to determine a L-submanifold embedded into $\Sigma_E$ we recall the example \ref{I-exemplum-mom1}[I]. Within the set of quadratic homogeneous functions on $\R^4$, the elements  $\{u_j\}_{j=1,\dots,3}$ 
\begin{align}
&u_1\,=\,\frac{1}{2}\,(q_1q_2+p_1p_2), \nn \\ 
&u_2\,=\,\frac{1}{2}\,(q_1p_2-q_2p_1), \nn \\
&u_3\,=\,\frac{1}{2}\,(q_1^2+p_1^2-q_2^2-p_2^2)
\label{eqsy8-b}
\end{align}
defined in \eqref{I-eqsy8}[I] provide the commutant of the Hamiltonian $H$, i.e. 
\beq
\label{commu1}
\{H,u_j\}=0
\eeq so we can define the submanifolds
\beq
\label{commu2}
\Gamma_j\,=\,\{m\,\in\,\R^4\,:\,H(m)=E, \,\,u_j=a_j\}
\eeq
for suitable constants $a_j$. It is clear from \eqref{commu1}  that each $\Gamma_j$ is a solution for the generalised HJ problem corresponding to $H$. Moreover, it is immediate to see that, for a fixed $j$,  the functions $u_k$ with $k\neq j$ generate (the corresponding Hamiltonian vector fields $X_{u_j}$ are in \eqref{I-eqsy9}[I]) canonical symmetries for the problem. 

By noticing that the constants of the motion $\varphi$ in \eqref{exL5} and $\varphi'$ in \eqref{exL5-1} are transformed under the Legendre map into  the function $u_1$ and $u_3$, we see that the generalised approach allows to recover within the HJ formalism the analogue of the notion of Newtonoid symmetries for the harmonic oscillator within the Lagrangian formalism (see the example \ref{exemplum-Lag2}). 
 
\end{example}

We conclude this section by recalling that different notions of generalised solutions for the HJ problem have been developed. We refer the reader to \cite{ghjt06, ghjt16}, where a solution to the generalised HJ problem in $T^*Q$ is given by any 1-form $\alpha=\alpha_j\dd q^j$ on $Q$ such that its range as a section of $T^*Q$, i.e. the submanifold $\Sigma_{\alpha}$ given by the conditions $$f_j^{(\alpha)}=p_j-\alpha_j(q)=0$$ is invariant under the flow generated by the Hamiltonian vector field $X_H$. Such notion clearly allows to extend the class of solutions given by \eqref{hj11}-\eqref{hj12}. 
A (generalised) solution $\alpha=\alpha_j\dd q^j$ is complete if it depends on the parameters $u\in U\subset \R^N$ such that the map (enlarging the class of solutions given by \eqref{2m.3})  $$\alpha\,:\,Q\times U\to T^*Q$$ is a local diffeomorphism. 
We nonetheless underline that the analysis in \cite{ghjt06, ghjt16} allows to study the HJ (generalised) problem  
also on $TQ$ when the dynamics has a Lagrangian formulation, even in the case the Lagrangian is singular.

\subsection{Reduction in the Hamilton-Jacobi formalism}
\label{sub:rhj}
Following our description of the HJ equation within the formalism of the jet bundles, it is clear that reduction procedures for p.d.e.'s are quite different from the general reduction scheme we have described for ode's, i.e. for vector fields. Moreover, as pointed by  Dirac in the quote we reported, a solution of the HJ equation gives a family of solutions for a classical dynamics. These profound differences induce us to   
 limit  ourselves to describe  specific examples which focus on 
interesting aspects of the reduction procedure for the HJ equation.  

\begin{example}
\label{ehj1}
We start by considering the dynamics described by the Hamiltonian 
\beq
\label{22m.1}
H\,=\,\frac{1}{2}(p_x^2+p_y^2)+V(x)
\eeq
on $(T^*Q, \omega_Q)$ with $Q=\R^2$. The time evolution is given by combining two independent motions. It is clear that one has \begin{align*}&p_y(t)=k, \\ &y(t)=y(0)+tk\end{align*} with constant $k\in\R$. Recalling the general theory on the solutions of the HJ in terms of characteristics, we see from \eqref{2m.2} that, if $W$ is a solution to the HJ p.d.e. associated to \eqref{22m.1}, then 
$$
\frac{\del W}{\del y}=k
$$
and this reads 
$$
W(x,y)=ky+\tilde{W}(x),
$$
 with $\tilde W(x)$ solving the HJ equation (which is now an ordinary differential equation) 
$$
\frac{1}{2}((\frac{\dd\tilde W}{\dd x})^2+k^2)+V(x)=E.
$$
associated to the Hamiltonian 
$$
\tilde H=\frac{1}{2}(p_x^2+k^2)+V(x)
$$
on $T^*\tilde Q$, with $\tilde Q$ the quotient of $Q$ by the vector field $\del/\del y$.  The reduced Hamiltonian $\tilde H$ depends, as we pointed out in the general description of symplectic reduction, by the value $k=p_y$ of the invariant function on $T^*Q$. 
\end{example}
This example suggests that the usual method of solving the HJ equation by separation of variables, when the Hamiltonian has so called cyclic coordinates, can be seen as an example of symplectic reduction on a phase space. A generalisation of this procedure is not completely straightforward. 
In particular, we notice that, even when the map $\phi$ on $T^*Q$  (see \eqref{18.10}) is a symmetry for the HJ equation associated to the Hamiltonian $H$, the possibility of defining a HJ equation on a reduced carrier manifold depends on the possibility that such a reduced manifold has an exact  cotangent bundle structure. We already discussed about this problem when describing the reduction within the Lagrangian formalism in section \ref{sss:red}, rephrasing what we discussed in section \ref{I-subsub:liesym}[I].  In the following lines we  describe a reduction procedure in a less specific example.

\begin{example}
\label{ehj2}
Consider the dynamics on the phase space $(T^*Q,\omega_Q)$ with $Q=\R^2$ given by the Hamiltonian
\beq
\label{22m.3}
H\,=\,\frac{1}{2}(p_x^2+p_y^2)+V(x-y),
\eeq
i.e. a dynamics driven by a force depending by the generalised positions $(x,y)$ only through the difference $x-y$. The additive group $(\R, +)$ acts upon $Q$ via 
\beq
\label{29m2}
x\mapsto x+s, \qquad y\mapsto y+s
\eeq
(here $s\in\R$  parametrises the group) with infinitesimal generator $X_Q=\del_x+\del_y$ on $Q$. Its lifted action on $T^*Q$ is Hamiltonian, with  $X=\del_x+\del_y$ on $T^*Q$ and $i_X\omega_Q=\dd u$ with $u=p_x+p_y$. This action gives an equivariant momentum map $\mu\,:\,T^*Q\,\to\,\R$, which can be written as
$$
\mu\,:\,(x,y,p_x,p_y)\qquad\mapsto\qquad u=p_x+p_y.
$$
The symplectic reduction driven by an equivariant momentum map $\mu$ has been described at length. The condition 
$$
p_x+p_y=k
$$
 defines a coisotropic submanifold $i_k:N_k\hookrightarrow T^*Q$, and the Hamiltonian vector field $X_H$ is tangent to it. The vector field $X$ on $N_k$ gives the kernel of the closed $i_k^*\omega_Q$ 2-form on $N_k$. The quotient $N_k/X$ turns to be diffeomorphic to $T^*(Q/X_Q)$.  
 
 We now describe such a quotient in terms of an adapted coordinate system on $Q$. We aim at introducing a diffeomorphism $\phi\,:\,Q\,\to\,Q$ written as 
 \beq
 \label{22m.4}
 \phi: (x,y)\quad\mapsto\quad(a,q)
 \eeq
 such that $\phi_*(X_Q)=\del_a$. It is evident that this only condition does not uniquely select $\phi$. We indeed consider begin by considering 
 \beq
 \label{22m.5}
 a=\frac{1}{2}(x+y),\qquad q=\frac{1}{2}(x-y)
 \eeq
 so to have
 $$
 \phi_*(\frac{\del}{\del x}+\frac{\del}{\del y})=\frac{\del}{\del a}, \qquad \phi_*(\frac{\del}{\del x}-\frac{\del}{\del y})=\frac{\del}{\del q}.
 $$ 
The cotangent lift of such a diffeomorphism on $Q$ provides a canonical diffeomorphism $\Phi$ on $T^*Q$, whose coordinate expression for the fiber variables is   
$$
p=p_x-p_y, \qquad u=p_x+p_y
$$
with clearly 
$$
\Phi_*(\dd x\wedge\dd p_y+\dd y\wedge\dd p_y)=\dd q\wedge\dd p+\dd a\wedge \dd u.
$$
The reduced phase space $T^*(Q/X_Q)$ is diffeomorphic to $T^*\tilde Q$ with $\tilde Q\simeq \R$. A (global) coordinate chart on $T^*\tilde Q$ is given by $(q,p)$ with symplectic structure $\omega_{\tilde Q}=\dd q\wedge \dd p$.
We write the Hamiltonian in the adapted coordinate system as
$$
\Phi^{-1*}(H)\,=\,\frac{1}{2}\left(\frac{(p+u)^2}{4}+\frac{(p-u)^2}{4}\right)+V(q)
$$
and its reduction to $N_{k}/X$ as 
$$
\tilde H=\frac{1}{2}\left(\frac{(p+k)^2}{4}+\frac{(p-k)^2}{4}\right)+V(q).
$$
As we showed in the previous example, such reduced Hamiltonian $\tilde H$ depends on the real value $k$ which is a fixed point for the momentum map $\mu$, and reads a HJ equation on $\tilde Q$ which is again an ordinary differential equation, namely 
\beq
\label{22m.2}
\frac{1}{8}\left((\frac{\dd \tilde W}{\dd q}+k)^2+(\frac{\dd \tilde W}{\dd q}-k)^2\right)+V(q)=E.
\eeq
It becomes immediate to see that, if $\tilde W$ solves \eqref{22m.2}, then  a solution $W$ to the HJ equation on $Q$ associated to the Hamiltonian $H$ in \eqref{22m.3} can be written as 
$$
W=ak+\tilde W(q)\qquad\Rightarrow\qquad \Phi^*(W)=\frac{k}{2}(x+y)+\tilde W(x-y).
$$
It is clear that this class of solutions depends on the choice of the diffeomorphism $\phi$ on $Q$ in \eqref{22m.4}-\eqref{22m.5}. 

Different choices of coordinates on $Q$ adapted to the quotient $Q/X_Q$ will give different classes of solutions for the HJ equation. Defining a coordinate system on $Q$ which is adapted to the quotient by $X_Q$ amounts to define a connection with respect to the fibration given by the quotient $\pi:Q\to Q/X_Q$. At each point in $Q$ the vertical subspace is given by $X_Q$, while the horizontal complement is suitably selected as $\alpha\del_x+\beta\del_y$ so that the diffeomorphism $\phi$ in \eqref{22m.4} gives
\begin{align}
&\phi_*(\frac{\del}{\del x}+\frac{\del}{\del y})=\frac{\del}{\del a}, \nn \\
&\phi_*(\alpha\frac{\del}{\del x}+\beta\frac{\del}{\del y})=\frac{\del}{\del q} \nn 
\end{align}
with $\alpha\neq\beta\in\mathcal F(Q)$. The corresponding canonical lift provides 
\begin{align}
&u=p_x+p_y, \label{29m1}\\
&p=\alpha\,p_x+\beta\,p_y \nn 
\end{align}
for the fiber coordinates, so that again $\Phi_*(\dd x\wedge\dd p_y+\dd y\wedge\dd p_y)=\dd q\wedge\dd p+\dd a\wedge \dd u$. The first relation \eqref{29m1} shows that the symplectic reduction is based on the invariance of the momentum $u$. When the the horizontal subspace spanned by $\del/\del q$ is selected with $\alpha\neq\beta\in\R$, i.e. the connection is invariant under the action \eqref{29m2} of the translation  group, then one has  
$$
q=\frac{1}{\beta-\alpha}(y-x), \qquad \qquad a=\frac{1}{\beta-\alpha}(\beta x-\alpha y).
$$
In such a case, one has
$$
\Phi^{-1*}(H)\,=\,\frac{1}{2}\left((\frac{\beta u-p}{\beta-\alpha})^2+
(\frac{p-\alpha u}{\beta-\alpha})^2\right)+V(q)
$$
and its reduction to $N_k/X$ as 
$$
\tilde H\,=\,\frac{1}{2}\left((\frac{\beta k-p}{\beta-\alpha})^2+
(\frac{p-\alpha k}{\beta-\alpha})^2\right)+V(q), 
$$
giving the HJ equation 
\beq
\label{29m3}
\frac{1}{2(\beta-\alpha)^2}\left((\beta k-\frac{\dd \tilde W}{\dd q})^2+ (\frac{\del\tilde W}{\dd q}-\alpha k)^2\right)+V(q)=E
\eeq
for $\tilde W(q)$. If $\tilde W$ solves \eqref{29m3} on $T^*\tilde Q$, then
$$
W=ak+\tilde W
$$
solves the HJ p.d.e. corresponding to $H$ on $T^*Q$.

\end{example}
The path described in the previous example can  generalised to higher dimensional phase spaces and to Hamiltonian which are not directly of the mechanical type, i.e. given by the sum of a non degenerate quadratic form for the momenta and a potential energy term depending on the position. An interesting analysis is in \cite{ge}, while in \cite{vaqu}  the ambiguity in the choice of a suitable coordinate system on $Q$ is formulated as a reconstruction problem where the introduction of  a connection on $Q/G$ provides a magnetic term 2-form to the canonical symplectic structure on the reduced phase space. This paper also studies how a reduction procedure for the HJ equation allows to have generalised solutions which are not   transversal Lagrangian submanifolds in $T^*Q$. 

We close our presentation of examples on the reduction of the HJ equation by presenting the analysis of the reduction of the free motion into an interacting one.
\begin{example}
\label{ex18.1}
In section \ref{I-sss:free}[I] we have described the free motion in three dimensions by considering the configuration space $Q=\R^3$ as the set of symmetric $2\times2$ matrices $X=X^T$ and the dynamics described by the Lagrangian
$$
\cL=\frac{1}{2} {\rm Tr}(\dot X\dot X)
$$ 
on $TQ$. The corresponding Hamiltonian on $T^*Q$ is $$
H=\frac{1}{2} {\rm Tr}(PP),
$$ with symplectic structure ($P_a^T=P_a$ on $T^*Q$)
$$
\omega_Q\,=\,\dd X^a\wedge\dd P_a.
$$
The explicit solutions of the dynamics are given by 
$$
X(t)=X(0)+tP,
$$
so that we can write a solution of the corresponding HJ equation for $t>0$ upon integrating the action functional along such solutions (see \eqref{15.1}):
\beq
\label{18.20}
S(t,X,X_0)\,=\,\frac{1}{2t}{\rm Tr}(X-X_0)^2.
\eeq
Since the angular momentum $M=[X,\dot X]$ is conserved along the motion, the condition 
$$
{\rm Tr} \,M^2=2L^2
$$
fixes an invariant manifold $\Sigma$ for the evolution which is not a vector subspace. The corresponding reduced dynamics gives, as described in \ref{I-sss:free}[I], Calogero-Moser dynamics on a symplectic $T^*(Q/G)$ (with $G$ the rotation group acting on $Q$) and the restriction of $S$ to the solutions giving 
$$
{\rm Tr}(X^2X_0^2-(XX_0)^2)=2L^2
$$
is proven (see \cite{ccgm}) to solve the HJ equation associated to the reduced dynamics.

\end{example} 

\section{Concluding comments}
\label{ende}
We have come to an end of our journey. Through the first and the second part of this paper we have declined the notion of symmetry within the Poisson, the Hamiltonian,  the Lagrangian and the Hamilton-Jacobi pictures of classical dynamics, and the reason by which we used the word  \emph{pictures} is far from being accidental, since we have presented such formalisms in relations with the Heisenberg, the Schr\"odinger, the Dirac \emph{pictures} for quantum mechanics. Our aim has been mainly to describe how the notion of symmetry allows for interesting reductions of a given dynamics on a carrier space, and how such reduction deeply depends on the pictures we consider, since it needs to be compatible with the geometric structures characterising the picture itself. In particular, since linearity and unitarity are  structural aspects of quantum mechanical dynamics, we have characterised linear (classical) dynamics (among them, free dynamics), and presented finite rank quantum systems in terms of unitary vector fields on a carrier manifold whose Hilbert  structure we have studied in terms of compatible metric and symplectic tensors.  We have showed how linear dynamics  may give rise to different non linear reduced dynamics. This suggests a way to analyse how interactions can geometrically formulated via a reduction procedure, while, somehow reversing such path, one can study how an unfolding procedure may provide linearised dynamics which can be indeed be quantised within the Heisenberg-Dirac picture, or within the Schr\"odinger picture, starting from the Hamilton-Jacobi formalism.   

Along this research line, we leave to a future work an intrinsic description of reduction and unfolding  in quantum mechanics within the formalism of finite and infinite dimensional Hilbert manifolds, and a geometric description of the Lagrangian formalism in quantum mechanics within a groupoidal approach, evolving \cite{hola}.  


\end{document}